\documentclass[prl,letterpaper,english,reprint,nofootinbib,aps,superscriptaddress,showpacs,showkeys]{revtex4-1}

\usepackage{babel,calc,amsmath,amsthm,amssymb,graphicx,subfigure,xcolor,comment}
\usepackage{mathdots}
\usepackage[T1]{fontenc}
\usepackage[unicode=true]{hyperref}
\hypersetup{
	colorlinks=true,       		
	linkcolor=red,          	
	citecolor=blue,             
	urlcolor=blue,          
}

\usepackage{booktabs}

\usepackage{makecell}
\usepackage{braket}
\usepackage[normalem]{ulem}

\newif\ifdebug

\debugtrue

\ifdebug
\definecolor{zhliu}{rgb}{0.48, 0.12, 0}

\newcommand{\note}[1]{\textcolor{zhliu}{#1}}
\newcommand\delete{\bgroup\markoverwith{\textcolor{zhliu}{\rule[0.5ex]{2pt}{0.8pt}}}\ULon}

\else

\newcommand{\note}[1]{\ignorespaces}
\newcommand{\delete}[1]{\ignorespaces}

\fi

\begin{document}

	\renewcommand{\figurename}{Fig.}
	
	\title{The Iteration Formula of $(n,2,d)$ Full-correlated Multi-component Bell Function and Its Applications}

	\author{Hui-Xian Meng}
   \affiliation{School of Mathematics and Physics, North China Electric Power University, Beijing 102206,
      People's Republic of China}

 \author{Yu Zhang}
   \affiliation{School of Mathematics and Physics, North China Electric Power University, Beijing 102206,
      People's Republic of China}

\author{Xing-Yan Fan}
   \affiliation{Theoretical Physics Division, Chern Institute of Mathematics, Nankai University, Tianjin 300071,
      People's Republic of China}

   \author{Jie Zhou}
      \affiliation{Theoretical Physics Division, Chern Institute of Mathematics, Nankai University, Tianjin 300071,
      People's Republic of China}
      \affiliation{College of Physics and Materials Science, Tianjin Normal University, Tianjin 300387, People's Republic of China}

   \author{Wei-Min Shang}
      \affiliation{Theoretical Physics Division, Chern Institute of Mathematics, Nankai University, Tianjin 300071,
      People's Republic of China}
      \affiliation{School of Science, Tianjin Chengjian University, Tianjin 300384, People's Republic of China}

   \author{Jing-Ling~Chen}
   \email{chenjl@nankai.edu.cn}
   \affiliation{Theoretical Physics Division, Chern Institute of Mathematics, Nankai University, Tianjin 300071,
      People's Republic of China}
	
	\date{\today}

\begin{abstract}
    It is very difficult and important to construct Bell inequalities for $n$-partite, $k$-settings of measurement, and $d$-dimensional $(n,k,d)$ systems. Inspired by the iteration formula form of the Mermin-Ardehali-Belinski{\u{\i}}-Klyshko (MABK) inequality, we generalize the multi-component correlation functions for bipartite $d$-dimensional systems to $n$-partite ones, and construct the corresponding Bell inequality. The Collins-Gisin-Linden-Massar-Popescu inequality can be reproduced by this way. The most important result is that for prime $d$ the general Bell function in full-correlated multi-component correlation function form for $(n,2,d)$ systems can be reformulated in iteration formula by two full-correlated multi-component Bell functions for $(n-1,2,d)$ systems. As applications, we recover the MABK inequality and the most robust coincidence Bell inequalities for $(3,2,3),(4,2,3),(5,2,3),$ and $(3,2,5)$ Bell scenarios with this iteration formula. This implies that the iteration formula is an efficient way of constructing multi-partite Bell inequalities.
 In addition,  we  also give some new Bell inequalities with the same robustness but inequivalent to the known ones.
\end{abstract}
	
	
\maketitle

\section{Introduction}

Bell nonlocality is an inherent attribute of quantum mechanics, and clearly demonstrated by the fact that some  statistical predictions of quantum mechanics can violate the  constraints  imposed by the local hidden variable
(LHV) models, known as  Bell inequalities \cite{RMP,Nonlocality}.  These inequalities are of paramount importance in the conceptual foundations of
quantum mechanics. For the simplest composite quantum system,
namely, a system of two two-dimensional particles (or two qubits), the original Bell inequality was presented in 1964 \cite{Bell2}. Subsequently,
 the  Clauser-Horne-Shimony-Holt (CHSH) inequality was formulated for allowing more flexibility in local measurement configurations \cite{CHSH}. Since then, the investigation of Bell inequalities turn to more complicated situations \cite{
M,A,BK,Bell's theorem,CGLMP,qutrits1,Coincidence,Multicomponent,Chen324,Mermin3,Platonic}, namely more observers, or more measurement settings for each observer, or higher dimensional systems. The  research topic of Bell inequalities \cite{Questions} includes but not limits to the construction of  generic Bell inequalities \cite{structure,L. B. Fu,SLK,SCL,Avenues,XYFan}, tight Bell inequalities \cite{Tightn22,Lluis Masanes,Tightn2d,completed}, relevant Bell inequalities \cite{relevant22,relevantk23}, and robust Bell inequalities for detecting the resource states with noises \cite{3n2,n23}, and their applications \cite{Quantum cryptography,QKD,3bit2,Strong nonlocality}.

In this paper, we restrict the number of measurement settings for per observer to two. Mermin,
Belinskii, and Klyshko separately generalized the CHSH
inequality to  $n$ ($n>2$) particles, and obtained the Mermin-Ardehali-Belinski{\u{\i}}-Klyshko (MABK) inequality both in correlation functions and joint probabilities forms \cite{M,A,BK}. To the best of our knowledge, it is the most robust inequality for revealing the nonlocal property of the $n$-qubit maximally entangled state with the white noise. For bipartite $d$-dimensional systems, Collins {\it et al.} suggested the Collins-Gisin-Linden-Massar-Popescu (CGLMP) inequality in joint probabilities form \cite{CGLMP}, which can reduce to the CHSH inequality in the case $d=2$.
To generalize the CGLMP inequality to three particles for $d = 3$, Chen {\it et al.} constructed a Bell inequality in coincidence probabilities \cite{Coincidence}. Whereafter, Meng {\it et al.} introduced the general formula of coincidence Bell inequalities for arbitrary $(n,2,3)$-scenario \cite{n23}, i.e., $n$-party, $2$-setting, and $3$-dimensional scenario. Deriving from the most robust $(n-1)$-party coincidence ones, they established tight  $(4,2,3)$ and $(5,2,3)$  coincidence Bell inequalities. As far as we know, the above Bell inequalities are the most robust ones for the corresponding scenarios in the sense that they can detect the nonlocal  property of the noisy $n$-qudit Greenberger-Horne-Zeilinger (GHZ) state with the most noises.

For arbitrary high-dimensional systems, Son {\it et al.} suggested a generic Bell inequality
and its variant, which was called the Son-Lee-Kim (SLK) inequality \cite{SLK},  could reduce to the CGLMP inequality by choosing appropriate coefficients \cite{SCL}.  In the SLK inequality, the measurement outcomes of the observables were taken as the $d$th roots
of unity over the complex field. Chen {\it et al.}  introduced $d$ $N$-dimensional unit vectors as the measurement outcomes of the observables with $N=d-1$ \cite{Multicomponent}, where these vectors satisfied
$
\sum_{k=0}^{d-1}{\rm \mathbf{v}}_k={\rm \mathbf{0}},\ {\rm \mathbf{v}}_i\cdot{\rm \mathbf{v}}_j=-\frac{1}{N},i\neq j, i,j=0,1,...,d-1,$
and developed Bell inequality  for bipartite $d$-dimensional systems  with the multi-component correlation functions.

In this paper, we present the generic Bell function for $n$-partite $d$-dimensional systems with  full-correlated multi-component correlation functions, and prove that for prime $d$ it can be simplified to an iteration formula  with two $(n-1,2,d)$ full-correlated multi-component Bell functions. Inspired by the representation of projectors with  the observable, we use the first component of the multi-component Bell functions to construct the general $(n,2,d)$ Bell inequality. In fact, we give a kind of way that construct $(n,2,d)$ Bell inequality with $(n-1,2,d)$ ones using the iteration formula for prime $d$. By taking appropriate coefficients, it can be reformulated to the MABK inequality, the CGLMP inequality, the most robust $(n,2,3)$ coincidence Bell inequalities \cite{n23}, and the most robust $(3,2,5)$ Bell inequality presented in \cite{Chen324}. With the iteration formula, we have not found more robust Bell inequality. Nevertheless, we  give ones with the same robustness but inequivalent to the known ones.

\section{The iteration formula of $(n,2,d)$ full-correlated multi-component Bell function}

The Bell-type scenario considered in this paper is the  $(n,2,d)$-scenario: There are $n$ space-separated observers labeled by $A,\cdots,$$B,C$, the $i$-th observer $X$ carries out two settings referred to $X_{0},X_{1}$, and the results of  measurement $X_{j}$ are denoted by  $d$ $N$-dimensional  unit vectors \cite{Multicomponent}:
\begin{eqnarray}{\label{v0}}
& &{\rm \mathbf{v}}_0=(1,0,0,...,0),\nonumber\\
& &{\rm \mathbf{v}}_1=\left(-\frac{1}{N},\frac{\sqrt{N^2-1}}{N},0,...,0\right),\nonumber\\
& &{\rm \mathbf{v}}_2
=\left(-\frac{1}{N},-\frac{1}{N}\sqrt{\frac{N+1}{N-1}},
\sqrt{\frac{(N-2)(N+1)}{N(N-1)}},0,...,0\right),\nonumber\\
& &...,\nonumber\\
& &{\rm \mathbf{v}}_{N-1}=\left(-\frac{1}{N},-\frac{1}{N}\sqrt{\frac{N+1}{N-1}},
-\frac{1}{N}\sqrt{\frac{(N+1)N}{(N-1)(N-2)}},\right.\nonumber\\
& &\left. ...,
-\frac{1}{N}\sqrt{\frac{(N+1)N}{3\times 2}},\frac{1}{N}\sqrt{\frac{(N+1)N}{2\times 1}}\right),\nonumber\\
& &{\rm \mathbf{v}}_{N}=\left(-\frac{1}{N},-\frac{1}{N}\sqrt{\frac{N+1}{N-1}},
-\frac{1}{N}\sqrt{\frac{(N+1)N}{(N-1)(N-2)}},\right.\nonumber\\
& &\left. ...,
-\frac{1}{N}\sqrt{\frac{(N+1)N}{3\times 2}},-\frac{1}{N}\sqrt{\frac{(N+1)N}{2\times 1}}\right),
\end{eqnarray}
with $N=d-1$, which satisfy the following properties:
\begin{eqnarray}{\label{v1}}
\sum_{k=0}^{d-1}{\rm \mathbf{v}}_k={\rm \mathbf{0}},\ {\rm \mathbf{v}}_i\cdot{\rm \mathbf{v}}_j=-\frac{1}{N},i\neq j, i,j=0,1,...,d-1.\ \ \
\end{eqnarray}
For $d=2$, there are just two variables with only one component, i.e., ${\rm \mathbf{v}}_{0}=(1),{\rm \mathbf{v}}_{1}=(-1)$,
obtained from a measurement.
Then the observable $X$ with $d$ outcomes can be represented by
$
X=\sum_{i=0}^{N}{\rm \mathbf{v}}_i\Pi_i,
$
where $\{\Pi_i\}$ are rank one projectors on $d$-dimensional complex Hilbert space $\mathbb{C}^d$, and satisfy $\Pi_i\Pi_{i'}=\delta_{i,i'}\Pi_i$ and $\sum_{i}\Pi_i=I_{d}$, i.e., $\{\Pi_i\}$ corresponds to an orthonormal basis(ONB). If we use $x_{j}$ to denote the outcome ${\rm \mathbf{v}}_{j}$ of $X$, then $(2\times d)^n$ probability terms $P(A_{i}=a_{i},\cdots,B_{j}=b_{j},C_{k}=c_{k})$ are obtained.

For two observers, Alice and Bob, if their measurement results  $a$ and $b$ are less than $N$,
then  a vector ${\rm \mathbf{v}}_{a+b}$ is used for the correlation between Alice
and Bob (${\rm \mathbf{v}}_{a+b}$ understood as ${\rm \mathbf{v}}_{t}$ , where $t={\rm Mod}[a+b,d]$ is the reminder of $a+b$ modulus $d$), i.e.,
$
{\rm \mathbf{v}}_{a}\circ {\rm \mathbf{v}}_{b}={\rm \mathbf{v}}_{{\rm Mod}[a+b,d]}.
$ If there is no risk of confusion, we also write it as ${\rm \mathbf{v}}_{a}{\rm \mathbf{v}}_{b}={\rm \mathbf{v}}_{{\rm Mod}[a+b,d]}$ for short.
Based on this, Ref. \cite{Multicomponent}  constructed multi-component correlation
functions,
$
\vec{Q}_{ij}=\sum_{a_i,b_j=0}^{N}{\rm \mathbf{v}}_{a_i}{\rm \mathbf{v}}_{b_j}P(A_i=a_i,B_j=b_j)=\sum_{t=0}^{N}{\rm \mathbf{v}}_tP(A_i+B_j=t)=(Q_{ij}^{(0)},Q_{ij}^{(1)},...,Q_{ij}^{(d-2)}).$

In this paper, we generalize them to the  $n$-party case, and define the $n$-party multi-component correlation
functions as
\begin{eqnarray}{\label{Qn}}
& &\vec{Q}_{i\cdots j k}\nonumber\\
&=&\sum_{m,\cdots,n,s=0}^{N}{\rm \mathbf{v}}_{m}\cdots {\rm \mathbf{v}}_{n}{\rm \mathbf{v}}_{s}P(A_i=m,\cdots,B_j=n,C_k=s)\nonumber\\
&=&\sum_{t=0}^{N}{\rm \mathbf{v}}_tP(A_i+\cdots+B_j+C_k=t)\nonumber\\
&=&\left(Q_{ij\cdots k}^{(0)},Q_{i\cdots j k}^{(1)},...,Q_{i\cdots j k}^{(d-2)}\right),
\end{eqnarray}
where $P(A_i=m,\cdots,B_j=n,C_k=s)$ is the joint probability of $A_i$ obtaining
outcome $m$, $\cdots$, $B_j$ obtaining outcome $n$, and $C_k$ obtaining outcome $s$,
$P(A_i+\cdots+B_j+C_k=t)=\sum_{{\rm Mod}\left[m+\cdots+n+s,d\right]=t}P(A_i=m,\cdots,B_j=n,C_k=s),$
is just the {\it coincidence} probability using the same idea as in  Ref. \cite{Coincidence},
and $Q_{i\cdots j k}^{(l)}$ represents the $(l+1)$-th
component of the vector correlation function $\vec{Q}_{i\cdots j k}$.

The quantum prediction for the joint probability reads
$P(A_i=m,\cdots,B_j=n,C_k=s)=\langle\psi|\Pi_{A_i=m}\otimes\cdots\otimes\Pi_{B_j=n}\otimes\Pi_{C_k=s}|\psi\rangle,$
where $|\psi\rangle$ is the pure state in quantum system $\mathbb{C}^d$, $\Pi_{A_i=m}=\mathcal{U}_A^{\dag}|m\rangle\langle m|\mathcal{U}_A$ is the
projector of Alice for the $i$-th measurement, $\cdots$, $\Pi_{B_j=n}$, and $\Pi_{C_k=s}$ are
defined similarly.

A general full-correlated multi-component Bell function for $(n,2,d)$-scenario has the form
\begin{eqnarray}{\label{Ind}}
\mathcal{I}_{n,d}&=&\sum_{i,\cdots,j,k=0,1}\omega_{i,\cdots,j,k}A_i\cdots B_j C_k\nonumber\\
&=&\sum_{i,\cdots,j,k=0,1}\omega_{i,\cdots,j,l}\vec{Q}_{i\cdots j k},
\end{eqnarray}
where $\omega_{i,j,\cdots,k}$ is the real and linear combination of ${\rm \mathbf{v}}_{0},{\rm \mathbf{v}}_{1},\cdots,{\rm \mathbf{v}}_{N}$.
For prime $d$, projectors $\Pi_i$ can be represented with the observable $X=\sum_{i=0}^{N}{\rm \mathbf{v}}_i\Pi_i$ as
$
{\rm \mathbf{v}}_0\Pi_i=\frac{1}{d}({\rm \mathbf{v}}_{0} I+{\rm \mathbf{v}}_{-i} X+
{\rm \mathbf{v}}_{-2i} X^2+\cdots+{\rm \mathbf{v}}_{-Ni} X^{N}),
$
where
$
X^j=\sum_{t=0}^{N}{\rm \mathbf{v}}_{{\rm Mod}[jt,d]}\Pi_{t},\ j=1,2,\cdots,N.
$
See Appendix A for the details of the proof. Inspired by this representation, we now  define a Bell quantity
$
\mathcal{B}_{n,d}=\sum_{i,j,\cdots,k=0,1}\omega_{i,j,\cdots,k}'Q_{ij\cdots k}^{(0)}
$
 as the first component of $\mathcal{I}_{n,d}$ for any $n\geq 2$  and $d\geq 2$,
where $\omega_{i,j,\cdots,k}'$ are real coefficients. Then it can be reduced to
\begin{eqnarray}{\label{GCI}}
\mathcal{B}_{n,d}=\sum_{i,\cdots,j,k}\sum_{r=0}^{N}\omega_{j,r}P\left(A_{i}+\cdots+B_{j}+C_{k}=r\right),\ \ \
\end{eqnarray}
which is just the general coincidence Bell quantity for $(n,2,d)$-scenario in probability form, as the coincidence $(n,2,3)$-Bell inequality  considered in Ref. \cite{Coincidence}, where $(i,\cdots,j,k)$ goes through all $2^n$ possible measurements for the $n$ observers, $\omega_{j,r}$'s are the real weight coefficients. Therefore, we obtain a coincidence Bell inequality
 $
 \mathcal{B}_{n,d}\overset{{\rm LHV}}{\leq}L,
$
with $L$ being the upper bound of the Bell quantity $\mathcal{B}_{n,d}$ in the LHV theory.

In Ref. \cite{Lluis Masanes}, Masanes defined the equivalence of Bell inequalities. Similarly, all these multi-component Bell functions $\mathcal{I}_{n,d}$  can also be grouped in families of equivalent classes. Two multi-component Bell functions are {\it equivalent} if we can
transform one into the other by composing the following symmetry transformations: relabeling of the party index,  the observable index in each particle, or the outcome index for one observable.  For instance,  the symmetry transformations on  $(2,2,d)$-scenario contain:
Party exchange (party symmetry): $A_iB_j\mapsto A_jB_i$;
Observable exchange (observable symmetry): $A_iB_j\mapsto A_{\bar{i}}B_{\bar{j}}$,  where $\bar{0}=1,\bar{1}=0$;
Relabeling of outcomes (outcome symmetry):
$A_iB_j\mapsto {\rm \mathbf{v}}_{a_i}A_i{\rm \mathbf{v}}_{b_j}B_j,$ where $a_i,b_j\in\{0,1,,\cdots,d-1\}$.
In addition, a Bell function is called {\it {symmetric}} if it is invariant under any party  symmetric transformation. For example, in the case
 $n=2,d=2$,
the multi-component Bell function
\begin{eqnarray}{\label{I22}}
\mathcal{I}_{2,2}&=&-\frac{1}{2}({\rm \mathbf{v}}_{1}A_0B_0+{\rm \mathbf{v}}_{1}A_0B_1+{\rm \mathbf{v}}_{1}A_1B_0-{\rm \mathbf{v}}_{1}A_1B_1)\nonumber\\
&=&-\frac{1}{2}({\rm \mathbf{v}}_{1}\vec{Q}_{00}+{\rm \mathbf{v}}_{1}\vec{Q}_{01}+{\rm \mathbf{v}}_{1}\vec{Q}_{10}-{\rm \mathbf{v}}_{1}\vec{Q}_{11}),
\end{eqnarray}
is  symmetric, and the corresponding Bell inequality is
$
\mathcal{B}_{2,2}=\frac{1}{2}[P(A_0+B_0=0)-P(A_0+B_0=1)+P(A_0+B_1=0)-P(A_0+B_1=1)+P(A_1+B_0=0)-P(A_1+B_0=1)-P(A_1+B_1=0)+P(A_1+B_1=1)]\overset{{\rm{LHV}}}{\leq}1,
$
which is just the joint-probability form of the CHSH inequality ignoring a constant.

By choosing appropriate coefficients, there exists multi-component Bell function $\mathcal{I}_{2,d}$ for which the corresponding Bell inequality is equivalent to the CGLMP inequality. See Example 1.

{\bf  Example. 1}--- For $n=2$ and $d\geq 2$,  when we take the multi-component Bell function as
\begin{eqnarray}\label{I2d}
\mathcal{I}_{2,d}&=&\frac{1}{d}\left[-\sum_{k=1}^{d-1}k{\rm \mathbf{v}}_{d-k}A_0B_0-\sum_{k=1}^{d-1}k{\rm \mathbf{v}}_{k}(A_0B_1+A_1B_0)\right.\nonumber\\
& &\left.+\sum_{k=1}^{d-1}k{\rm \mathbf{v}}_{k}A_1B_1\right],
\end{eqnarray}
the corresponding Bell inequality is
$\mathcal{B}_{2,d}=-\frac{1}{d-1}\sum_{k=1}^{d-1}k*(P(A_0+B_0=k)+
P(A_0+B_1=-k))-\frac{1}{d-1}\sum_{k=1}^{d-1}k*(P(A_1+B_0=-k)-P(A_1+B_1=-k))+1\overset{{\rm LHV}}{\leq}1,$
which is equivalent to the CGLMP inequality, and also equivalent to the Bell inequality
$
\mathcal{B}_{d}=\sum_{k=0}^{N-1}\sqrt{\frac{(N+1-k)(N-k)}{(N+1)N}}\mathcal{B}^{(k)}\overset{{\rm LHV}}{\leq}2
$
in Ref. \cite{Multicomponent} ignoring a constant,
where $\mathcal{B}^{(0)}=Q_{00}^{(0)}+Q_{01}^{(0)}-Q_{10}^{(0)}+Q_{11}^{(0)},
\mathcal{B}^{(k)}=Q_{00}^{(k)}-Q_{01}^{(k)}-Q_{10}^{(k)}+Q_{11}^{(k)},k\neq 0.$
The proof is presented in Appendix B.

If we use $\Pi_{k_0}$  and $\Pi'_{k_1}$ as the projectors for the measurement outcome ${\rm \mathbf{v}}_{k_0}$ of $C_0$  and the measurement outcome ${\rm \mathbf{v}}_{k_1}$ of $C_1$ respectively,
then
$C_0=\sum_{k_0=0}^{N}{\rm \mathbf{v}}_{k_0}\Pi_{k_0},C_1=\sum_{k_1=0}^{N}{\rm \mathbf{v}}_{k_1}\Pi'_{k_1},$
with $\{\Pi_{k_0}\},\{\Pi'_{k_1}\}$ being ONBs. Due to
\begin{eqnarray}
&&C_0=\sum_{k_0=0}^{N}{\rm \mathbf{v}}_{k_0}\Pi_{k_0}\left(\sum_{k_1=0}^{N}\Pi'_{k_1}\right)=\sum_{k_0,k_1}{\rm \mathbf{v}}_{k_0}\Pi_{k_0}\Pi'_{k_1},\nonumber\\
&&C_1=\sum_{k_1=0}^{N}{\rm \mathbf{v}}_{k_1}\left(\sum_{k_0=0}^{N}\Pi_k\right)\Pi'_{k_1}=\sum_{k_0,k_1}{\rm \mathbf{v}}_{k_1}\Pi_{k_0}\Pi'_{k_1},
\end{eqnarray}
if we take
\begin{eqnarray}{\label{In-1}}
\mathcal{I}_{n-1,d}^{k_0,k_1}&=&\sum_{i,\cdots,j=0,1}\omega_{i,\cdots,j,0} {\rm \mathbf{v}}_{k_0} A_i\cdots B_j\nonumber\\
& &+\sum_{i,\cdots,j=0,1}\omega_{i,\cdots,j,1} {\rm \mathbf{v}}_{k_1} A_i\cdots B_j,
\end{eqnarray}
then
 the general Bell function $\mathcal{I}_{n,d}$ can be reformulated as
\begin{eqnarray}{\label{Ind2}}
\mathcal{I}_{n,d}=\sum_{k_0,k_1=0}^{d-1} \mathcal{I}_{n-1,d}^{k_0,k_1}\Pi_{k_0}\Pi'_{k_1},
\end{eqnarray}
where $\mathcal{I}_{n-1,d}^{k_0,k_1}$ is a full-correlated multi-component Bell function for $(n-1,2,d)$-scenario. Inspired by this observation, we can present the following iteration formula. The proof will be given in Appendix C.

{\bf Theorem. 1}--- For prime $d$,  the Bell function $\mathcal{I}_{n,d}$
 can be reduced to
\begin{eqnarray}{\label{Ind3}}
& &\mathcal{I}_{n,d}=\frac{1}{d}\left(
                \begin{array}{cc}
                  \mathcal{I}_{n-1,d}^{0,0} & \mathcal{I}_{n-1,d}^{0,1} \\
                \end{array}
              \right)\nonumber\\
                     & &\left(
                       \begin{array}{cc}
                         -\sum_{k=1}^{d-1}(d-k){\rm \mathbf{v}}_{k} & -\sum_{k=1}^{d-1}k{\rm \mathbf{v}}_k \\
                         -\sum_{k=1}^{d-1}k{\rm \mathbf{v}}_k & \sum_{k=1}^{d-1}k{\rm \mathbf{v}}_k \\
                       \end{array}
                     \right)\left(
                              \begin{array}{c}
                                C_0 \\
                                C_1 \\
                              \end{array}
                            \right),
\end{eqnarray}
where $\mathcal{I}_{n-1,d}^{0,0}$ and $\mathcal{I}_{n-1,d}^{0,1}$ are full-correlated multi-component Bell functions for $(n-1,2,d)$-scenario.

 For example, in the case $d=2$, if we take $\mathcal{I}_{2,2}^{0,0}$ as the Bell function $\mathcal{I}_{2,2}$ in Example 1, $\mathcal{I}_{n-1,2}^{0,0}=\mathcal{I}_{n-1,2}$ for $n\geq 3$,
and $\mathcal{I}_{n-1,2}^{0,1}$ obtained from $\mathcal{I}_{n-1,2}^{0,0}$ by swapping the two measurements for each party, i.e., $X^{(k)}_0\leftrightarrow X^{(k)}_1$ for $k=1,2,\cdots,n-1$, then
\begin{eqnarray}{\label{In2}}
\mathcal{I}_{n,2}
          &=&\frac{1}{2}\left(
                \begin{array}{cc}
                  \mathcal{I}_{n-1,2}^{0,0} & \mathcal{I}_{n-1,2}^{0,1} \\
                \end{array}
              \right)\left(
                       \begin{array}{cc}
                         -{\rm \mathbf{v}}_1 & -{\rm \mathbf{v}}_1 \\
                        -{\rm \mathbf{v}}_1 & {\rm \mathbf{v}}_1 \\
                       \end{array}
                     \right)\left(
                              \begin{array}{c}
                                C_0 \\
                                C_1 \\
                              \end{array}
                            \right).\ \ \ \
\end{eqnarray}
Interestingly, the corresponding Bell inequality is nothing but the MABK
inequality, the most robust one for the $(n,2,2)$-scenario.

Based on this iteration formula, we can construct Bell function $\mathcal{I}_{n,d}$ with $\mathcal{I}_{n-1,d}^{0,0}$ and $\mathcal{I}_{n-1,d}^{0,1}$. This means that we give a way of constructing new $(n,2,d)$ Bell inequalities with $(n-1,2,d)$ ones. In the following, we use one of the most robust $(n-1)$-party full-correlated multi-component Bell function as $\mathcal{I}_{n-1,d}^{0,0}$, and
make  $\mathcal{I}_{n-1,d}^{0,1}$ run over the ones equivalent to the most robust $(n-1)$-party multi-component Bell function, then we can obtain
the most robust  $n$-party $\mathcal{I}_{n,d}$ with the iteration formula, i.e., the most robust Bell inequality $\mathcal{B}_{n,d}\overset{{\rm LHV}}{\leq}L$ for the $n$-qudit system.

\section{The most robust $(n,2,d)$-scenario full-correlated Bell inequalities}

We focus on the most robust $\mathcal{I}_{n,d}$ for the mixture of the $n$-qudit GHZ state $|\Psi\rangle$  for prime $d\geq 3$ and the white noise. That is, let $v_c$ be the lowest critical visibility, when $v>v_c$, the noisy state $\rho=v |\Psi\rangle\langle \Psi|+(1-v)\;\mathbb{I}/d^n$ would violate the $n$-qudit Bell inequality $\mathcal{B}_{n,d}\overset{{\rm LHV}}{\leq}L$. If the lowest critical visibility $v_c$ is lower, the Bell inequality will detect more nonlocal noisy state $\rho$, i.e., it is more robust. Here $\mathbb{I}$ denotes the unit matrix for the corresponding quantum system, and $\mathbb{I}/d^n$ represents the white noise for the $n$-qudit system. For each observer, the two
von Neumann measuring apparatus are confined to the unbiased symmetric $(d\times 2)$-port beam splitters, which can be used to maximally violate the coincidence Bell inequalities for the quantum systems of qudits \cite{CGLMP,Coincidence,n23,Optimal measurements}, and are experimentally realizable  \cite{splitter1,splitter2}.

The density matrix of the noisy $n$-qudit GHZ state is given by
$\rho=v |\Psi\rangle\langle \Psi|+(1-v)\; \mathbb{I}/d^n$, with $v\in[0,1]$, and the pure state
$
|\Psi\rangle=\frac{1}{\sqrt{d}}(|0\rangle^{\otimes n}+|1\rangle^{\otimes n}+\cdots+|d-1\rangle^{\otimes n})
$
is the so-called $n$-qudit GHZ state. Then the inequality (\ref{GCI}) can  detect the nonlocality of $\rho$ if and only if
$v\times NL_{|\Psi\rangle}+(1-v)\times NL_{\mathbb{I}}>L$, i.e.,
$v>v_{c}=(L-NL_{\mathbb{I}})/(NL_{|\Psi\rangle}-NL_{\mathbb{I}}),$ where $NL_{|\Psi\rangle}$ and $NL_{\mathbb{I}}$ are the maximal quantum violations of $\mathcal{B}_{n,d}$ for  the
  state $|\Psi\rangle$ and the state $\mathbb{I}/d^n$, respectively.
Hence, the parameter $v_{c}$ exactly reflects the ability of the inequality $\mathcal{B}_{n,d}\overset{{\rm LHV}}{\leq}L$ to detect nonlocal $\rho$. It is just the meaning of the critical visibility. Apparently, two equivalent Bell inequalities have the same visibility.
If a Bell inequality has a lower critical visibility, then for more noise $(\mathbb{I}/d^n)$ it still can detect the nonlocality of $\rho$, which means that it is {\it more robust}.

The
action of the unbiased symmetric $(d\times 2)$-port beam splitters in the computational basis is as
follows: firstly, a phase factor is applied depending on the
initial state, i.e., $|j\rangle\mapsto {\rm e}^{{\rm i}\phi_j}|j\rangle$; following this, a quantum Fourier transform (QFT) is performed and the resulting state is measured in the computational basis.
Therefore, any of these measurements is defined by a
$d$-phase vector $\vec{\phi}=(\phi_0,\phi_1,\cdots,\phi_{d-1})$ and the corresponding unitary transformation,
$
[U_{{\rm QFT}}U(\vec{\phi})]_{ij}=\frac{1}{\sqrt{d}}{\rm e}^{{\rm i}\frac{2\pi}{d}(i-1)(j-1)}{\rm e}^{{\rm i}\phi_{i-1}}.
$
Given a
measurement apparatus for $n$-parties
specified by the $d$-phase vectors
$\vec{\phi}^{(X)}=(\phi^{(X)}_0,\phi^{(X)}_1,\cdots,\phi^{(X)}_{d-1}),$ and an
initial state $|\Phi\rangle\in (\mathbb{C}^d)^{\otimes n}$, the probability of obtaining the outcome
$(a_i,\cdots,b_j,c_k)$ is
$P(A_i=a_i,\cdots,B_j=b_j,C_k=c_k)
=|\langle a_i\cdots b_jc_k|U_{{\rm QFT}}U(\vec{\phi}^{(A_i)})\otimes\cdots \otimes U_{{\rm QFT}}U(\vec{\phi}^{(B_j)})\otimes$ $U_{{\rm QFT}}U(\vec{\phi}^{(C_k)})|\Phi\rangle|^2.$
The phase vectors can be changed by the observers, which represent the local macroscopic parameters
available to them. For the coincidence terms appearing in the Bell inequality and the maximally entangled state, the direct calculation yields
\begin{widetext}
\begin{eqnarray}{\label{prob.}}
&&P\left(A_{i}+\cdots+B_{j}+C_{k}=r\right)=\frac{d+
2\sum_{t=0}^{d-1}\sum_{s=t+1}^{d-1}\cos{\left(\phi^{(A_{i})}_s-\phi^{(A_{i})}_{t}+\cdots+\phi^{(B_{j})}_s-\phi^{(B_{j})}_{t}+\phi^{(C_{k})}_s-\phi^{(C_{k})}_{t}
+\frac{2(s-t)r\pi}{d}\right)}}{d^2}.\nonumber\\
&&
\end{eqnarray}
\end{widetext}

For prime $d$, the first component of ${\rm \mathbf{v}}_kX$ is $(dP(X=d-k)-1)/(d-1)$, which yields that in this case the Bell function $\mathcal{I}_{n,d}$ corresponds to $\mathcal{B}_{n,d}$ one by one. Hence, if the Bell inequality is the most robust, we also say that the corresponding  multi-component Bell function is the robust one.
 Next, based on the Bell function $\mathcal{I}_{2,d}$ in Example 1, which is the  most robust one for the $(n,2,d)$-scenario, we use the iteration formula (\ref{Ind3}) to present the most robust  $n$-party $\mathcal{I}_{n,d}$, and then obtain the robust Bell inequality $\mathcal{B}_{n,d}\overset{{\rm LHV}}{\leq}L$ for $n$-party and qudit systems.

{\bf Example. 2}--- In this example, we focus on the case $d=3$.

 (a) When $n=3$, take $\mathcal{I}_{2,3}^{0,0}$ as $\mathcal{I}_{2,3}$ in Example 1. Let $\mathcal{I}_{2,3}^{0,1}$ run over the equivalent class of $\mathcal{I}_{2,3}$, which has $54$ elements. There are four Bell functions $\mathcal{I}_{3,3}(1),\mathcal{I}_{3,3}(2),\mathcal{I}_{3,3}(3),\mathcal{I}_{3,3}(4)$, for which the corresponding Bell inequalities has the lowest critical visibility $v_c=0.6$.
For example, in the case that
$
 \mathcal{I}_{2,3}^{0,1}=\frac{1}{3}[(-{\rm \mathbf{v}}_{1}-2{\rm \mathbf{v}}_{2})A_0B_0+(-2{\rm \mathbf{v}}_{1}-{\rm \mathbf{v}}_{2})A_0B_1
+
 ({\rm \mathbf{v}}_{1}-2{\rm \mathbf{v}}_{2})A_1B_0+(-{\rm \mathbf{v}}_{1}+{\rm \mathbf{v}}_{2})A_1B_1],
$
 we obtain
 \begin{eqnarray}{\label{I331}}
 \mathcal{I}_{3,3}(1)&=&\frac{1}{3}[{\rm \mathbf{v}}_{0}A_0B_0C_0
 -{\rm \mathbf{v}}_{1}A_0B_0C_1+2{\rm \mathbf{v}}_{0}A_0B_1C_0\nonumber\\
 & &+{\rm \mathbf{v}}_{1}A_0B_1C_1
 -{\rm \mathbf{v}}_{2}A_1B_0C_0+{\rm \mathbf{v}}_{0}A_1B_0C_1\nonumber\\
 & &+{\rm \mathbf{v}}_{2}A_1B_1C_0-
{\rm \mathbf{v}}_{0}A_1B_1C_1],
 \end{eqnarray}
the corresponding Bell inequality of which is
$
\mathcal{B}_{3,3}(1)=\frac{1}{2}[P(A_0+B_0+C_0=0)-P(A_0+B_0+C_1=2)+2P(A_0+B_1+C_0=0)+P(A_0+B_1+C_1=2)-P(A_1+B_0+C_0=1)
+P(A_1+B_0+C_1=0)+P(A_1+B_1+C_0=1)-P(A_1+B_1+C_1=0)-1]\overset{{\rm LHV}}{\leq}1.
$
When the measurements are restricted to unbiased symmetric six-port beam splitters \cite{Coincidence}, $NL_{|\Psi\rangle}=5/3,NL_{\mathbb{I}}=0$. Then $v_{c}=(L-NL_{\mathbb{I}})/(NL_{|\Psi\rangle}-NL_{\mathbb{I}})=0.6$. It is the most robust one for $(3,2,3)$-scenario obtained by the iteration formula (\ref{Ind3}). Here, we list the sequential operations used for obtaining  $\mathcal{I}_{2,3}^{0,1}$:
 (i) replacing the outcome $a_1$ by ${\rm Mod}[a_1+2,3]$; (ii) exchanging the label of measurements $B_0,B_1$, i.e., $B_0\leftrightarrow B_1$.  For the sake of narrative, we write the above procedure
 as
 $\mathcal{I}_{2,3}^{0,1}=\mathcal{I}_{2,3}^{0,0}/.\{a_1\rightarrow {\rm Mod}[a_1+2,3],B_0\leftrightarrow B_1\}.$
 In fact, $\mathcal{I}_{3,3}(2),\mathcal{I}_{3,3}(3),$ and $\mathcal{I}_{3,3}(4)$  are equivalent to $\mathcal{I}_{3,3}(1)$. See Appendix D for more details.

There are $648$ elements in the equivalent class of $\mathcal{I}_{3,3}(1)$, among which
 we find a party symmetric $\mathcal{I}_{3,3}(5)=\frac{1}{3}[{\rm \mathbf{v}}_{0}A_0B_0C_0+2{\rm \mathbf{v}}_{0}A_1B_1C_1]
 -{\rm \mathbf{v}}_{1}(A_0B_0C_1+A_0B_1C_0+A_1B_0C_0)+{\rm \mathbf{v}}_{2}(A_0B_1C_1+A_1B_0C_1+A_1B_1C_0)].
$
The equivalence implies that $\mathcal{I}_{3,3}(5)$'s Bell inequality also has the same lowest critical visibility $v_c=0.6$. In fact, the Bell inequality
 is just the most robust coincidence Bell inequality in Ref. \cite{Coincidence}. When $c_1=c_2={\rm \mathbf{v}}_{0}$ and $c_1={\rm \mathbf{v}}_{0},c_2={\rm \mathbf{v}}_{1}$, it reduces to $(\mathcal{I}_{2,3}(5))^{0,0}$ and $(\mathcal{I}_{2,3}(5))^{0,1}$ respectively. For both of them, the corresponding Bell inequalities are equivalent to the CGLMP inequality for $d=3$.

(b) For $n=4$, we take  $\mathcal{I}_{3,3}^{0,0}=\mathcal{I}_{3,3}(3)$, which is outcome symmetric with $\mathcal{I}_{3,3}(5)$ and reduces to $\mathcal{I}_{2,3}$ when $c_1=c_2={\rm \mathbf{v}}_{0}$, and allow $\mathcal{I}_{3,3}^{0,1}$ to go through the $648$ ones equivalent to $\mathcal{I}_{3,3}(3)$. Then we obtain $16$ most robust $\mathcal{I}_{4,3}$, for which the corresponding Bell inequalities have the same lowest critical visibility $v_c=0.5$.  {\bf By the equivalence checking, we find that the following four ones are not equivalent to each other,}
\begin{eqnarray}
\mathcal{I}_{4,3}(i)=\sum_{i,j,k,l=0}^{1}\omega_{i,j,k,l}^{(i)}A_iB_jC_kD_l,\ i=1,2,3,4,
\end{eqnarray}
with
\begin{widetext}
\begin{eqnarray}
& &\left(\omega_{0,0,0,0}^{(1)},\omega_{0,0,0,1}^{(1)},\omega_{0,0,1,0}^{(1)},\omega_{0,0,1,1}^{(1)},
\omega_{0,1,0,0}^{(1)},\omega_{0,1,0,1}^{(1)},\omega_{0,1,1,0}^{(1)},\omega_{0,1,1,1}^{(1)},\omega_{1,0,0,0}^{(1)},\omega_{1,0,0,1}^{(1)},\omega_{1,0,1,0}^{(1)},\omega_{1,0,1,1}^{(1)},
\omega_{1,1,0,0}^{(1)},\omega_{1,1,0,1}^{(1)},\right.\nonumber\\
& &\left.\omega_{1,1,1,0}^{(1)},\omega_{1,1,1,1}^{(1)}\right)=
\frac{1}{9}(-4{\rm \mathbf{v}}_{1}+{\rm \mathbf{v}}_{2},{\rm \mathbf{v}}_{1}-4{\rm \mathbf{v}}_{2},
-2{\rm \mathbf{v}}_{1}-{\rm \mathbf{v}}_{2},-{\rm \mathbf{v}}_{1}+{\rm \mathbf{v}}_{2},
{\rm \mathbf{v}}_{1}-{\rm \mathbf{v}}_{2},-{\rm \mathbf{v}}_{1}-2{\rm \mathbf{v}}_{2},
-{\rm \mathbf{v}}_{1}-2{\rm \mathbf{v}}_{2},-2{\rm \mathbf{v}}_{1}-{\rm \mathbf{v}}_{2},\nonumber\\
& &
{\rm \mathbf{v}}_{1}-{\rm \mathbf{v}}_{2},-{\rm \mathbf{v}}_{1}-2{\rm \mathbf{v}}_{2},
-{\rm \mathbf{v}}_{1}-2{\rm \mathbf{v}}_{2},-2{\rm \mathbf{v}}_{1}-{\rm \mathbf{v}}_{2},
2{\rm \mathbf{v}}_{1}+{\rm \mathbf{v}}_{2},{\rm \mathbf{v}}_{1}-{\rm \mathbf{v}}_{2},
4{\rm \mathbf{v}}_{1}+5{\rm \mathbf{v}}_{2},-4{\rm \mathbf{v}}_{1}+{\rm \mathbf{v}}_{2}),\nonumber\\
& &\left(\omega_{0,0,0,0}^{(2)},\omega_{0,0,0,1}^{(2)},\omega_{0,0,1,0}^{(2)},\omega_{0,0,1,1}^{(2)},
\omega_{0,1,0,0}^{(2)},\omega_{0,1,0,1}^{(2)},\omega_{0,1,1,0}^{(2)},\omega_{0,1,1,1}^{(2)},\omega_{1,0,0,0}^{(2)},\omega_{1,0,0,1}^{(2)},\omega_{1,0,1,0}^{(2)},\omega_{1,0,1,1}^{(2)},
\omega_{1,1,0,0}^{(2)},\omega_{1,1,0,1}^{(2)},\right.\nonumber\\
& &\left.\omega_{1,1,1,0}^{(2)},\omega_{1,1,1,1}^{(2)}\right)=
\frac{1}{9}(-{\rm \mathbf{v}}_{1}-2{\rm \mathbf{v}}_{2},-2{\rm \mathbf{v}}_{1}-{\rm \mathbf{v}}_{2},
-2{\rm \mathbf{v}}_{1}-{\rm \mathbf{v}}_{2},-{\rm \mathbf{v}}_{1}+{\rm \mathbf{v}}_{2},
{\rm \mathbf{v}}_{1}-{\rm \mathbf{v}}_{2},-{\rm \mathbf{v}}_{1}-2{\rm \mathbf{v}}_{2},
-4{\rm \mathbf{v}}_{1}+{\rm \mathbf{v}}_{2},{\rm \mathbf{v}}_{1}-4{\rm \mathbf{v}}_{2},\nonumber\\
& &
{\rm \mathbf{v}}_{1}-{\rm \mathbf{v}}_{2},-{\rm \mathbf{v}}_{1}-2{\rm \mathbf{v}}_{2},
-{\rm \mathbf{v}}_{1}-2{\rm \mathbf{v}}_{2},-2{\rm \mathbf{v}}_{1}-{\rm \mathbf{v}}_{2},
2{\rm \mathbf{v}}_{1}+{\rm \mathbf{v}}_{2},{\rm \mathbf{v}}_{1}-{\rm \mathbf{v}}_{2},
4{\rm \mathbf{v}}_{1}+5{\rm \mathbf{v}}_{2},-4{\rm \mathbf{v}}_{1}+{\rm \mathbf{v}}_{2}),\nonumber\\
& &\left(\omega_{0,0,0,0}^{(3)},\omega_{0,0,0,1}^{(3)},\omega_{0,0,1,0}^{(3)},\omega_{0,0,1,1}^{(3)},
\omega_{0,1,0,0}^{(3)},\omega_{0,1,0,1}^{(3)},\omega_{0,1,1,0}^{(3)},\omega_{0,1,1,1}^{(3)},\omega_{1,0,0,0}^{(3)},\omega_{1,0,0,1}^{(3)},\omega_{1,0,1,0}^{(3)},\omega_{1,0,1,1}^{(3)},
\omega_{1,1,0,0}^{(3)},\omega_{1,1,0,1}^{(3)},\right.\nonumber\\
& &\left.\omega_{1,1,1,0}^{(3)},\omega_{1,1,1,1}^{(3)}\right)=
\frac{1}{3}(-{\rm \mathbf{v}}_{1},-{\rm \mathbf{v}}_{2},
{\rm \mathbf{v}}_{0},{\rm \mathbf{v}}_{2},
{\rm \mathbf{v}},{\rm \mathbf{0}},
{\rm \mathbf{v}}_{0},{\rm \mathbf{0}},
{\rm \mathbf{0}},-{\rm \mathbf{v}}_{2},
{\rm \mathbf{0}},{\rm \mathbf{v}}_{0},
-{\rm \mathbf{v}}_{2},-{\rm \mathbf{v}}_{0},
{\rm \mathbf{v}}_{2},{\rm \mathbf{v}}_{2}),\nonumber\\
& &\left(\omega_{0,0,0,0}^{(4)},\omega_{0,0,0,1}^{(4)},\omega_{0,0,1,0}^{(4)},\omega_{0,0,1,1}^{(4)},
\omega_{0,1,0,0}^{(4)},\omega_{0,1,0,1}^{(4)},\omega_{0,1,1,0}^{(4)},\omega_{0,1,1,1}^{(4)},\omega_{1,0,0,0}^{(4)},\omega_{1,0,0,1}^{(4)},\omega_{1,0,1,0}^{(4)},\omega_{1,0,1,1}^{(4)},
\omega_{1,1,0,0}^{(4)},\omega_{1,1,0,1}^{(4)},\right.\nonumber\\
& &\left.\omega_{1,1,1,0}^{(4)},\omega_{1,1,1,1}^{(4)}\right)=
\frac{1}{3}({\rm \mathbf{v}}_{0},{\rm \mathbf{0}},
{\rm \mathbf{v}}_{0},{\rm \mathbf{v}}_{2},
-{\rm \mathbf{v}}_{2},{\rm \mathbf{0}},
-{\rm \mathbf{v}}_{1},-{\rm \mathbf{v}}_{2},
-{\rm \mathbf{v}}_{2},{\rm \mathbf{0}},
-{\rm \mathbf{v}}_{1},-{\rm \mathbf{v}}_{2},
{\rm \mathbf{v}}_{1},{\rm \mathbf{v}}_{2},
{\rm \mathbf{0}},{\rm \mathbf{v}}_{2}).
\end{eqnarray}
\end{widetext}
Since the first component of ${\rm \mathbf{v}}_{k}X$ is $\frac{3}{2}P(X=3-k)-\frac{1}{2}$, the four corresponding Bell inequalities can be listed in Appendix D.
For these four Bell inequalities, $NL_{|\Psi\rangle}=2,NL_{\mathbb{I}}=0$, and then $v_{c}=(L-NL_{\mathbb{I}})/(NL_{|\Psi\rangle}-NL_{\mathbb{I}})=0.5$.

The fist $(4,2,3)$-scenario Bell inequality in Ref. \cite{n23} corresponds to
$\mathcal{I}_{4,3}(17)=\mathcal{I}_{4,3}(3)/.\{a_1\rightarrow {\rm Mod}[a_1+2,3],b_1\rightarrow {\rm Mod}[b_1+2,3],c_1\rightarrow {\rm Mod}[c_1+1,3]\},
$
i.e., it is equivalent to $\mathcal{I}_{4,3}(3)$.
When $c_1=c_2={\rm \mathbf{v}}_{0}$ and $c_1={\rm \mathbf{v}}_{0},c_2={\rm \mathbf{v}}_{1}$, $\mathcal{I}_{4,3}(17)$ reduces to $(\mathcal{I}_{3,3}(17))^{0,0}=\mathcal{I}_{3,3}(3)/.\{ABCD\rightarrow CBAD,b_1\rightarrow {\rm Mod}[b_1+2,3]\}$, and
$(\mathcal{I}_{3,3}(17))^{0,1}=\mathcal{I}_{3,3}(3)/.\{a_1\rightarrow {\rm Mod}[a_1+2,3],b_1\rightarrow {\rm Mod}[b_1+2,3],c_0\rightarrow {\rm Mod}[c_0+2,3],B_0\leftrightarrow B_1,C_0\leftrightarrow C_1\}$, respectively.
That is to say, both of them are equivalent to $\mathcal{I}_{3,3}$.

The second $(4,2,3)$-scenario Bell inequality with probability form in Ref. \cite{n23} corresponds to
$
\mathcal{I}_{4,3}(18)=\mathcal{I}_{4,3}(2)/.\{ABCD\rightarrow CBAD,b_1\rightarrow {\rm Mod}[b_1+2,3]\},
$
i.e., it is equivalent to $\mathcal{I}_{4,3}(2)$.
When $c_1=c_2={\rm \mathbf{v}}_{0}$ and $c_1={\rm \mathbf{v}}_{0},c_2={\rm \mathbf{v}}_{1}$, it reduces to $(\mathcal{I}_{3,3}(18))^{0,0}=(\mathcal{I}_{3,3}(17))^{0,0}$ and $(\mathcal{I}_{2,3}(18))^{0,1}=\mathcal{I}_{3,3}(3)/.\{a_1\rightarrow {\rm Mod}[a_1+2,3],c_1\rightarrow {\rm Mod}[c_1+1,3],B_0\leftrightarrow B_1\}$, respectively. Hence, both of them are
 equivalent to $\mathcal{I}_{3,3}$.

(d) For $n=5$, let $\mathcal{I}_{4,3}^{0,1}$ run over all ones equivalent to $\mathcal{I}_{4,3}(i),i=1,2,3,4$, and $\mathcal{I}_{4,3}^{0,0}=\mathcal{I}_{4,3}(j)$ for $j=1,2,3,4$. Then for the most robust $\mathcal{I}_{5,3}$, the corresponding Bell inequalities have the lowest critical visibility $v_c\approx 0.488756$. In the following Table I, we list the number of $\mathcal{I}_{5,3}$ with the lowest $v_c\approx 0.488756$ for each case, where $\mathcal{I}_{4,3}^{0,1}\sim \mathcal{I}_{4,3}(i)$ means that $\mathcal{I}_{4,3}^{0,1}$ runs over all ones equivalent to $\mathcal{I}_{4,3}(i),i=1,2,3,4$.
\begin{widetext}
\begin{center}\begin{table}[h]
\caption{The number of $\mathcal{I}_{5,3}$ with the lowest $v_c\approx 0.488756$\label{table1}}
\begin{tabular}{|c|c|c|c|c|}
  \hline
    The number of $\mathcal{I}_{5,3}$ with the lowest $v_c\approx 0.488756$ & $\mathcal{I}_{4,3}^{0,1}\sim \mathcal{I}_{4,3}(1)$ & $\mathcal{I}_{4,3}^{0,1}\sim \mathcal{I}_{4,3}(2)$ & $\mathcal{I}_{4,3}^{0,1}\sim \mathcal{I}_{4,3}(3)$ & $\mathcal{I}_{4,3}^{0,1}\sim \mathcal{I}_{4,3}(4)$\\
   \hline
   $\mathcal{I}_{4,3}^{0,0}=\mathcal{I}_{4,3}(1)$ & 12 & 12 & 12 & 12\\
   \hline
   $\mathcal{I}_{4,3}^{0,0}=\mathcal{I}_{4,3}(2)$ & 8 & 16 & 8 & 16\\
   \hline
   $\mathcal{I}_{4,3}^{0,0}=\mathcal{I}_{4,3}(3)$ & 5 & 6  & 0 & 0\\
   \hline
   $\mathcal{I}_{4,3}^{0,0}=\mathcal{I}_{4,3}(4)$ & 4 & 5  & 0 & 0\\
  \hline
\end{tabular}
\end{table}\end{center}
\end{widetext}
For example,
let $\mathcal{I}_{4,3}^{0,1}$ run over all ones equivalent to $\mathcal{I}_{4,3}(1)$, and $\mathcal{I}_{4,3}^{0,0}=\mathcal{I}_{4,3}(1)$, then the number of $\mathcal{I}_{5,3}$ with the lowest $v_c\approx 0.488756$ is $12$, among which one is $\mathcal{I}_{5,3}(1)$. Although the verification of equivalence of $\mathcal{I}_{5,3}$ is very difficult limited by our computer, we can also list another $\mathcal{I}_{5,3}(2)$ with lowest $v_c\approx 0.488756$ which is not equivalent to $\mathcal{I}_{5,3}(1)$. $\mathcal{I}_{5,3}(2)$ is obtained when
 $\mathcal{I}_{4,3}^{0,1}$ runs over all ones equivalent to $\mathcal{I}_{4,3}(2)$, and $\mathcal{I}_{4,3}^{0,0}=\mathcal{I}_{4,3}(2)$. In the following, we list $\mathcal{I}_{5,3}(1)$ and $\mathcal{I}_{5,3}(2)$,
 \begin{eqnarray}
\mathcal{I}_{5,3}(i)=\sum_{i,j,k,l,m=0}^{1}\omega_{i,j,k,l,m}^{(i)}A_iB_jC_kD_lE_m,\ i=1,2,\ \ \ \
\end{eqnarray}
with
\begin{widetext}
\begin{eqnarray}
& &\left(\omega_{0,0,0,0,0}^{(1)},\omega_{0,0,0,0,1}^{(1)},\omega_{0,0,0,1,0}^{(1)},\omega_{0,0,0,1,1}^{(1)},
\omega_{0,0,1,0,0}^{(1)},\omega_{0,0,1,0,1}^{(1)},\omega_{0,0,1,1,0}^{(1)},\omega_{0,0,1,1,1}^{(1)},\omega_{0,1,0,0,0}^{(1)},\omega_{0,1,0,0,1}^{(1)},
\omega_{0,1,0,1,0}^{(1)},\omega_{0,1,0,1,1}^{(1)},\right.\nonumber\\
& &\omega_{0,1,1,0,0}^{(1)},\omega_{0,1,1,0,1}^{(1)},\omega_{0,1,1,1,0}^{(1)},\omega_{0,1,1,1,1}^{(1)},
\omega_{1,0,0,0,0}^{(1)},\omega_{1,0,0,0,1}^{(1)},\omega_{1,0,0,1,0}^{(1)},\omega_{1,0,0,1,1}^{(1)},
\omega_{1,0,1,0,0}^{(1)},\omega_{1,0,1,0,1}^{(1)},\omega_{1,0,1,1,0}^{(1)},\omega_{1,0,1,1,1}^{(1)},\nonumber\\
& &\left.\omega_{1,1,1,0,0}^{(1)},\omega_{1,1,1,0,1}^{(1)},\omega_{1,1,1,1,0}^{(1)},\omega_{1,1,1,1,1}^{(1)}\right)=
\frac{1}{9}(-4{\rm \mathbf{v}}_{1},{\rm \mathbf{v}}_{2},-2{\rm \mathbf{v}}_{2},{\rm \mathbf{v}}_{1}-2{\rm \mathbf{v}}_{2},-2{\rm \mathbf{v}}_{2},-2{\rm \mathbf{v}}_{1}+{\rm \mathbf{v}}_{2},
-{\rm \mathbf{v}}_{0},-2{\rm \mathbf{v}}_{1},-2{\rm \mathbf{v}}_{2},-{\rm \mathbf{v}}_{0},\nonumber\\
& &2{\rm \mathbf{v}}_{0},{\rm \mathbf{v}}_{1},2{\rm \mathbf{v}}_{0},{\rm \mathbf{v}}_{1},-2{\rm \mathbf{v}}_{1},-{\rm \mathbf{v}}_{2},
{\rm \mathbf{v}}_{1},-{\rm \mathbf{v}}_{2},-{\rm \mathbf{v}}_{2},{\rm \mathbf{v}}_{0},-{\rm \mathbf{v}}_{2},{\rm \mathbf{v}}_{0},{\rm \mathbf{v}}_{0},-{\rm \mathbf{v}}_{1},-{\rm \mathbf{v}}_{2},-2{\rm \mathbf{v}}_{0},-2{\rm \mathbf{v}}_{0},-{\rm \mathbf{v}}_{1}-3{\rm \mathbf{v}}_{2},-2{\rm \mathbf{v}}_{0},2{\rm \mathbf{v}}_{1}+3{\rm \mathbf{v}}_{2},\nonumber\\
& &
-4{\rm \mathbf{v}}_{1},{\rm \mathbf{v}}_{2}),\nonumber\\
& &\left(\omega_{0,0,0,0,0}^{(2)},\omega_{0,0,0,0,1}^{(2)},\omega_{0,0,0,1,0}^{(2)},\omega_{0,0,0,1,1}^{(2)},
\omega_{0,0,1,0,0}^{(2)},\omega_{0,0,1,0,1}^{(2)},\omega_{0,0,1,1,0}^{(2)},\omega_{0,0,1,1,1}^{(2)},\omega_{0,1,0,0,0}^{(2)},\omega_{0,1,0,0,1}^{(2)},\omega_{0,1,0,1,0}^{(2)},
\omega_{0,1,0,1,1}^{(2)},\right.\nonumber\\
& &\omega_{0,1,1,0,0}^{(2)},\omega_{0,1,1,0,1}^{(2)},\omega_{0,1,1,1,0}^{(2)},\omega_{0,1,1,1,1}^{(2)},
\omega_{1,0,0,0,0}^{(2)},\omega_{1,0,0,0,1}^{(2)},\omega_{1,0,0,1,0}^{(2)},\omega_{1,0,0,1,1}^{(2)},
\omega_{1,0,1,0,0}^{(2)},\omega_{1,0,1,0,1}^{(2)},\omega_{1,0,1,1,0}^{(2)},\omega_{1,0,1,1,1}^{(2)},\nonumber\\
& &\left.\omega_{1,1,1,0,0}^{(2)},\omega_{1,1,1,0,1}^{(2)},\omega_{1,1,1,1,0}^{(2)},\omega_{1,1,1,1,1}^{(2)}\right)=
\frac{1}{9}(-{\rm \mathbf{v}}_{1},-2{\rm \mathbf{v}}_{2},{\rm \mathbf{v}}_{2},2{\rm \mathbf{v}}_{0},-4{\rm \mathbf{v}}_{1}-{\rm \mathbf{v}}_{2},2{\rm \mathbf{v}}_{1},2{\rm \mathbf{v}}_{1},-3{\rm \mathbf{v}}_{1}+{\rm \mathbf{v}}_{2},-2{\rm \mathbf{v}}_{1},-{\rm \mathbf{v}}_{0},2{\rm \mathbf{v}}_{0},
\nonumber\\
& &{\rm \mathbf{v}}_{1},-{\rm \mathbf{v}}_{1}+3{\rm \mathbf{v}}_{2},-3{\rm \mathbf{v}}_{1}-2{\rm \mathbf{v}}_{2},-2{\rm \mathbf{v}}_{2},
{\rm \mathbf{v}}_{1}-2{\rm \mathbf{v}}_{2},
{\rm \mathbf{v}}_{1},-{\rm \mathbf{v}}_{2},-{\rm \mathbf{v}}_{2},{\rm \mathbf{v}}_{0},{\rm \mathbf{v}}_{1}-2{\rm \mathbf{v}}_{2},
-2{\rm \mathbf{v}}_{1},{\rm \mathbf{v}}_{1}+3{\rm \mathbf{v}}_{2},
-3{\rm \mathbf{v}}_{1}-4{\rm \mathbf{v}}_{2},-{\rm \mathbf{v}}_{2},-2{\rm \mathbf{v}}_{0},\nonumber\\
& &{\rm \mathbf{v}}_{0},2{\rm \mathbf{v}}_{1},4{\rm \mathbf{v}}_{1}+3{\rm \mathbf{v}}_{2},2{\rm \mathbf{v}}_{2},
2{\rm \mathbf{v}}_{2},-4{\rm \mathbf{v}}_{1}-{\rm \mathbf{v}}_{2}).
\end{eqnarray}
\end{widetext}
In fact, party and observable symmetric operations  do not change the coefficients of full-correlated terms, while outcome symmetric operations change $\omega_{i,j,\cdots,k}$ to $\omega_{i,j,\cdots,k}{\rm \mathbf{v}}_{1}$ or $\omega_{i,j,\cdots,k}{\rm \mathbf{v}}_{2}$. Note that there exists  coefficient $-4{\rm \mathbf{v}}_{1}$ in $\mathcal{I}_{5,3}(1)$, but no  $-4{\rm \mathbf{v}}_{1}$, $-4{\rm \mathbf{v}}_{2}$ and $-4{\rm \mathbf{v}}_{0}=4{\rm \mathbf{v}}_{1}+4{\rm \mathbf{v}}_{2}$ appears in $\mathcal{I}_{5,3}(2)$. {\bf Hence, $\mathcal{I}_{5,3}(1)$ and $\mathcal{I}_{5,3}(2)$ is not equivalent to each other.} See Appendix D for the corresponding Bell inequalities.

{\bf Example. 3}--- In this example, we focus on the case $d=5$,

(a) For $n=3$, when we take $\mathcal{I}_{2,5}^{0,0}$ as $\mathcal{I}_{2,5}(1)=\mathcal{I}_{2,5}$ in Example 1, and let $\mathcal{I}_{2,5}^{0,1}$ go through all ones equivalent to $\mathcal{I}_{2,5}$, the most robust $\mathcal{I}_{3,5}$ obtained by (\ref{Ind3}) has the same robustness as $\mathcal{I}_{2,5}$. That is, the corresponding Bell inequalities have the same
lowest critical visibility $v_c\approx 0.687157$ as the CGLMP inequality for $d=5$.  Employing the roots method of Bell functions as in \cite{root method}, we  find that the Bell  quantity $\mathcal{B}_{2,5}$ of $\mathcal{I}_{2,5}$ has three deterministic values $-\frac{3}{2},-\frac{1}{4},1$ in LHV  theory. Hence, we study the generic form of  full-correlated $(2,2,5)$ multi-component Bell function
\begin{eqnarray}{\label{I25}}
\mathcal{I}_{2,5}&=&\left(\sum_{k=1}^{4}f_{1k}{\rm \mathbf{v}}_{k}\right)A_0B_0+\left(\sum_{k=1}^{4}f_{2k}{\rm \mathbf{v}}_{k}\right)A_0B_1\nonumber\\
& &+
\left(\sum_{k=1}^{4}g_{1k}{\rm \mathbf{v}}_{k}\right)A_1B_0+\left(\sum_{k=1}^{4}g_{2k}{\rm \mathbf{v}}_{k}\right)A_1B_1,\ \ \
\end{eqnarray}
which satisfies that its corresponding Bell quantity has three deterministic values $-\frac{3}{2},-\frac{1}{4},1$ in LHV  theory and its corresponding Bell inequality has the same lowest critical visibility $v_c\approx 0.687157$. Finally, we  find only one another
\begin{eqnarray}{\label{I252}}
\mathcal{I}_{2,5}(2)&=&\frac{3{\rm \mathbf{v}}_{1}+{\rm \mathbf{v}}_{2}-{\rm \mathbf{v}}_{3}+2{\rm \mathbf{v}}_{4}}{5}A_0B_0\nonumber\\
& &+
\frac{-3{\rm \mathbf{v}}_{1}-{\rm \mathbf{v}}_{2}-4{\rm \mathbf{v}}_{3}-2{\rm \mathbf{v}}_{4}}{5}(A_0B_1+A_1B_0)\nonumber\\
& &+\frac{-2{\rm \mathbf{v}}_{1}+{\rm \mathbf{v}}_{2}-{\rm \mathbf{v}}_{3}-3{\rm \mathbf{v}}_{4}}{5}A_1B_1,
\end{eqnarray}
which is not equivalent to $\mathcal{I}_{2,5}(1)$.
{\bf Hence, we find one $(2,2,5)$-scenario Bell inequality which is as robust as the CGLMP inequality for $d=5$, but not equivalent to it.}

For $n=3$, if we take $\mathcal{I}_{2,5}^{0,0}=\mathcal{I}_{2,5}(1)=\mathcal{I}_{2,5}$, and  let $\mathcal{I}_{2,5}^{0,1}$ go through all ones equivalent to $\mathcal{I}_{2,5}(2)$, then we can obtain the most robust $\mathcal{I}_{3,5}(1)$ by (\ref{Ind3}), for which the corresponding Bell inequality  has the lowest critical threshold $v_c\approx 0.595047$. In addition, when we take $\mathcal{I}_{2,5}^{0,0}=\mathcal{I}_{2,5}(2)$, and let $\mathcal{I}_{2,5}^{0,1}$ go through all ones equivalent to $\mathcal{I}_{2,5}$ or $\mathcal{I}_{2,5}(2)$, the most robust ones for the three-partite system have the same robustness as $\mathcal{I}_{3,5}(1)$, i.e., their corresponding Bell inequalities also  have the same lowest critical threshold $v_c\approx 0.595047$.
By the checking of equivalence, these most robust Bell functions for the three-partite system are equivalent to each other. Moreover, among them there exist a party symmetric one $\mathcal{I}_{3,5}(1)$,
\begin{eqnarray}{\label{I35}}
& &\mathcal{I}_{3,5}(1)=\frac{3{\rm \mathbf{v}}_{1}+2{\rm \mathbf{v}}_{2}+2{\rm \mathbf{v}}_{3}+3{\rm \mathbf{v}}_{4}}{5}A_0B_0C_0\nonumber\\
& &+
\frac{-{\rm \mathbf{v}}_{2}-3{\rm \mathbf{v}}_{3}-{\rm \mathbf{v}}_{4}}{5}(A_0B_0C_1+A_0B_1C_0+A_1B_0C_0)\nonumber\\
& &
+\frac{-3{\rm \mathbf{v}}_{1}-{\rm \mathbf{v}}_{3}-{\rm \mathbf{v}}_{4}}{5}(A_0B_1C_1+A_1B_0C_1+A_1B_1C_0)\nonumber\\
& &+
\frac{{\rm \mathbf{v}}_{1}+{\rm \mathbf{v}}_{2}-2{\rm \mathbf{v}}_{4}}{5}A_1B_1C_1,
\end{eqnarray}
whose
corresponding Bell inequality is just the one in Ref. \cite{Chen324}. See more details in Appendix D.

For $n=4$, when we take $\mathcal{I}_{3,5}^{0,0}=\mathcal{I}_{3,5}(1)$, and let $\mathcal{I}_{3,5}^{0,1}$ go through all ones (there are $1250$ elements) equivalent to $\mathcal{I}_{3,5}(1)$, the most robust $\mathcal{I}_{4,5}$ obtained by (\ref{Ind3})  also has the same robustness as $\mathcal{I}_{3,5}(1)$, i.e., the corresponding Bell inequalities have the same lowest critical threshold $v_c\approx 0.595047$ as $\mathcal{B}_{3,5}(1)$.

When we focus on the case $d=7$.

(a) For $n=3$, when we take $\mathcal{I}_{2,7}^{0,0}=\mathcal{I}_{2,7}(1)=\mathcal{I}_{2,7}$, which appears in Example 1, and let $\mathcal{I}_{2,7}^{0,1}$ go through all ones equivalent to $\mathcal{I}_{2,7}$, the most robust $\mathcal{I}_{3,7}$ obtained by (\ref{Ind3}) has the same robustness as $\mathcal{I}_{2,7}$. That is, the corresponding Bell inequalities have the same
lowest critical visibility $v_c\approx 0.683256$7 as the CGLMP inequality for $d=7$.
  Considering that the Bell  quantity $\mathcal{B}_{2,7}$ of $\mathcal{I}_{2,7}$ has three deterministic values $-\frac{4}{3},-\frac{1}{6},1$ in LHV  theory, we study all full-correlated and party symmetric
\begin{eqnarray}{\label{I27}}
\mathcal{I}_{2,7}&=&\left(\sum_{k=1}^{6}f_{1k}{\rm \mathbf{v}}_{k}\right)A_0B_0+\left(\sum_{k=1}^{6}f_{2k}{\rm \mathbf{v}}_{k}\right)(A_0B_1+A_1B_0)\nonumber\\
& &+\left(\sum_{k=1}^{6}g_{2k}{\rm \mathbf{v}}_{k}\right)A_1B_1,
\end{eqnarray}
with the following properties: (i) its corresponding Bell quantity has three deterministic values $-\frac{4}{3},-\frac{1}{6},1$ in LHV  theory; (ii) its corresponding Bell inequality has the same robustness as $\mathcal{B}_{2,7}$, i.e., $v_c\approx 0.683256$; (iii) it is not equivalent to $\mathcal{I}_{2,7}(1)$.
The reasons that we focus on party symmetric ones are (i) including all party unsymmetrical  one, we need more than one year to give all of them with these three deterministic values $-\frac{4}{3},-\frac{1}{6},1$ in LHV  theory; (ii) all the most robust $\mathcal{I}_{2,d}$ for prime $d\leq 5$ are party symmetric.
At last, we find  another three ones
\begin{eqnarray}{\label{I272}}
& &\mathcal{I}_{2,7}(2)=\frac{2{\rm \mathbf{v}}_{1}-3{\rm \mathbf{v}}_{2}-{\rm \mathbf{v}}_{3}+{\rm \mathbf{v}}_{4}-4{\rm \mathbf{v}}_{5}-2{\rm \mathbf{v}}_{6}}{7}A_0B_0\nonumber\\
& &+
\frac{-2{\rm \mathbf{v}}_{1}-4{\rm \mathbf{v}}_{2}+{\rm \mathbf{v}}_{3}-{\rm \mathbf{v}}_{4}-3{\rm \mathbf{v}}_{5}-5{\rm \mathbf{v}}_{6}}{7}(A_0B_1+A_1B_0)\nonumber\\
& &+\frac{2{\rm \mathbf{v}}_{1}-3{\rm \mathbf{v}}_{2}-{\rm \mathbf{v}}_{3}+{\rm \mathbf{v}}_{4}-4{\rm \mathbf{v}}_{5}-2{\rm \mathbf{v}}_{6}}{7}A_1B_1,\nonumber\\
& &\mathcal{I}_{2,7}(3)=\frac{-2{\rm \mathbf{v}}_{1}-4{\rm \mathbf{v}}_{2}+{\rm \mathbf{v}}_{3}-{\rm \mathbf{v}}_{4}-3{\rm \mathbf{v}}_{5}-5{\rm \mathbf{v}}_{6}}{7}A_0B_0\nonumber\\
& &+
\frac{-5{\rm \mathbf{v}}_{1}-3{\rm \mathbf{v}}_{2}-{\rm \mathbf{v}}_{3}+{\rm \mathbf{v}}_{4}-4{\rm \mathbf{v}}_{5}-2{\rm \mathbf{v}}_{6}}{7}(A_0B_1+A_1B_0)\nonumber\\
& &+\frac{-2{\rm \mathbf{v}}_{1}+3{\rm \mathbf{v}}_{2}+{\rm \mathbf{v}}_{3}-{\rm \mathbf{v}}_{4}-3{\rm \mathbf{v}}_{5}+2{\rm \mathbf{v}}_{6}}{7}A_1B_1,\nonumber\end{eqnarray}
\begin{eqnarray}
& &\mathcal{I}_{2,7}(4)=\frac{-4{\rm \mathbf{v}}_{1}-{\rm \mathbf{v}}_{2}+2{\rm \mathbf{v}}_{3}-2{\rm \mathbf{v}}_{4}+{\rm \mathbf{v}}_{5}-3{\rm \mathbf{v}}_{6}}{7}A_0B_0\nonumber\\
& &+
\frac{4{\rm \mathbf{v}}_{1}+{\rm \mathbf{v}}_{2}+5{\rm \mathbf{v}}_{3}+2{\rm \mathbf{v}}_{4}+6{\rm \mathbf{v}}_{5}+3{\rm \mathbf{v}}_{6}}{7}(A_0B_1+A_1B_0)\nonumber\\
& &+\frac{3{\rm \mathbf{v}}_{1}+6{\rm \mathbf{v}}_{2}+2{\rm \mathbf{v}}_{3}+5{\rm \mathbf{v}}_{4}+{\rm \mathbf{v}}_{5}+4{\rm \mathbf{v}}_{6}}{7}A_1B_1.
\end{eqnarray}
{\bf Hence, we find other three $(2,2,7)$-scenario Bell inequalities, which are as robust as the CGLMP inequality for $d=7$, but not equivalent to it.} See Appendix D for their corresponding Bell inequalities.

For $n=3$, when we take $\mathcal{I}_{2,7}^{0,0}=\mathcal{I}_{2,7}(i),i=1,2,3,4$, and let $\mathcal{I}_{2,7}^{0,1}$ go through all ones equivalent to all $\mathcal{I}_{2,7}(i),i=1,2,3,4$, the most robust $\mathcal{I}_{3,7}$ obtained by (\ref{Ind3}) also has the same robustness as $\mathcal{I}_{2,7}$.
	
\section{Discussion}

In this paper, we generalize the multi-component correlation functions for bipartite $d$-dimensional systems to $n$-partite $d$-dimensional systems, and construct the corresponding Bell inequality. The CGLMP inequality can be obtained by this way through choosing appropriate coefficients. We also simplify the general $(n,2,d)$ full-correlated multi-component Bell function to an iteration formula  for prime $d$ with two $(n-1,2,d)$ full-correlated multi-component Bell functions. In fact, we give a kind of efficient way that constructs $(n,2,d)$ Bell inequality with $(n-1,2,d)$ ones using the iteration formula for prime $d$. By taking appropriate coefficients, it can be reformulated to the MABK inequality,  the most robust $(n,2,3)$ coincidence Bell inequalities, and the most robust $(3,2,5)$ Bell inequality presented in \cite{Chen324}. With the iteration formula, we also  give Bell inequalities with the same robustness but inequivalent to the known ones. In the future, we wish to generalize the above results to multi-settings and hope to obtain more robust Bell inequality for $n$-partite $d$-dimensional systems. Moreover, the Bell inequality on multipartite, multi-settings, and high-dimensional systems has widely application, such as the construction of Hardy-type paradox  \cite{Hardy1,Hardy3,Hardy4,Hardy's paradox1,Hardy2,Stronger Hardy-type paradox}, quantum communication \cite{Quantum Communication}, quantum computing \cite{quantum computing}, and quantum network \cite{machine learning,Quantum Networks1,Quantum Networks2,Quantum Networks3,Quantum Networks4}. We hope that the iteration formula of generic Bell inequalities can promote the development of related fields.
	
\begin{acknowledgments}

\section{Acknowledgements}
	
	H. X. M. was supported by the National Natural Science Foundations of China (Grant No. 11901317), the Project funded by
China Postdoctoral Science Foundation (Grant No. 2020M680480), the Fundamental Research Funds for the Central Universities (Grant No. 2023MS078), and  the Beijing Natural Science Foundation(Grant No.1232021).
J. L. C. was supported by the National Natural Science Foundations of China (Grant Nos. 12275136 and 12075001), the 111 Project of B23045. X. Y. F. was supported by the Nankai Zhide Foundations.
	\end{acknowledgments}

\onecolumngrid

\appendix

\section{The proof of projectors' representation with the observable}

{\bf Observation 1} If the observable $X$ can be represented by $
X=\sum_{i=0}^{N}{\rm \mathbf{v}}_i\Pi_i,
$
where  $\{\Pi_i\}$ corresponds to an orthonormal basis(ONB), then
$
{\rm \mathbf{v}}_0\Pi_i=\frac{{\rm \mathbf{v}}_{0} I+{\rm \mathbf{v}}_{-i} X+
{\rm \mathbf{v}}_{-2i} X^2+\cdots+{\rm \mathbf{v}}_{-(d-1)i} X^{d-1}}{d},\ i=0,1,\cdots,d-1.
$

{\bf Proof} To represent the projectors with the observable, we need to solve the equations:
\begin{eqnarray}
& &{\rm \mathbf{v}}_0 \Pi_0+{\rm \mathbf{v}}_0 \Pi_1+{\rm \mathbf{v}}_0 \Pi_2+\cdots +{\rm \mathbf{v}}_0 \Pi_{d-1}={\rm \mathbf{v}}_0 I,\nonumber\\
& &{\rm \mathbf{v}}_0 \Pi_0+{\rm \mathbf{v}}_1 \Pi_1+{\rm \mathbf{v}}_2 \Pi_2+\cdots+{\rm \mathbf{v}}_{d-1} \Pi_{d-1}=X,\nonumber\\
& &{\rm \mathbf{v}}_0 \Pi_0+{\rm \mathbf{v}}_2 \Pi_1+{\rm \mathbf{v}}_4 \Pi_2+\cdots+{\rm \mathbf{v}}_{{\rm Mod}[2(d-1),d]} \Pi_{d-1}=X^2,\nonumber\\
& &{\rm \mathbf{v}}_0 \Pi_0+{\rm \mathbf{v}}_3 \Pi_1+{\rm \mathbf{v}}_{{\rm Mod}[3\times 2,d]} \Pi_2+{\rm \mathbf{v}}_{{\rm Mod}[3\times 3,d]} \Pi_3+\cdots+{\rm \mathbf{v}}_{{\rm Mod}[3(d-1),d]} \Pi_{d-1}=X^3,\cdots\nonumber\\
& &{\rm \mathbf{v}}_0 \Pi_0+{\rm \mathbf{v}}_{{\rm Mod}[(d-1),d]} \Pi_1+{\rm \mathbf{v}}_{{\rm Mod}[2(d-1),d]} \Pi_2+{\rm \mathbf{v}}_{{\rm Mod}[3(d-1),d]} \Pi_3+\cdots+{\rm \mathbf{v}}_{{\rm Mod}[(d-1)(d-1),d]} \Pi_{d-1}=X^{d-1}.
\end{eqnarray}
which can be described in matrix language as,
\begin{eqnarray}{\label{A2}}
\left(
  \begin{array}{ccccc}
    {\rm \mathbf{v}}_{0} & {\rm \mathbf{v}}_{0} & {\rm \mathbf{v}}_{0} & \cdots & {\rm \mathbf{v}}_{0}   \\
    {\rm \mathbf{v}}_{0} & {\rm \mathbf{v}}_{1} & {\rm \mathbf{v}}_{2} & \cdots & {\rm \mathbf{v}}_{d-1} \\
    {\rm \mathbf{v}}_{0} & {\rm \mathbf{v}}_{2} & {\rm \mathbf{v}}_{4} & \cdots & {\rm \mathbf{v}}_{{\rm Mod}[2(d-1),d]} \\
    \vdots & \vdots & \vdots & \vdots & \vdots  \\
    {\rm \mathbf{v}}_{0} & {\rm \mathbf{v}}_{{\rm Mod}[(d-1),d]} & {\rm \mathbf{v}}_{{\rm Mod}[2(d-1),d]} & \cdots & {\rm \mathbf{v}}_{{\rm Mod}[(d-1)(d-1),d]}  \\
  \end{array}
\right)\left(
  \begin{array}{c}
    \Pi_0  \\
    \Pi_1  \\
    \Pi_2  \\
    \vdots  \\
    \Pi_{d-1}  \\
  \end{array}
\right)=\left(
  \begin{array}{c}
    {\rm \mathbf{v}}_{0}I \\
     X\\
     X^2\\
    \vdots  \\
      X^{d-1} \\
  \end{array}
\right).
\end{eqnarray}
 For prime $d$,
\begin{eqnarray}{\label{A3}}
&&\left(
  \begin{array}{ccccc}
    {\rm \mathbf{v}}_{0} & {\rm \mathbf{v}}_{0} & {\rm \mathbf{v}}_{0} & \cdots & {\rm \mathbf{v}}_{0} \\
    {\rm \mathbf{v}}_{0} & {\rm \mathbf{v}}_{-1} & {\rm \mathbf{v}}_{-2} & \cdots & {\rm \mathbf{v}}_{-d+1} \\
    {\rm \mathbf{v}}_{0} & {\rm \mathbf{v}}_{-2} & {\rm \mathbf{v}}_{-4} & \cdots & {\rm \mathbf{v}}_{{\rm Mod}[-2(d-1),d]} \\
    \vdots & \vdots & \vdots & \vdots & \vdots \\
    {\rm \mathbf{v}}_{0} & {\rm \mathbf{v}}_{{\rm Mod}[-(d-1),d]} & {\rm \mathbf{v}}_{{\rm Mod}[-2(d-1),d]} & \cdots & {\rm \mathbf{v}}_{{\rm Mod}[-(d-1)(d-1),d]} \\
  \end{array}
\right)\nonumber\\
&&\left(
  \begin{array}{cccccc}
    {\rm \mathbf{v}}_{0} & {\rm \mathbf{v}}_{0} & {\rm \mathbf{v}}_{0} & \cdots & {\rm \mathbf{v}}_{0} & {\rm \mathbf{v}}_{0}I \\
    {\rm \mathbf{v}}_{0} & {\rm \mathbf{v}}_{1} & {\rm \mathbf{v}}_{2} & \cdots & {\rm \mathbf{v}}_{d-1} & X\\
    {\rm \mathbf{v}}_{0} & {\rm \mathbf{v}}_{2} & {\rm \mathbf{v}}_{4} & \cdots & {\rm \mathbf{v}}_{{\rm Mod}[2(d-1),d]} & X^2\\
    \vdots & \vdots & \vdots & \vdots & \vdots & \vdots \\
    {\rm \mathbf{v}}_{0} & {\rm \mathbf{v}}_{{\rm Mod}[(d-1),d]} & {\rm \mathbf{v}}_{{\rm Mod}[2(d-1),d]} & \cdots & {\rm \mathbf{v}}_{{\rm Mod}[(d-1)(d-1),d]} &  X^{d-1} \\
  \end{array}
\right)\nonumber
\end{eqnarray}
\begin{eqnarray}
&=&\left(
     \begin{array}{cccccc}
       d{\rm \mathbf{v}}_{0} & {\rm \mathbf{0}} & {\rm \mathbf{0}} & \cdots & {\rm \mathbf{0}} & {\rm \mathbf{v}}_{0}I+X+X^2+\cdots+X^{d-1} \\
       {\rm \mathbf{0}} & d{\rm \mathbf{v}}_{0} & {\rm \mathbf{0}} & \cdots & {\rm \mathbf{0}} & {\rm \mathbf{v}}_{0}I+{\rm \mathbf{v}}_{-1}X+{\rm \mathbf{v}}_{-2}X^2
+\cdots+{\rm \mathbf{v}}_{-(d-1)}X^{d-1} \\
       {\rm \mathbf{0}} & {\rm \mathbf{0}} & d{\rm \mathbf{v}}_{0} & \cdots & {\rm \mathbf{0}} & {\rm \mathbf{v}}_{0}I+{\rm \mathbf{v}}_{-2}X+{\rm \mathbf{v}}_{-4}X^2
+\cdots+{\rm \mathbf{v}}_{-2(d-1)}X^{d-1} \\
       \vdots & \vdots & \vdots & \ddots & \vdots & \vdots \\
       {\rm \mathbf{0}} & {\rm \mathbf{0}} &  \cdots & {\rm \mathbf{0}} & d{\rm \mathbf{v}}_{0} & {\rm \mathbf{v}}_{0}I+{\rm \mathbf{v}}_{-(d-1)}X+{\rm \mathbf{v}}_{-2(d-1)}X^2
+\cdots+{\rm \mathbf{v}}_{-(d-1)(d-1)}X^{d-1} \\
     \end{array}
   \right)
\end{eqnarray}
holds. Hence, when we left multiply ($\circ$) the first matrix in (\ref{A2}) to the equation (\ref{A3}), we obtain
\begin{eqnarray}
\left(
  \begin{array}{c}
    d\Pi_0  \\
    d\Pi_1  \\
    d\Pi_2  \\
    \vdots  \\
    d\Pi_{d-1}  \\
  \end{array}
\right)=\left(
     \begin{array}{c}
     {\rm \mathbf{v}}_{0}I+X+X^2+\cdots+X^{d-1} \\
        {\rm \mathbf{v}}_{0}I+{\rm \mathbf{v}}_{-1}X+{\rm \mathbf{v}}_{-2}X^2
+\cdots+{\rm \mathbf{v}}_{-(d-1)}X^{d-1} \\
        {\rm \mathbf{v}}_{0}I+{\rm \mathbf{v}}_{-2}X+{\rm \mathbf{v}}_{-4}X^2
+\cdots+{\rm \mathbf{v}}_{-2(d-1)}X^{d-1} \\
        \vdots \\
         {\rm \mathbf{v}}_{0}I+{\rm \mathbf{v}}_{-(d-1)}X+{\rm \mathbf{v}}_{-2(d-1)}X^2
+\cdots+{\rm \mathbf{v}}_{-(d-1)(d-1)}X^{d-1} \\
     \end{array}
   \right)
\end{eqnarray}
which implies that
\begin{eqnarray}
{\rm \mathbf{v}}_0\Pi_i=\frac{{\rm \mathbf{v}}_{0} I+{\rm \mathbf{v}}_{-i} X+
{\rm \mathbf{v}}_{-2i} X^2+\cdots+{\rm \mathbf{v}}_{-(d-1)i} X^{d-1}}{d},\ i=0,1,\cdots,d-1.
\end{eqnarray}
$\Box$

\section{ The proof of Example 1}

{\bf Example 1} For $n=2$ and $d\geq 2$,  when we take the multi-component Bell function as
\begin{eqnarray}\label{AI2d}
\mathcal{I}_{2,d}=\frac{1}{d}\left[-\sum_{k=1}^{d-1}k{\rm \mathbf{v}}_{d-k}A_0B_0-\sum_{k=1}^{d-1}k{\rm \mathbf{v}}_{k}(A_0B_1+A_1B_0)+\sum_{k=1}^{d-1}k{\rm \mathbf{v}}_{k}A_1B_1\right],
\end{eqnarray}
the corresponding Bell inequality is
\begin{eqnarray}\mathcal{B}_{2,d}=-\frac{1}{d-1}\sum_{k=1}^{d-1}k(P(A_0+B_0=k)+
P(A_0+B_1=-k)+P(A_1+B_0=-k)-P(A_1+B_1=-k))\overset{{\rm LHV}}{\leq}1,\end{eqnarray}
which is equivalent to the CGLMP inequality for two-party and $d$-dimensional systems \cite{CGLMP}, and also equivalent to the Bell inequality
$
\mathcal{B}_{d}=\sum_{k=0}^{N-1}\sqrt{\frac{(N+1-k)(N-k)}{(N+1)N}}\mathcal{B}^{(k)}\overset{{\rm LHV}}{\leq}2,
$
in Ref. \cite{Multicomponent} ignoring a constant,
where $\mathcal{B}^{(0)}=Q_{00}^{(0)}+Q_{01}^{(0)}-Q_{10}^{(0)}+Q_{11}^{(0)},
\mathcal{B}^{(k)}=Q_{00}^{(k)}-Q_{01}^{(k)}-Q_{10}^{(k)}+Q_{11}^{(k)},k\neq 0.$

{\bf Proof} For $n=2$ and $d\geq 2$, the multi-component Bell function is
\begin{eqnarray}
\mathcal{I}_{2,d}&=&-\frac{1}{d}\left(\sum_{k=1}^{d-1}k{\rm \mathbf{v}}_{d-k}\right)A_1B_1-\frac{1}{d}\left(\sum_{k=1}^{d-1}k{\rm \mathbf{v}}_{k}\right)(A_1B_2+A_2B_1)+\frac{1}{d}\left(\sum_{k=1}^{d-1}k{\rm \mathbf{v}}_{k}\right)A_2B_2\nonumber\\
&=&-\frac{1}{d}\sum_{k=1}^{d-1}k\sum_{k'=0}^{d-1}({\rm \mathbf{v}}_{d-k+k'}P(A_1+B_1=k')+{\rm \mathbf{v}}_{k+k'}(P(A_1+B_2=k')+P(A_2+B_1=k'))-{\rm \mathbf{v}}_{k+k'}P(A_2+B_2=k')).\nonumber\\
& &
\end{eqnarray}
Its corresponding Bell quantity is
\begin{eqnarray}
\mathcal{B}_{2,d}&=&-\frac{1}{d}\sum_{k=1}^{d-1}k\left(P(A_1+B_1=k)-\frac{1}{d-1}\sum_{k'\neq k}P(A_1+B_1=k')\right)\nonumber\\
& &-\frac{1}{d}\sum_{k=1}^{d-1}k\left(P(A_1+B_2=d-k)+P(A_2+B_1=d-k)-\frac{1}{d-1}\sum_{k'\neq d-k}(P(A_1+B_2=k')+P(A_2+B_1=k'))\right)\nonumber\\
& &+\frac{1}{d}\sum_{k=1}^{d-1}k\left(P(A_2+B_2=d-k)-\frac{1}{d-1}\sum_{k'\neq d-k}P(A_2+B_2=k')\right)\nonumber\\
&=&-\frac{1}{d}\sum_{k=1}^{d-1}k\left(\frac{d}{d-1}P(A_1+B_1=k)-\frac{1}{d-1}\right)+\frac{1}{d}\sum_{k=1}^{d-1}k\left(\frac{d}{d-1}P(A_2+B_2=d-k)-\frac{1}{d-1}\right)\nonumber\\
& &-\frac{1}{d}\sum_{k=1}^{d-1}k\left(\frac{d}{d-1}(P(A_1+B_2=d-k)+P(A_2+B_1=d-k))-\frac{2}{d-1}\right)\nonumber\\
&=&-\frac{1}{d-1}\sum_{k=1}^{d-1}k(P(A_1+B_1=k)+P(A_1+B_2=d-k)+P(A_2+B_1=d-k)-P(A_2+B_2=d-k))+1.
\end{eqnarray}
When we replace $b_i$ by $d-b_i$, $\mathcal{B}_{2,d}$ becomes
\begin{eqnarray}\label{CGLMP}
\mathcal{B}_{2,d}'&=&-\frac{1}{d-1}\sum_{k=1}^{d-1}k\{[P(A_1+B_1=-k)+P(A_1+B_2=k)]+[P(A_2+B_1=k)-P(A_2+B_2=k)]\}+1\nonumber\\
&=&\sum_{k=0}^{[d/2]-1}\left(\frac{1}{2}-\frac{k}{d-1}\right)(P(A_1+B_1=-k)-P(A_1+B_1=k+1)+P(A_2+B_2=-k-1)-[P(A_2+B_2=k))\nonumber\\
& &+\sum_{k=0}^{[d/2]-1}\left(\frac{1}{2}-\frac{k}{d-1}\right)(P(A_1+B_2=k)+P(A_2+B_1=k)-P(A_1+B_2=-k-1)-P(A_2+B_1=-k-1)),\nonumber\\
&\overset{{\rm LHV}}{\leq}& 1.
\end{eqnarray}
Then we replace $a_1$ by $d-a_1$ and $a_2$ by $d-(a_2+1)$ respectively, and obtain $\mathcal{B}_{2,d}'$
 \begin{eqnarray}\label{CGLMP}
& &\sum_{k=0}^{[d/2]-1}\left(\frac{1}{2}-\frac{k}{d-1}\right)(P(A_1=B_1+k)-P(A_1=B_1-k-1)+P(A_2=B_2+k)-[P(A_2=B_2-k-1))\nonumber\\
& &+\sum_{k=0}^{[d/2]-1}\left(\frac{1}{2}-\frac{k}{d-1}\right)(P(B_2=A_1+k)+P(B_1=A_2+k+1)-P(B_2=A_1-k-1)-P(B_1=A_2-k))\overset{{\rm LHV}}{\leq} 1,\ \ \ \
\end{eqnarray}
which is  the  CGLMP inequality ignoring a constant.
In a word, the Bell inequality $\mathcal{B}_{2,d}\overset{{\rm LHV}}{\leq}1$ is outcome symmetric with the CGLMP inequality.

The Bell quantity $\mathcal{B}_{d}$ in Ref. \cite{Multicomponent} is
\begin{eqnarray}
\mathcal{B}_{d}=\sum_{k=0}^{N-1}\sqrt{\frac{(N+1-k)(N-k)}{(N+1)N}}\mathcal{B}^{(k)},
\end{eqnarray}
where $\mathcal{B}^{(0)}=Q_{11}^{(0)}+Q_{12}^{(0)}-Q_{21}^{(0)}+Q_{22}^{(0)},
\mathcal{B}^{(k)}=Q_{11}^{(k)}-Q_{12}^{(k)}-Q_{21}^{(k)}+Q_{22}^{(k)},k\neq 0.$
Due to
\begin{eqnarray}
& &{\rm \mathbf{v}}_0=(1,0,0,...,0),\nonumber\\
& &{\rm \mathbf{v}}_1=\left(-\frac{1}{N},\frac{\sqrt{N^2-1}}{N},0,...,0\right),\nonumber\\
& &{\rm \mathbf{v}}_2
=\left(-\frac{1}{N},-\frac{1}{N}\sqrt{\frac{N+1}{N-1}},
\sqrt{\frac{(N-2)*(N+1)}{N*(N-1)}},0,0,...,0\right),...,\nonumber\\
& &{\rm \mathbf{v}}_{N-1}=\left(-\frac{1}{N},-\frac{1}{N}\sqrt{\frac{N+1}{N-1}},
-\frac{1}{N}\sqrt{\frac{(N+1)*N}{(N-1)*(N-2)}},...,
-\frac{1}{N}\sqrt{\frac{(N+1)*N}{3*2}},\frac{1}{N}\sqrt{\frac{(N+1)*N}{2*1}}\right),\nonumber\\
& &{\rm \mathbf{v}}_{N}=\left(-\frac{1}{N},-\frac{1}{N}\sqrt{\frac{N+1}{N-1}},
-\frac{1}{N}\sqrt{\frac{(N+1)*N}{(N-1)*(N-2)}},...,
-\frac{1}{N}\sqrt{\frac{(N+1)*N}{3*2}},-\frac{1}{N}\sqrt{\frac{(N+1)*N}{2*1}}\right),
\end{eqnarray}
and
\begin{eqnarray}
\vec{Q}_{ij}=\sum_{k=0}^{d-1}{\rm \mathbf{v}}_kP(A_i+B_j=k)=\left(Q_{ij}^{(0)},Q_{ij}^{(1)},...,Q_{ij}^{(d-2)}\right),
\end{eqnarray}
we obtain
\begin{eqnarray}
& &Q_{ii}^{(0)}+\sum_{k=1}^{N-1}\sqrt{\frac{(N+1-k)(N-k)}{(N+1)N}}Q_{ii}^{(k)}\nonumber\\
&=&P(A_i+B_i=0)-\frac{1}{N}\sum_{k=1}^{d-1}P(A_i+B_i=k)+\frac{N-1}{N}P(A_i+B_i=1)-\frac{1}{N}\sum_{k=2}^{d-1}P(A_i+B_i=k)\nonumber\\
& &+\frac{N-2}{N}P(A_i+B_i=2)-\frac{1}{N}\sum_{k=3}^{d-1}P(A_i+B_i=k)+\cdots+\frac{1}{N}P(A_i+B_i=d-2)-\frac{1}{N}P(A_i+B_i=d-1)\nonumber\\
&=&1-\frac{2}{N}P(A_i+B_i=1)-\frac{4}{N}P(A_i+B_i=2)-\cdots+\left(-1-\frac{d-3}{N}\right)P(A_i+B_i=d-2)\nonumber\\
& &+\left(-1-\frac{d-1}{N}\right)P(A_i+B_i=d-1)\nonumber\\
&=&1-\frac{2}{N}\sum_{k=1}^{d-1}kP(A_i+B_i=k)
\end{eqnarray}
for $i=1,2$, and
\begin{eqnarray}
& &Q_{12}^{(0)}-\sum_{k=1}^{N-1}\sqrt{\frac{(N+1-k)(N-k)}{(N+1)N}}Q_{12}^{(k)}=1-\frac{2}{N}\sum_{k=1}^{d-1}kP(A_1+B_2=-k)\\
& &-Q_{21}^{(0)}-\sum_{k=1}^{N-1}\sqrt{\frac{(N+1-k)(N-k)}{(N+1)N}}Q_{21}^{(k)}=-1+\frac{2}{N}\sum_{k=1}^{d-1}kP(A_2+B_1=k).
\end{eqnarray}
Therefore,
\begin{eqnarray}
\mathcal{B}_{d}=2-\frac{2}{N}\sum_{k=1}^{d-1}k(P(A_1+B_1=k)+P(A_2+B_2=k)+P(A_1+B_2=-k)-P(A_2+B_1=k)).
\end{eqnarray}
When we replace $a_i$ by $-a_i$ and $b_i$ by $-b_i$ respectively, and exchange $B_1$ and $B_2$, then $\frac{\mathcal{B}_{d}}{2}$ becomes $\mathcal{B}_{2,d}$. $\Box$

\section{The proof of Theorem 1}

{\bf Theorem 1} For prime $d$,  the Bell function $\mathcal{I}_{n,d}$
 can be reduced to
\begin{eqnarray}
\mathcal{I}_{n,d}=\frac{1}{d}\left(
                \begin{array}{cc}
                  \mathcal{I}_{n-1,d}^{0,0} & \mathcal{I}_{n-1,d}^{0,1} \\
                \end{array}
              \right)\left(
                       \begin{array}{cc}
                         -\sum_{k=1}^{d-1}(d-k){\rm \mathbf{v}}_{k} & -\sum_{k=1}^{d-1}k{\rm \mathbf{v}}_k \\
                         -\sum_{k=1}^{d-1}k{\rm \mathbf{v}}_k & \sum_{k=1}^{d-1}k{\rm \mathbf{v}}_k \\
                       \end{array}
                     \right)\left(
                              \begin{array}{c}
                                C_0 \\
                                C_1 \\
                              \end{array}
                            \right),
\end{eqnarray}
where $\mathcal{I}_{n-1,d}^{0,0}$ and $\mathcal{I}_{n-1,d}^{0,1}$ are full-correlated multi-component Bell functions for $(n-1,2,d)$-scenario.

{\bf Proof} Since
\begin{eqnarray}
\mathcal{I}_{n-1,d}^{k_0,k_1}=\mathcal{I}_{n,d}|_{C_0=
{\rm \mathbf{v}}_{k_0},C_1={\rm \mathbf{v}}_{k_1}}=\sum_{i,\cdots,j=0,1}\omega_{i,\cdots,j,0} {\rm \mathbf{v}}_{k_0} A_i\cdots B_j+\sum_{i,\cdots,j=0,1}\omega_{i,\cdots,j,1} {\rm \mathbf{v}}_{k_1} A_i\cdots B_j,
\end{eqnarray}
holds,
 we have
\begin{eqnarray}
\mathcal{I}_{n-1,d}^{k_0,k_1}={\rm \mathbf{v}}_{k_0} \mathcal{I}_{n,d}|_{C_0={\rm \mathbf{v}}_{0},C_1={\rm \mathbf{v}}_{k_1- k_0}}
=
{\rm \mathbf{v}}_{k_0}\mathcal{I}_{n,d}^{0,{\rm Mod}[k_1- k_0,d]},
\end{eqnarray}
and then
\begin{eqnarray}
\mathcal{I}_{n,d}&=&\mathcal{I}_{n-1,d}^{0,0}({\rm \mathbf{v}}_0\Pi_0\Pi'_0+{\rm \mathbf{v}}_1\Pi_1\Pi'_1+\cdots+
{\rm \mathbf{v}}_{d-1}\Pi_{d-1}\Pi'_{d-1})
+\mathcal{I}_{n-1,d}^{0,1}({\rm \mathbf{v}}_0\Pi_0\Pi'_1+{\rm \mathbf{v}}_1\Pi_1\Pi'_2+
\cdots+{\rm \mathbf{v}}_{d-1}\Pi_{d-1}\Pi'_{0})+\nonumber\\
& &\cdots\nonumber\\
& &+\mathcal{I}_{n-1,d}^{0,d-1}({\rm \mathbf{v}}_0\Pi_0\Pi'_{d-1}+{\rm \mathbf{v}}_1\Pi_1\Pi'_0+\cdots+{\rm \mathbf{v}}_{d-1}\Pi_{d-1}\Pi'_{d-2}).\nonumber\\
& &
\end{eqnarray}
By the representation of projectors with the observable, we have
\begin{eqnarray}
{\rm \mathbf{v}}_0\Pi_i\Pi'_j=\frac{(\sum_{k=0}^{d-1}{\rm \mathbf{v}}_{-ki} (C_0)^k) (\sum_{k=0}^{d-1}{\rm \mathbf{v}}_{-kj} (C_1)^k)}{d^2}=\frac{\sum_{k,k'}{\rm \mathbf{v}}_{-(ki+k'j)} (C_0)^k(C_1)^{k'}}{d^2},
\end{eqnarray}
and then
\begin{eqnarray}
{\rm \mathbf{v}}_0\Pi_0\Pi'_l+{\rm \mathbf{v}}_1\Pi_1\Pi'_{l+1}+\cdots+{\rm \mathbf{v}}_{d-1}\Pi_{d-1}\Pi'_{l-1}
&=&\sum_{i=0}^{d-1}{\rm \mathbf{v}}_i\Pi_i\Pi'_{{\rm Mod}[i+l,d]}\nonumber\\
&
=&\frac{\sum_{i,k,k'=0,...,d-1}{\rm \mathbf{v}}_i{\rm \mathbf{v}}_{-(ki+k'(i+l))} (C_0)^k (C_1)^{k'}}{d^2}\nonumber\\
&
=&\frac{\sum_{r=0}^{2(d-1)}\sum_{k={\rm Max}[r-d+1,0]}^{{\rm Min}[r,d-1]}{\rm \mathbf{v}}_{-kl}\sum_{i}{\rm \mathbf{v}}_{i(1-r)} (C_0)^{r-k} (C_1)^{k}}{d^2}\nonumber\\
&=&\frac{d({\rm \mathbf{v}}_{0} (C_0)+{\rm \mathbf{v}}_{-l} (C_1)+\sum_{k=2}^{d-1}{\rm \mathbf{v}}_{-kl} (C_0)^{d+1-k} (C_1)^{k})}{d^2}
\nonumber\\
&
=&\frac{{\rm \mathbf{v}}_{0} C_0+{\rm \mathbf{v}}_{-l} C_1+
\sum_{k=2}^{d-1}{\rm \mathbf{v}}_{(k-1)l} (C_0)^{k} (C_1)^{d+1-k}}{d}.
\end{eqnarray}
Hence,
\begin{eqnarray}\label{Ind}
\mathcal{I}_{n,d}&=&\sum_{l=0}^{d-1}\mathcal{I}_{n-1,d}^{0,l}\sum_{i=0}^{d-1}{\rm \mathbf{v}}_i\Pi_i\Pi'_{{\rm Mod}[i+l,d]}\nonumber\\
&=&\frac{1}{d}\sum_{l=0}^{d-1}\mathcal{I}_{n-1,d}^{0,l}C_0+\frac{1}{d}\sum_{l=0}^{d-1}{\rm \mathbf{v}}_{-l}\mathcal{I}_{n-1,d}^{0,l}C_1
+\frac{1}{d}\sum_{k=2}^{d-1}\sum_{l=0}^{d-1}({\rm \mathbf{v}}_{(k-1)l}\mathcal{I}_{N-1,2,d}^{0,l})(C_0)^{2}(C_1)^{d+1-k}\ \ \
\end{eqnarray}
for $d\geq 3$, and
\begin{eqnarray}
\mathcal{I}_{n,2}&=&\frac{1}{2}(\mathcal{I}_{n-1,2}^{0,0}+\mathcal{I}_{n-1,2}^{0,1})C_0+
\frac{1}{2}(\mathcal{I}_{n-1,2}^{0,0}+{\rm \mathbf{v}}_{1}\mathcal{I}_{n-1,2}^{0,1})C_1\nonumber\\
&=&\frac{1}{2}\left(
                \begin{array}{cc}
                  \mathcal{I}_{n-1,2}^{0,0} & \mathcal{I}_{n-1,2}^{0,1} \\
                \end{array}
              \right)\left(
                       \begin{array}{cc}
                         {\rm \mathbf{v}}_{0} & {\rm \mathbf{v}}_0 \\
                        {\rm \mathbf{v}}_0 & {\rm \mathbf{v}}_1 \\
                       \end{array}
                     \right)\left(
                              \begin{array}{c}
                                C_0 \\
                                C_1 \\
                              \end{array}
                            \right)=\frac{1}{2}\left(
                \begin{array}{cc}
                  \mathcal{I}_{n-1,2}^{0,0} & \mathcal{I}_{n-1,2}^{0,1} \\
                \end{array}
              \right)\left(
                       \begin{array}{cc}
                         -{\rm \mathbf{v}}_1 & -{\rm \mathbf{v}}_1 \\
                        -{\rm \mathbf{v}}_1 & {\rm \mathbf{v}}_1 \\
                       \end{array}
                     \right)\left(
                              \begin{array}{c}
                                C_0 \\
                                C_1 \\
                              \end{array}
                            \right).
\end{eqnarray}
In the following, we focus on the cases $d\geq 3$. Since $C_0^kC_1^{d+1-k},k=2,3,...,d-1$ can not appear in quantum mechanic, we have
\begin{eqnarray}
\sum_{l=0}^{d-1}{\rm \mathbf{v}}_{(k-1)l}\mathcal{I}_{n-1,d}^{0,l}={\rm \mathbf{0}},k=2,3,...,d-1.
\end{eqnarray}
By solving these equations, we have
\begin{eqnarray}
&&\mathcal{I}_{n-1,d}^{0,2}=-{\rm \mathbf{v}}_{1}\mathcal{I}_{n-1,d}^{0,0}+({\rm \mathbf{v}}_{0}+{\rm \mathbf{v}}_{1})\mathcal{I}_{n-1,d}^{0,1},\nonumber\\
&&\mathcal{I}_{n-1,d}^{0,3}=-({\rm \mathbf{v}}_{1}+{\rm \mathbf{v}}_{2})\mathcal{I}_{n-1,d}^{0,0}+({\rm \mathbf{v}}_{0}+{\rm \mathbf{v}}_{1}+{\rm \mathbf{v}}_{2}),\nonumber\\
&&\cdots,\nonumber\\
&&\mathcal{I}_{n-1,d}^{0,[d/2]}=-\sum_{k=1}^{[d/2]-1}{\rm \mathbf{v}}_{k}\mathcal{I}_{n-1,d}^{0,0}+\sum_{k=0}^{[d/2]-1}{\rm \mathbf{v}}_{k}\mathcal{I}_{n-1,d}^{0,1},\nonumber\\
&&\mathcal{I}_{n-1,d}^{0,[d/2]+1}=-\sum_{k=1}^{[d/2]}{\rm \mathbf{v}}_{k}\mathcal{I}_{n-1,d}^{0,0}-\sum_{k=[d/2]+1}^{d-1}{\rm \mathbf{v}}_{k}\mathcal{I}_{n-1,d}^{0,1},\nonumber\\
&&\mathcal{I}_{n-1,d}^{0,[d/2]+2}=\sum_{k=[d/2]+2}^{d}{\rm \mathbf{v}}_{k}\mathcal{I}_{n-1,d}^{0,0}-\sum_{k=[d/2]+2}^{d-1}{\rm \mathbf{v}}_{k}\mathcal{I}_{n-1,d}^{0,1},\nonumber\\
&&\mathcal{I}_{n-1,d}^{0,[d/2]+3}=\sum_{k=[d/2]+3}^{d}{\rm \mathbf{v}}_{k}\mathcal{I}_{n-1,d}^{0,0}-\sum_{k=[d/2]+2}^{d-1}{\rm \mathbf{v}}_{k}\mathcal{I}_{n-1,d}^{0,1},\nonumber\\
&&\cdots,\nonumber\\
&&\mathcal{I}_{n-1,d}^{0,d-1}=({\rm \mathbf{v}}_{d-1}+{\rm \mathbf{v}}_{0})\mathcal{I}_{n-1,d}^{0,0}-{\rm \mathbf{v}}_{d-1}\mathcal{I}_{n-1,d}^{0,1}.
\end{eqnarray}
Take them into (\ref{Ind}), we obtain
\begin{eqnarray}
\mathcal{I}_{n,d}&=&\frac{1}{d}\sum_{l=0}^{d-1}\mathcal{I}_{n-1,d}^{0,l}C_0
+\frac{1}{d}\sum_{l=0}^{d-1}{\rm \mathbf{v}}_{-l}\mathcal{I}_{n-1,d}^{0,l}C_1\nonumber\\
&=&\frac{1}{d}\left([d/2]{\rm \mathbf{v}}_{0}-\sum_{k=1}^{[d/2]}([d/2]+1-k){\rm \mathbf{v}}_{k}+\sum_{k=[d/2]+2}^{d-1}(k-[d/2]-1){\rm \mathbf{v}}_{k}\right)\mathcal{I}_{n-1,7}^{0,0}C_0\nonumber\\
& &+
\frac{1}{d}\left([d/2]{\rm \mathbf{v}}_{0}+\sum_{k=1}^{[d/2]-1}([d/2]-k){\rm \mathbf{v}}_{k}-\sum_{k=[d/2]+1}^{d-1}(k-[d/2]){\rm \mathbf{v}}_{k}\right)\mathcal{I}_{n-1,7}^{0,1}C_0\nonumber\\
& &+
\frac{1}{d}\left([d/2]{\rm \mathbf{v}}_{0}+\sum_{k=1}^{[d/2]-1}([d/2]-k){\rm \mathbf{v}}_{k}-\sum_{k=1}^{[d/2]}([d/2]+1-k){\rm \mathbf{v}}_{-k}\right)\mathcal{I}_{n-1,7}^{0,0}C_1\nonumber\\
& &+\frac{1}{d}\left(-[d/2]{\rm \mathbf{v}}_{0}-\sum_{k=1}^{[d/2]-1}([d/2]-k){\rm \mathbf{v}}_{k}+\sum_{k=1}^{[d/2]}([d/2]+1-k){\rm \mathbf{v}}_{-k}\right)\mathcal{I}_{n-1,7}^{0,1}C_1\nonumber\\
&=&-\frac{1}{d}\sum_{k=1}^{d-1}(d-k){\rm \mathbf{v}}_{k} \mathcal{I}_{n-1,d}^{0,0}C_0 -\frac{1}{d}\sum_{k=1}^{d-1}k{\rm \mathbf{v}}_{k} \mathcal{I}_{n-1,d}^{0,1}X^{(n)}_1-\frac{1}{d}\sum_{k=1}^{d-1}k{\rm \mathbf{v}}_{k} \mathcal{I}_{n-1,d}^{0,0}X^{(n)}_2+\frac{1}{d}\sum_{k=1}^{d-1}k{\rm \mathbf{v}}_{k} \mathcal{I}_{n-1,d}^{0,1}C_1\nonumber\\
&=&-\frac{1}{d}\left(
                 \begin{array}{cc}
                   \mathcal{I}_{n-1,d}^{0,0} & \mathcal{I}_{n-1,d}^{0,1} \\
                 \end{array}
               \right)\left(
                        \begin{array}{cc}
                          \sum_{k=1}^{d-1}k{\rm \mathbf{v}}_{d-k} & \sum_{k=1}^{d-1}k{\rm \mathbf{v}}_{k} \\
                          \sum_{k=1}^{d-1}k{\rm \mathbf{v}}_{k} & -\sum_{k=1}^{d-1}k{\rm \mathbf{v}}_{k} \\
                        \end{array}
                      \right)\left(
                               \begin{array}{c}
                                 C_0 \\
                                C_1 \\
                               \end{array}
                             \right).
\end{eqnarray}
 $\Box$

 \section{The most robust $(n,2,d)$-scenario full-correlated Bell inequalities}

 In Example 2,  when we take
 \begin{eqnarray}\mathcal{I}_{2,3}^{0,1}&=&\frac{1}{3}[(-{\rm \mathbf{v}}_{1}-2{\rm \mathbf{v}}_{2})A_0B_0+({\rm \mathbf{v}}_{1}-{\rm \mathbf{v}}_{2})A_0B_1+
 (-2{\rm \mathbf{v}}_{1}-{\rm \mathbf{v}}_{2})A_1B_0+(-{\rm \mathbf{v}}_{1}+{\rm \mathbf{v}}_{2})A_1B_1]\nonumber\\
 &=&\mathcal{I}_{2,3}^{0,0}/.\{b_1\rightarrow {\rm Mod}[b_1+2,3],A_0\leftrightarrow A_1\},\end{eqnarray}
 we have
 \begin{eqnarray}\mathcal{I}_{3,3}(2)=\frac{{\rm \mathbf{v}}_{0}(A_0B_0C_0+A_0B_1C_1+2A_1B_0C_0-A_1B_1C_1)
 +{\rm \mathbf{v}}_{1}(-A_0B_0C_1+A_1B_0C_1)+{\rm \mathbf{v}}_{2}(-A_0B_1C_0
 +A_1B_1C_0)}{3}.\ \ \ \ \ \end{eqnarray}
 For the corresponding Bell inequality $\mathcal{B}_{3,3}(2)\overset{{\rm LHV}}{\leq}1$,  we have  $NL_{|\Psi\rangle}=5/3,NL_{\mathbb{I}}=0$, and then  $v_{c}=(L-NL_{\mathbb{I}})/(NL_{|\Psi\rangle}-NL_{\mathbb{I}})=0.6$. Since
 \begin{eqnarray}
 \mathcal{I}_{3,3}(2)=\mathcal{I}_{3,3}(1)/.\{a_1\rightarrow {\rm Mod}[a_1+1,3],b_0\rightarrow {\rm Mod}[b_0+2,3],A_0\leftrightarrow A_1,B_0\leftrightarrow B_1\},
 \end{eqnarray}
we obtain that $\mathcal{I}_{3,3}(2)$ is equivalent to $\mathcal{I}_{3,3}(1)$. In the following case,
 \begin{eqnarray}\mathcal{I}_{2,3}^{0,1}&=&\frac{1}{3}[(-{\rm \mathbf{v}}_{1}-2{\rm \mathbf{v}}_{2})A_0B_0+({\rm \mathbf{v}}_{1}-{\rm \mathbf{v}}_{2})A_0B_1+
 ({\rm \mathbf{v}}_{1}-{\rm \mathbf{v}}_{2})A_1B_0+(-{\rm \mathbf{v}}_{1}-2{\rm \mathbf{v}}_{2})A_1B_1]\nonumber\\
 &=&\mathcal{I}_{2,3}^{0,0}/.\{a_1\rightarrow {\rm Mod}[a_1+2,3],b_0\rightarrow {\rm Mod}[b_0+1,3],B_0\leftrightarrow B_1\},\end{eqnarray}
the direct computation recovers
 \begin{eqnarray}\mathcal{I}_{3,3}(3)&=&\frac{{\rm \mathbf{v}}_{0}(A_0B_0C_0+A_0B_1C_1+A_1B_0C_1)
 +{\rm \mathbf{v}}_{1}(-A_0B_0C_1+A_1B_1C_0)+{\rm \mathbf{v}}_{2}(-A_0B_1C_0
 -A_1B_0C_0+
 2A_1B_1C_1)}{3}\nonumber\\
 &=&\mathcal{I}_{3,3}(1)/.\{a_1\rightarrow {\rm Mod}[a_1+2,3],b_1\rightarrow {\rm Mod}[b_1+2,3],c_1\rightarrow {\rm Mod}[c_1+1,3],A_0\leftrightarrow A_1,C_0\leftrightarrow C_1\},\end{eqnarray}
 which implies that $\mathcal{I}_{3,3}(3)$ is equivalent to $\mathcal{I}_{3,3}(1)$. Then  the corresponding Bell inequality $\mathcal{B}_{3,3}(3)\overset{{\rm LHV}}{\leq}1$ has $v_{c}=0.6$. In the case that
 \begin{eqnarray}\mathcal{I}_{2,3}^{0,1}&=&\frac{1}{3}[(2{\rm \mathbf{v}}_{1}+{\rm \mathbf{v}}_{2})A_0B_0+({\rm \mathbf{v}}_{1}-{\rm \mathbf{v}}_{2})A_0B_1+
 ({\rm \mathbf{v}}_{1}-{\rm \mathbf{v}}_{2})A_1B_0+(-{\rm \mathbf{v}}_{1}+{\rm \mathbf{v}}_{2})A_1B_1]\nonumber\\
 &=&\mathcal{I}_{2,3}^{0,0}/.\{a_0\rightarrow {\rm Mod}[a_0+1,3],a_1\rightarrow {\rm Mod}[a_1+2,3],B_0\leftrightarrow B_1\},\end{eqnarray}
 we have
  \begin{eqnarray}\mathcal{I}_{3,3}(4)&=&\frac{{\rm \mathbf{v}}_{2}(A_0B_0C_0-A_0B_1C_0-A_1B_0C_0+A_1B_1C_0)
 +{\rm \mathbf{v}}_{0}(2A_0B_0C_1+A_0B_1C_1
+A_1B_0C_1-A_1B_1C_1)}{3},\nonumber\\
&=&\mathcal{I}_{3,3}(1)/.\{a_1\rightarrow {\rm Mod}[a_1+1,3],c_1\rightarrow {\rm Mod}[c_1+1,3],B_0\leftrightarrow B_1,C_0\leftrightarrow C_1\},\end{eqnarray}
which says that $\mathcal{I}_{3,3}(4)$ is equivalent to $\mathcal{I}_{3,3}(1)$. Hence,
 the corresponding Bell inequality $\mathcal{B}_{3,3}(4)\overset{{\rm LHV}}{\leq}1$ has $v_{c}=0.6$.

In the equivalent class of $\mathcal{I}_{3,3}(1)$, we find one party symmetric multi-component Bell function $\mathcal{I}_{3,3}(5)$,
\begin{eqnarray}
 \mathcal{I}_{3,3}(5)&=&\frac{{\rm \mathbf{v}}_{0}A_0B_0C_0+2{\rm \mathbf{v}}_{0}A_1B_1C_1
 -{\rm \mathbf{v}}_{1}(A_0B_0C_1+A_0B_1C_0+A_1B_0C_0)+{\rm \mathbf{v}}_{2}(A_0B_1C_1+A_1B_0C_1+A_1B_1C_0)}{3}\nonumber\\
 &=&\mathcal{I}_{3,3}(1)/.\{b_1\rightarrow {\rm Mod}[b_1+1,3],c_0\rightarrow {\rm Mod}[c_0+2,3],A_0\leftrightarrow A_1,C_0\leftrightarrow C_1\}.
 \end{eqnarray}
For $\mathcal{I}_{3,3}(5)$, the corresponding Bell inequality is
   \begin{eqnarray}\label{323-1}
&&P(A_0+B_0+C_0=0)+P(A_0+B_1+C_1=0)+P(A_1+B_0+C_1=0)+P(A_1+B_1+C_0=1)\nonumber\\
&&+2P(A_1+B_1+C_1=2)-P(A_1+B_0+C_0=2)-P(A_0+B_1+C_0=2)-P(A_0+B_0+C_1=1)\overset{{\rm{LHV}}}{\leq}3,
\end{eqnarray}
which is just the most robust coincidence Bell inequality in Ref. \cite{Coincidence}.
When $c_1=c_2={\rm \mathbf{v}}_{0}$ and $c_1={\rm \mathbf{v}}_{0},c_2={\rm \mathbf{v}}_{1}$, it reduces to $(\mathcal{I}_{2,3}(5))^{0,0}$ and $(\mathcal{I}_{2,3}(5))^{0,1}$, respectively. Since
\begin{eqnarray}
 \mathcal{I}_{2,3}(5)^{0,0}&=&\frac{(-2{\rm \mathbf{v}}_{1}-{\rm \mathbf{v}}_{2})A_0B_0 +(-{\rm \mathbf{v}}_{1}+{\rm \mathbf{v}}_{2})(A_0B_1+A_1B_0)+(-2{\rm \mathbf{v}}_{1}-{\rm \mathbf{v}}_{2})A_1B_1
}{3}\nonumber\\
&=&\mathcal{I}_{2,3}/.\{a_1\rightarrow {\rm Mod}[a_1+2,3],b_1\rightarrow {\rm Mod}[b_1+2,3]\},\nonumber\\
\mathcal{I}_{2,3}(5)^{0,1}&=&\frac{(-{\rm \mathbf{v}}_{1}-2{\rm \mathbf{v}}_{2})A_0B_0 +(-2{\rm \mathbf{v}}_{1}-{\rm \mathbf{v}}_{2})(A_0B_1+A_1B_0)+(2{\rm \mathbf{v}}_{1}+{\rm \mathbf{v}}_{2})A_1B_1
}{3}\nonumber\\
&=&\mathcal{I}_{2,3}/.\{a_1\rightarrow {\rm Mod}[a_1+1,3],B_0\leftrightarrow B_1\},
  \end{eqnarray}
 the corresponding Bell inequalities are equivalent to the CGLMP inequality for $d=3$.

For $n=4$, we take  $\mathcal{I}_{3,3}^{0,0}=\mathcal{I}_{3,3}(3)$, and let $\mathcal{I}_{3,3}^{0,1}$ go through the $648$ ones equivalent to $\mathcal{I}_{3,3}(3)$. Then we obtain $16$ most robust $\mathcal{I}_{4,3}$, whose coincidence Bell inequality has the same $v_c=0.5$. In the mian text, we have listed four of them with the property that they are not equivalent to each other, whose corresponding Bell inequalities are
  \begin{eqnarray}
\mathcal{B}_{4,3}(1)&=&\frac{1}{6}[P(A_0+B_0+C_0+D_0=1)+P(A_0+B_0+C_0+D_1=2)+P(A_0+B_0+C_1+D_0=0))\nonumber\\
& &+P(A_0+B_0+C_1+D_1=1)+P(A_0+B_1+C_0+D_0=2)+P(A_0+B_1+C_0+D_1=0)\nonumber\\
& &+P(A_0+B_1+C_1+D_0=0)+P(A_0+B_1+C_1+D_1=0)+P(A_1+B_0+C_0+D_0=2)\nonumber\\
& &+P(A_1+B_0+C_0+D_1=0)+P(A_1+B_0+C_1+D_0=0)+P(A_1+B_0+C_1+D_1=0)\nonumber\\
& &+P(A_1+B_1+C_0+D_0=1)+P(A_1+B_1+C_0+D_1=2)+P(A_1+B_1+C_1+D_0=1)\nonumber\\
& &+P(A_1+B_1+C_1+D_1=1)]-
\frac{1}{6}[4P(A_0+B_0+C_0+D_0=2)+4P(A_0+B_0+C_0+D_1=1)\nonumber\\
& &+P(A_0+B_0+C_1+D_0=1)+P(A_0+B_0+C_1+D_1=2)+P(A_0+B_1+C_0+D_0=1)\nonumber\\
& &+P(A_0+B_1+C_0+D_1=1)+P(A_0+B_1+C_1+D_0=1)+P(A_0+B_1+C_1+D_1=1)\nonumber\\
& &+P(A_1+B_0+C_0+D_0=1)+P(A_1+B_0+C_0+D_1=1)+P(A_1+B_0+C_1+D_0=1)\nonumber\\
& &+P(A_1+B_0+C_1+D_1=1)+P(A_1+B_1+C_0+D_0=0)+P(A_1+B_1+C_0+D_1=1)\nonumber\\
& &+4P(A_1+B_1+C_1+D_0=0)+4P(A_1+B_1+C_1+D_1=2)]+\frac{2}{3}\overset{{\rm LHV}}{\leq}1,\nonumber\\
\mathcal{B}_{4,3}(2)&=&\frac{1}{6}[P(A_0+B_0+C_0+D_0=0)+P(A_0+B_0+C_0+D_1=0)+P(A_0+B_0+C_1+D_0=0)\nonumber\\
& &+P(A_0+B_0+C_1+D_1=1))+P(A_0+B_1+C_0+D_0=2)+P(A_0+B_1+C_0+D_1=0)\nonumber\\
& &+P(A_0+B_1+C_1+D_0=1)+P(A_0+B_1+C_1+D_1=2)+P(A_1+B_0+C_0+D_0=2)\nonumber\\
& &+P(A_1+B_0+C_0+D_1=0)+P(A_1+B_0+C_1+D_0=0)+P(A_1+B_0+C_1+D_1=0)\nonumber\\
& &+P(A_1+B_1+C_0+D_0=1)+P(A_1+B_1+C_0+D_1=2)+P(A_1+B_1+C_1+D_0=1)\nonumber\\
& &
+P(A_1+B_1+C_1+D_1=1)]-
\frac{1}{6}[P(A_0+B_0+C_0+D_0=1)+P(A_0+B_0+C_0+D_1=1)\nonumber\\
& &+P(A_0+B_0+C_1+D_0=1)+P(A_0+B_0+C_1+D_1=2)+P(A_0+B_1+C_0+D_0=1)\nonumber\\
& &+P(A_0+B_1+C_0+D_1=1)+4P(A_0+B_1+C_1+D_0=2)+4P(A_0+B_1+C_1+D_1=1)\nonumber\\
& &+P(A_1+B_0+C_0+D_0=1)+P(A_1+B_0+C_0+D_1=1)+P(A_1+B_0+C_1+D_0=1)\nonumber\\
& &+P(A_1+B_0+C_1+D_1=1)+P(A_1+B_1+C_0+D_0=0)+P(A_1+B_1+C_0+D_1=1)\nonumber\\
& &+4P(A_1+B_1+C_1+D_0=0)+4P(A_1+B_1+C_1+D_1=2)]+\frac{2}{3}\overset{{\rm LHV}}{\leq}1,\nonumber\\
\mathcal{B}_{4,3}(3)&=&\frac{1}{2}[-P(A_0+B_0+C_0+D_0=2)-P(A_0+B_0+C_0+D_1=1)+P(A_0+B_0+C_1+D_0=0)\nonumber\\
& &+P(A_0+B_0+C_1+D_1=1)-P(A_0+B_1+C_0+D_0=1)+P(A_0+B_1+C_1+D_0=0)\nonumber\\
& &-P(A_1+B_0+C_0+D_1=1)+P(A_1+B_0+C_1+D_1=0)-P(A_1+B_1+C_0+D_0=1)\nonumber\\
& &-P(A_1+B_1+C_0+D_1=0)+P(A_1+B_1+C_1+D_0=1)+P(A_1+B_1+C_1+D_1=1)]\overset{{\rm LHV}}{\leq}1,\nonumber\\
\mathcal{B}_{4,3}(4)&=&\frac{1}{2}[P(A_0+B_0+C_0+D_0=0)+P(A_0+B_0+C_1+D_0=0)+P(A_0+B_0+C_1+D_1=1)\nonumber\\
& &-P(A_0+B_1+C_0+D_0=1)-P(A_0+B_1+C_1+D_0=2)-P(A_0+B_1+C_1+D_1=1)\nonumber\\
& &-P(A_1+B_0+C_0+D_0=1)-P(A_1+B_0+C_1+D_0=2)-P(A_1+B_0+C_1+D_1=1)\nonumber\\
& &+P(A_1+B_1+C_0+D_0=2)+P(A_1+B_1+C_0+D_1=1)+P(A_1+B_1+C_1+D_1=1)]\overset{{\rm LHV}}{\leq}1.
\end{eqnarray}
Now, we list the other $\mathcal{I}_{4,3}(i)$ in the following,
\begin{eqnarray}
\mathcal{I}_{4,3}(5)
&=&\frac{-{\rm \mathbf{v}}_{1}-2{\rm \mathbf{v}}_{2}}{9}A_0B_0C_0D_0+
\frac{-2{\rm \mathbf{v}}_{1}-{\rm \mathbf{v}}_{2}}{9}A_0B_0C_0D_1+
\frac{-2{\rm \mathbf{v}}_{1}-{\rm \mathbf{v}}_{2}}{9}A_0B_0C_1D_0+
\frac{-{\rm \mathbf{v}}_{1}+{\rm \mathbf{v}}_{2}}{9}A_0B_0C_1D_1\nonumber\\
& &+\frac{{\rm \mathbf{v}}_{1}-{\rm \mathbf{v}}_{2}}{9}A_0B_1C_0D_0+
\frac{-{\rm \mathbf{v}}_{1}-2{\rm \mathbf{v}}_{2}}{9}A_0B_1C_0D_1+
\frac{-{\rm \mathbf{v}}_{1}-2{\rm \mathbf{v}}_{2}}{9}A_0B_1C_1D_0+
\frac{-2{\rm \mathbf{v}}_{1}-{\rm \mathbf{v}}_{2}}{9}A_0B_1C_1D_1\nonumber\\
& &+\frac{{\rm \mathbf{v}}_{1}-{\rm \mathbf{v}}_{2}}{9}A_1B_0C_0D_0+
\frac{-{\rm \mathbf{v}}_{1}-2{\rm \mathbf{v}}_{2}}{9}A_1B_0C_0D_1+
\frac{-4{\rm \mathbf{v}}_{1}+{\rm \mathbf{v}}_{2}}{9}A_1B_0C_1D_0+
\frac{{\rm \mathbf{v}}_{1}-4{\rm \mathbf{v}}_{2}}{9}A_1B_0C_1D_1\nonumber\end{eqnarray}
\begin{eqnarray}
& &+\frac{2{\rm \mathbf{v}}_{1}+{\rm \mathbf{v}}_{2}}{9}A_1B_1C_0D_0+
\frac{{\rm \mathbf{v}}_{1}-{\rm \mathbf{v}}_{2}}{9}A_1B_1C_0D_1+
\frac{4{\rm \mathbf{v}}_{1}+5{\rm \mathbf{v}}_{2}}{9}A_1B_1C_1D_0+
\frac{-4{\rm \mathbf{v}}_{1}+{\rm \mathbf{v}}_{2}}{9}A_1B_1C_1D_1\nonumber\\
&=&\mathcal{I}_{4,3}(2)/.\{ABCD\rightarrow BACD\},\nonumber\\
\mathcal{I}_{4,3}(6)
&=&\frac{-{\rm \mathbf{v}}_{1}-2{\rm \mathbf{v}}_{2}}{9}A_0B_0C_0D_0+
\frac{-2{\rm \mathbf{v}}_{1}-{\rm \mathbf{v}}_{2}}{9}A_0B_0C_0D_1+
\frac{-2{\rm \mathbf{v}}_{1}-{\rm \mathbf{v}}_{2}}{9}A_0B_0C_1D_0+
\frac{-{\rm \mathbf{v}}_{1}+{\rm \mathbf{v}}_{2}}{9}A_0B_0C_1D_1\nonumber\\
& &+\frac{{\rm \mathbf{v}}_{1}-{\rm \mathbf{v}}_{2}}{9}A_0B_1C_0D_0+
\frac{-{\rm \mathbf{v}}_{1}-2{\rm \mathbf{v}}_{2}}{9}A_0B_1C_0D_1+
\frac{-{\rm \mathbf{v}}_{1}-2{\rm \mathbf{v}}_{2}}{9}A_0B_1C_1D_0+
\frac{-2{\rm \mathbf{v}}_{1}-{\rm \mathbf{v}}_{2}}{9}A_0B_1C_1D_1\nonumber\\
& &+\frac{{\rm \mathbf{v}}_{1}-{\rm \mathbf{v}}_{2}}{9}A_1B_0C_0D_0+
\frac{-{\rm \mathbf{v}}_{1}-2{\rm \mathbf{v}}_{2}}{9}A_1B_0C_0D_1+
\frac{-{\rm \mathbf{v}}_{1}-2{\rm \mathbf{v}}_{2}}{9}A_1B_0C_1D_0+
\frac{-2{\rm \mathbf{v}}_{1}-{\rm \mathbf{v}}_{2}}{9}A_1B_0C_1D_1\nonumber\\
& &+\frac{-{\rm \mathbf{v}}_{1}-5{\rm \mathbf{v}}_{2}}{9}A_1B_1C_0D_0+
\frac{4{\rm \mathbf{v}}_{1}+5{\rm \mathbf{v}}_{2}}{9}A_1B_1C_0D_1+
\frac{4{\rm \mathbf{v}}_{1}+5{\rm \mathbf{v}}_{2}}{9}A_1B_1C_1D_0+
\frac{-4{\rm \mathbf{v}}_{1}+{\rm \mathbf{v}}_{2}}{9}A_1B_1C_1D_1\nonumber\\
&=&\mathcal{I}_{4,3}(2)/.\{ABCD\rightarrow CABD,b_1\rightarrow {\rm Mod}[b_1+1,3],c_1\rightarrow {\rm Mod}[c_1+2,3]\},\nonumber\\
\mathcal{I}_{4,3}(7)&=&\frac{-{\rm \mathbf{v}}_{1}}{3}A_0B_0C_0D_0+\frac{-{\rm \mathbf{v}}_{2}}{3}A_0B_0C_0D_1
+\frac{{\rm \mathbf{v}}_{0}}{3}A_0B_0C_1D_0+\frac{{\rm \mathbf{v}}_{2}}{3}A_0B_0C_1D_1\nonumber\\
& &+\frac{-{\rm \mathbf{v}}_{2}}{3}A_0B_1C_0D_1+\frac{{\rm \mathbf{v}}_{0}}{3}A_0B_1C_1D_1+\frac{-{\rm \mathbf{v}}_{2}}{3}A_1B_0C_0D_0+
\frac{{\rm \mathbf{v}}_{0}}{3}A_1B_0C_1D_0\nonumber\\
& &+\frac{-{\rm \mathbf{v}}_{2}}{3}A_1B_1C_0D_0+\frac{-{\rm \mathbf{v}}_{0}}{3}A_1B_1C_0D_1+\frac{{\rm \mathbf{v}}_{2}}{3}A_1B_1C_1D_0+
\frac{{\rm \mathbf{v}}_{2}}{3}A_1B_1C_1D_1\nonumber\\
&=&\mathcal{I}_{4,3}(3)/.\{ABCD\rightarrow BACD\},\nonumber\\
\mathcal{I}_{4,3}(8)&=&\frac{-{\rm \mathbf{v}}_{1}}{3}A_0B_0C_0D_0+\frac{-{\rm \mathbf{v}}_{2}}{3}A_0B_0C_0D_1
+\frac{-{\rm \mathbf{v}}_{1}}{3}A_0B_0C_1D_1\nonumber\\
& &+\frac{-{\rm \mathbf{v}}_{2}}{3}A_0B_1C_0D_0+\frac{-{\rm \mathbf{v}}_{1}}{3}A_0B_1C_1D_0+\frac{-{\rm \mathbf{v}}_{2}}{3}A_0B_1C_1D_1
+\frac{{\rm \mathbf{v}}_{1}}{3}A_1B_0C_0D_0+
\frac{{\rm \mathbf{v}}_{0}}{3}A_1B_0C_0D_1+\frac{{\rm \mathbf{v}}_{0}}{3}A_1B_0C_1D_1\nonumber\\
& &+\frac{{\rm \mathbf{v}}_{1}}{3}A_1B_1C_0D_0+\frac{{\rm \mathbf{v}}_{2}}{3}A_1B_1C_1D_0+
\frac{{\rm \mathbf{v}}_{2}}{3}A_1B_1C_1D_1\nonumber\\
&=&\mathcal{I}_{4,3}(3)/.\{ABCD\rightarrow DCAB,d_1\rightarrow {\rm Mod}[d_1+2,3],C_0\leftrightarrow C_1,D_0\leftrightarrow D_1\},\nonumber\\
\mathcal{I}_{4,3}(9)&=&\frac{-{\rm \mathbf{v}}_{1}}{3}A_0B_0C_0D_0+\frac{-{\rm \mathbf{v}}_{2}}{3}A_0B_0C_0D_1
+\frac{-{\rm \mathbf{v}}_{1}}{3}A_0B_0C_1D_1\nonumber\\
& &+\frac{{\rm \mathbf{v}}_{1}}{3}A_0B_1C_0D_0+\frac{{\rm \mathbf{v}}_{0}}{3}A_0B_1C_0D_1+\frac{{\rm \mathbf{v}}_{0}}{3}A_0B_1C_1D_1
+\frac{-{\rm \mathbf{v}}_{2}}{3}A_1B_0C_0D_0+
\frac{-{\rm \mathbf{v}}_{1}}{3}A_1B_0C_1D_0+\frac{-{\rm \mathbf{v}}_{2}}{3}A_1B_0C_1D_1\nonumber\\
& &+\frac{{\rm \mathbf{v}}_{1}}{3}A_1B_1C_0D_0+\frac{{\rm \mathbf{v}}_{2}}{3}A_1B_1C_1D_0+
\frac{{\rm \mathbf{v}}_{2}}{3}A_1B_1C_1D_1\nonumber\\
&=&\mathcal{I}_{4,3}(3)/.\{ABCD\rightarrow DCBA,d_1\rightarrow {\rm Mod}[d_1+2,3],C_0\leftrightarrow C_1,D_0\leftrightarrow D_1\},\nonumber\\
\mathcal{I}_{4,3}(10)&=&\frac{-{\rm \mathbf{v}}_{1}}{3}A_0B_0C_0D_0+\frac{-{\rm \mathbf{v}}_{2}}{3}A_0B_0C_0D_1
+\frac{-{\rm \mathbf{v}}_{1}}{3}A_0B_0C_1D_0\nonumber\\
& &+\frac{-{\rm \mathbf{v}}_{2}}{3}A_0B_1C_0D_1+\frac{-{\rm \mathbf{v}}_{1}}{3}A_0B_1C_1D_0+\frac{-{\rm \mathbf{v}}_{2}}{3}A_0B_1C_1D_1
+\frac{{\rm \mathbf{v}}_{1}}{3}A_1B_0C_0D_0+\frac{{\rm \mathbf{v}}_{0}}{3}A_1B_0C_0D_1+
\frac{{\rm \mathbf{v}}_{0}}{3}A_1B_0C_1D_0\nonumber\\
& &+\frac{{\rm \mathbf{v}}_{1}}{3}A_1B_1C_0D_1+\frac{{\rm \mathbf{v}}_{2}}{3}A_1B_1C_1D_0+
\frac{{\rm \mathbf{v}}_{2}}{3}A_1B_1C_1D_1\nonumber\\
&=&\mathcal{I}_{4,3}(3)/.\{ABCD\rightarrow BCAD,a_1\rightarrow {\rm Mod}[a_1+1,3],c_1\rightarrow {\rm Mod}[c_1+2,3]\},\nonumber\\
\mathcal{I}_{4,3}(11)&=&\frac{-{\rm \mathbf{v}}_{1}}{3}A_0B_0C_0D_0+\frac{-{\rm \mathbf{v}}_{2}}{3}A_0B_0C_0D_1
+\frac{-{\rm \mathbf{v}}_{1}}{3}A_0B_0C_1D_0\nonumber\\
& &+\frac{{\rm \mathbf{v}}_{1}}{3}A_0B_1C_0D_0+\frac{{\rm \mathbf{v}}_{0}}{3}A_0B_1C_0D_1+\frac{{\rm \mathbf{v}}_{0}}{3}A_0B_1C_1D_0
+\frac{-{\rm \mathbf{v}}_{2}}{3}A_1B_0C_0D_1+
\frac{-{\rm \mathbf{v}}_{1}}{3}A_1B_0C_1D_0+
\frac{-{\rm \mathbf{v}}_{2}}{3}A_1B_0C_1D_1\nonumber\\
& &+\frac{{\rm \mathbf{v}}_{1}}{3}A_1B_1C_0D_1+\frac{{\rm \mathbf{v}}_{2}}{3}A_1B_1C_1D_0+
\frac{{\rm \mathbf{v}}_{2}}{3}A_1B_1C_1D_1\nonumber\\
&=&\mathcal{I}_{4,3}(3)/.\{ABCD\rightarrow ACBD,b_1\rightarrow {\rm Mod}[b_1+1,3],c_1\rightarrow {\rm Mod}[c_1+2,3]\},\nonumber\\
\mathcal{I}_{4,3}(12)&=&\frac{{\rm \mathbf{v}}_{0}}{3}A_0B_0C_0D_0
+\frac{-{\rm \mathbf{v}}_{1}}{3}A_0B_0C_1D_0\nonumber\\
& &+\frac{-{\rm \mathbf{v}}_{2}}{3}A_0B_1C_0D_0+\frac{{\rm \mathbf{v}}_{0}}{3}A_0B_1C_1D_0
+\frac{{\rm \mathbf{v}}_{1}}{3}A_1B_0C_0D_0+\frac{{\rm \mathbf{v}}_{0}}{3}A_1B_0C_0D_1+
\frac{-{\rm \mathbf{v}}_{1}}{3}A_1B_0C_1D_0+
\frac{-{\rm \mathbf{v}}_{2}}{3}A_1B_0C_1D_1\nonumber\\
& &+\frac{-{\rm \mathbf{v}}_{2}}{3}A_1B_1C_0D_0+\frac{-{\rm \mathbf{v}}_{0}}{3}A_1B_1C_0D_1+\frac{{\rm \mathbf{v}}_{2}}{3}A_1B_1C_1D_0+
\frac{{\rm \mathbf{v}}_{2}}{3}A_1B_1C_1D_1\nonumber\\
&=&\mathcal{I}_{4,3}(4)/.\{ABCD\rightarrow BCAD,a_1\rightarrow {\rm Mod}[a_1+1,3],c_1\rightarrow {\rm Mod}[c_1+2,3]\},\nonumber\end{eqnarray}
\begin{eqnarray}
\mathcal{I}_{4,3}(13)&=&\frac{{\rm \mathbf{v}}_{0}}{3}A_0B_0C_0D_0
+\frac{-{\rm \mathbf{v}}_{1}}{3}A_0B_0C_1D_0\nonumber\\
& &+\frac{{\rm \mathbf{v}}_{1}}{3}A_0B_1C_0D_0+\frac{{\rm \mathbf{v}}_{0}}{3}A_0B_1C_0D_1+\frac{-{\rm \mathbf{v}}_{1}}{3}A_0B_1C_1D_0
+\frac{-{\rm \mathbf{v}}_{2}}{3}A_0B_1C_1D_1
+\frac{-{\rm \mathbf{v}}_{2}}{3}A_1B_0C_0D_0+
\frac{{\rm \mathbf{v}}_{0}}{3}A_1B_0C_1D_0\nonumber\\
& &+\frac{-{\rm \mathbf{v}}_{2}}{3}A_1B_1C_0D_0+\frac{-{\rm \mathbf{v}}_{0}}{3}A_1B_1C_0D_1+\frac{{\rm \mathbf{v}}_{2}}{3}A_1B_1C_1D_0+
\frac{{\rm \mathbf{v}}_{2}}{3}A_1B_1C_1D_1\nonumber\\
&=&\mathcal{I}_{4,3}(4)/.\{ABCD\rightarrow ACBD,b_1\rightarrow {\rm Mod}[b_1+1,3],c_1\rightarrow {\rm Mod}[c_1+2,3]\},\nonumber\\
\mathcal{I}_{4,3}(14)&=&\frac{{\rm \mathbf{v}}_{0}}{3}A_0B_0C_0D_1
+\frac{{\rm \mathbf{v}}_{0}}{3}A_0B_0C_1D_0+\frac{{\rm \mathbf{v}}_{2}}{3}A_0B_0C_1D_1+\frac{-{\rm \mathbf{v}}_{2}}{3}A_0B_1C_0D_1\nonumber\\
& &+\frac{-{\rm \mathbf{v}}_{1}}{3}A_0B_1C_1D_0
+\frac{-{\rm \mathbf{v}}_{2}}{3}A_0B_1C_1D_1
+\frac{-{\rm \mathbf{v}}_{2}}{3}A_1B_0C_0D_1+
\frac{-{\rm \mathbf{v}}_{1}}{3}A_1B_0C_1D_0+
\frac{-{\rm \mathbf{v}}_{2}}{3}A_1B_0C_1D_1\nonumber\\
& &+\frac{{\rm \mathbf{v}}_{1}}{3}A_1B_1C_0D_1+\frac{{\rm \mathbf{v}}_{2}}{3}A_1B_1C_1D_0+
\frac{{\rm \mathbf{v}}_{2}}{3}A_1B_1C_1D_1\nonumber\\
&=&\mathcal{I}_{4,3}(4)/.\{a_1\rightarrow {\rm Mod}[a_1+2,3],b_1\rightarrow {\rm Mod}[b_1+2,3],c_1\rightarrow {\rm Mod}[c_1+1,3],
d_0\rightarrow {\rm Mod}[d_0+1,3],\nonumber\\
& &d_1\rightarrow {\rm Mod}[d_1+2,3],A_0\leftrightarrow A_1,B_0\leftrightarrow B_1,D_0\leftrightarrow D_1\},\nonumber\\
\mathcal{I}_{4,3}(15)&=&\frac{{\rm \mathbf{v}}_{0}}{3}A_0B_0C_0D_1
+\frac{-{\rm \mathbf{v}}_{1}}{3}A_0B_0C_1D_1\nonumber\\
& &+\frac{-{\rm \mathbf{v}}_{2}}{3}A_0B_1C_0D_1
+\frac{{\rm \mathbf{v}}_{0}}{3}A_0B_1C_1D_1+\frac{{\rm \mathbf{v}}_{1}}{3}A_1B_0C_0D_0
+\frac{{\rm \mathbf{v}}_{0}}{3}A_1B_0C_0D_1+
\frac{-{\rm \mathbf{v}}_{1}}{3}A_1B_0C_1D_0+
\frac{-{\rm \mathbf{v}}_{2}}{3}A_1B_0C_1D_1\nonumber\\
& &+\frac{-{\rm \mathbf{v}}_{2}}{3}A_1B_1C_0D_0+\frac{-{\rm \mathbf{v}}_{0}}{3}A_1B_1C_0D_1+\frac{{\rm \mathbf{v}}_{2}}{3}A_1B_1C_1D_0+
\frac{{\rm \mathbf{v}}_{2}}{3}A_1B_1C_1D_1\nonumber\\
&=&\mathcal{I}_{4,3}(4)/.\{ABCD\rightarrow BCDA,a_1\rightarrow {\rm Mod}[a_1+1,3],b_1\rightarrow {\rm Mod}[b_1+2,3],c_1\rightarrow {\rm Mod}[c_1+1,3],\nonumber\\
& &
d_0\rightarrow {\rm Mod}[d_0+1,3],A_0\leftrightarrow A_1\},\nonumber\\
\mathcal{I}_{4,3}(16)&=&\frac{{\rm \mathbf{v}}_{0}}{3}A_0B_0C_0D_1
+\frac{-{\rm \mathbf{v}}_{1}}{3}A_0B_0C_1D_1\nonumber\\
& &+\frac{{\rm \mathbf{v}}_{1}}{3}A_0B_1C_0D_0+\frac{{\rm \mathbf{v}}_{0}}{3}A_0B_1C_0D_1+\frac{-{\rm \mathbf{v}}_{1}}{3}A_0B_1C_1D_0
+\frac{-{\rm \mathbf{v}}_{2}}{3}A_0B_1C_1D_1+
\frac{-{\rm \mathbf{v}}_{2}}{3}A_1B_0C_1D_0+
\frac{{\rm \mathbf{v}}_{0}}{3}A_1B_0C_1D_1\nonumber\\
& &+\frac{-{\rm \mathbf{v}}_{2}}{3}A_1B_1C_0D_0+\frac{-{\rm \mathbf{v}}_{0}}{3}A_1B_1C_0D_1+\frac{{\rm \mathbf{v}}_{2}}{3}A_1B_1C_1D_0+
\frac{{\rm \mathbf{v}}_{2}}{3}A_1B_1C_1D_1\nonumber\\
&=&\mathcal{I}_{4,3}(4)/.\{ABCD\rightarrow ACDB,a_1\rightarrow {\rm Mod}[a_1+2,3],b_1\rightarrow {\rm Mod}[b_1+1,3],c_1\rightarrow {\rm Mod}[c_1+1,3],\nonumber\\
& &
d_0\rightarrow {\rm Mod}[d_0+1,3],B_0\leftrightarrow B_1\}.
\end{eqnarray}
Considering that $\mathcal{I}_{3,3}^{0,1}(i)$ is  important for constructing $\mathcal{I}_{4,3}(i)$, we also list $\mathcal{I}_{3,3}^{0,1}(i)$ for
readers' convenience,
   \begin{eqnarray}
 \mathcal{I}_{3,3}^{0,1}(1)&=&\frac{2{\rm \mathbf{v}}_{2}A_0B_0C_0
 +{\rm \mathbf{v}}_{0}A_0B_0C_1+{\rm \mathbf{v}}_{1}A_0B_1C_0-{\rm \mathbf{v}}_{2}A_0B_1C_1
 +{\rm \mathbf{v}}_{1}A_1B_0C_0-{\rm \mathbf{v}}_{2}A_1B_0C_1-{\rm \mathbf{v}}_{0}A_1B_1C_0+
 {\rm \mathbf{v}}_{1}A_1B_1C_1}{3}\nonumber\\
 &=&\mathcal{I}_{3,3}(3)/.\{a_1\rightarrow {\rm Mod}[a_1+2,3],b_0\rightarrow {\rm Mod}[b_0+1,3],c_1\rightarrow {\rm Mod}[c_1+1,3],A_0\leftrightarrow A_1,B_0\leftrightarrow B_1,C_0\leftrightarrow C_1\},\nonumber\\
 \mathcal{I}_{3,3}^{0,1}(2)&=&\frac{-{\rm \mathbf{v}}_{2}A_0B_0C_0
 +{\rm \mathbf{v}}_{0}A_0B_0C_1+{\rm \mathbf{v}}_{1}A_0B_1C_0+2{\rm \mathbf{v}}_{2}A_0B_1C_1
 +{\rm \mathbf{v}}_{1}A_1B_0C_0-{\rm \mathbf{v}}_{2}A_1B_0C_1-{\rm \mathbf{v}}_{0}A_1B_1C_0+
 {\rm \mathbf{v}}_{1}A_1B_1C_1}{3}\nonumber\\
 &=&\mathcal{I}_{3,3}(3)/.\{a_1\rightarrow {\rm Mod}[a_1+2,3],c_0\rightarrow {\rm Mod}[c_0+1,3],c_1\rightarrow {\rm Mod}[c_1+1,3],A_0\leftrightarrow A_1\},\nonumber\\
 \mathcal{I}_{3,3}^{0,1}(3)&=&\frac{{\rm \mathbf{v}}_{2}A_0B_0C_0
 +2{\rm \mathbf{v}}_{0}A_0B_0C_1-{\rm \mathbf{v}}_{2}A_0B_1C_0+{\rm \mathbf{v}}_{0}A_0B_1C_1
 -{\rm \mathbf{v}}_{0}A_1B_0C_0+{\rm \mathbf{v}}_{1}A_1B_0C_1+{\rm \mathbf{v}}_{0}A_1B_1C_0-
 {\rm \mathbf{v}}_{1}A_1B_1C_1}{3}\nonumber\\
 &=&\mathcal{I}_{3,3}(3)/.\{b_1\rightarrow {\rm Mod}[b_1+1,3],A_0\leftrightarrow A_1,B_0\leftrightarrow B_1\},\nonumber\\
 \mathcal{I}_{3,3}^{0,1}(4)&=&\frac{{\rm \mathbf{v}}_{0}A_0B_0C_0
 +2{\rm \mathbf{v}}_{0}A_0B_0C_1-{\rm \mathbf{v}}_{2}A_0B_1C_0+{\rm \mathbf{v}}_{2}A_0B_1C_1
 -{\rm \mathbf{v}}_{2}A_1B_0C_0+{\rm \mathbf{v}}_{2}A_1B_0C_1+{\rm \mathbf{v}}_{1}A_1B_1C_0-
 {\rm \mathbf{v}}_{1}A_1B_1C_1}{3}\nonumber\\
 &=&\mathcal{I}_{3,3}(3)/.\{a_1\rightarrow {\rm Mod}[a_1+2,3],b_1\rightarrow {\rm Mod}[b_1+2,3],c_0\rightarrow {\rm Mod}[c_0+1,3],A_0\leftrightarrow A_1,B_0\leftrightarrow B_1\},\nonumber\\
 \mathcal{I}_{3,3}^{0,1}(5)&=&\frac{-{\rm \mathbf{v}}_{2}A_0B_0C_0
 +{\rm \mathbf{v}}_{0}A_0B_0C_1+{\rm \mathbf{v}}_{1}A_0B_1C_0-{\rm \mathbf{v}}_{2}A_0B_1C_1
 +{\rm \mathbf{v}}_{1}A_1B_0C_0+2{\rm \mathbf{v}}_{2}A_1B_0C_1-{\rm \mathbf{v}}_{0}A_1B_1C_0+
 {\rm \mathbf{v}}_{1}A_1B_1C_1}{3}\nonumber\\
  &=&\mathcal{I}_{3,3}(3)/.\{b_0\rightarrow {\rm Mod}[b_0+1,3],B_0\leftrightarrow B_1\},\nonumber\\
  \mathcal{I}_{3,3}^{0,1}(6)&=&\frac{-{\rm \mathbf{v}}_{2}A_0B_0C_0
 +{\rm \mathbf{v}}_{0}A_0B_0C_1+{\rm \mathbf{v}}_{1}A_0B_1C_0-{\rm \mathbf{v}}_{2}A_0B_1C_1
 +{\rm \mathbf{v}}_{1}A_1B_0C_0-{\rm \mathbf{v}}_{2}A_1B_0C_1+2{\rm \mathbf{v}}_{0}A_1B_1C_0+
 {\rm \mathbf{v}}_{1}A_1B_1C_1}{3}\nonumber\\
  &=&\mathcal{I}_{3,3}(3)/.\{c_1\rightarrow {\rm Mod}[c_1+1,3],C_0\leftrightarrow C_1\},\nonumber\\
 \mathcal{I}_{3,3}^{0,1}(7)&=&\frac{{\rm \mathbf{v}}_{2}A_0B_0C_0
 +2{\rm \mathbf{v}}_{0}A_0B_0C_1-{\rm \mathbf{v}}_{0}A_0B_1C_0+{\rm \mathbf{v}}_{1}A_0B_1C_1
 -{\rm \mathbf{v}}_{2}A_1B_0C_0+{\rm \mathbf{v}}_{0}A_1B_0C_1+{\rm \mathbf{v}}_{0}A_1B_1C_0-
 {\rm \mathbf{v}}_{1}A_1B_1C_1}{3}\nonumber\\
  &=&\mathcal{I}_{3,3}(3)/.\{a_1\rightarrow {\rm Mod}[a_1+1,3],A_0\leftrightarrow A_1,B_0\leftrightarrow B_1\},\nonumber\end{eqnarray}
\begin{eqnarray}
 \mathcal{I}_{3,3}^{0,1}(8)&=&\frac{{\rm \mathbf{v}}_{2}A_0B_0C_0
 -{\rm \mathbf{v}}_{2}A_0B_0C_1-{\rm \mathbf{v}}_{2}A_0B_1C_0+{\rm \mathbf{v}}_{2}A_0B_1C_1
 +2{\rm \mathbf{v}}_{1}A_1B_0C_0+{\rm \mathbf{v}}_{1}A_1B_0C_1+{\rm \mathbf{v}}_{1}A_1B_1C_0-
 {\rm \mathbf{v}}_{1}A_1B_1C_1}{3}\nonumber\\
  &=&\mathcal{I}_{3,3}(3)/.\{b_0\rightarrow {\rm Mod}[b_0+2,3],c_1\rightarrow {\rm Mod}[c_1+2,3],B_0\leftrightarrow B_1,C_0\leftrightarrow C_1\},\nonumber\\
  \mathcal{I}_{3,3}^{0,1}(9)&=&\frac{{\rm \mathbf{v}}_{2}A_0B_0C_0
 -{\rm \mathbf{v}}_{2}A_0B_0C_1+2{\rm \mathbf{v}}_{1}A_0B_1C_0+{\rm \mathbf{v}}_{1}A_0B_1C_1
 -{\rm \mathbf{v}}_{2}A_1B_0C_0+{\rm \mathbf{v}}_{2}A_1B_0C_1+{\rm \mathbf{v}}_{1}A_1B_1C_0-
 {\rm \mathbf{v}}_{1}A_1B_1C_1}{3}\nonumber\\
  &=&\mathcal{I}_{3,3}(3)/.\{a_0\rightarrow {\rm Mod}[a_0+2,3],c_1\rightarrow {\rm Mod}[c_1+2,3],A_0\leftrightarrow A_1,C_0\leftrightarrow C_1\},\nonumber\\
 \mathcal{I}_{3,3}^{0,1}(10)&=&\frac{{\rm \mathbf{v}}_{2}A_0B_0C_0
 -{\rm \mathbf{v}}_{1}A_0B_0C_1-{\rm \mathbf{v}}_{0}A_0B_1C_0+{\rm \mathbf{v}}_{2}A_0B_1C_1
 +2{\rm \mathbf{v}}_{1}A_1B_0C_0+{\rm \mathbf{v}}_{0}A_1B_0C_1+{\rm \mathbf{v}}_{2}A_1B_1C_0-
 {\rm \mathbf{v}}_{1}A_1B_1C_1}{3}\nonumber\\
  &=&\mathcal{I}_{3,3}(3)/.\{c_0\rightarrow {\rm Mod}[c_0+2,3],c_1\rightarrow {\rm Mod}[c_1+2,3],B_0\leftrightarrow B_1,C_0\leftrightarrow C_1\},\nonumber\\
  \mathcal{I}_{3,3}^{0,1}(11)&=&\frac{{\rm \mathbf{v}}_{2}A_0B_0C_0
 -{\rm \mathbf{v}}_{1}A_0B_0C_1+2{\rm \mathbf{v}}_{2}A_0B_1C_0+{\rm \mathbf{v}}_{0}A_0B_1C_1
 -{\rm \mathbf{v}}_{0}A_1B_0C_0+{\rm \mathbf{v}}_{2}A_1B_0C_1+{\rm \mathbf{v}}_{2}A_1B_1C_0-
 {\rm \mathbf{v}}_{1}A_1B_1C_1}{3}\nonumber\\
  &=&\mathcal{I}_{3,3}(3)/.\{c_0\rightarrow {\rm Mod}[c_0+2,3],c_1\rightarrow {\rm Mod}[c_1+2,3],A_0\leftrightarrow A_1,C_0\leftrightarrow C_1\},\nonumber\\
  \mathcal{I}_{3,3}^{0,1}(12)&=&\frac{{\rm \mathbf{v}}_{0}A_0B_0C_0
 -{\rm \mathbf{v}}_{1}A_0B_0C_1-{\rm \mathbf{v}}_{2}A_0B_1C_0+{\rm \mathbf{v}}_{0}A_0B_1C_1
 +2{\rm \mathbf{v}}_{1}A_1B_0C_0+{\rm \mathbf{v}}_{2}A_1B_0C_1+{\rm \mathbf{v}}_{0}A_1B_1C_0-
 {\rm \mathbf{v}}_{1}A_1B_1C_1}{3}\nonumber\\
  &=&\mathcal{I}_{3,3}(3)/.\{a_1\rightarrow {\rm Mod}[a_1+2,3],b_1\rightarrow {\rm Mod}[b_1+2,3],c_1\rightarrow {\rm Mod}[c_1+1,3],B_0\leftrightarrow B_1,C_0\leftrightarrow C_1\},\nonumber\\
 \mathcal{I}_{3,3}^{0,1}(13)&=&\frac{{\rm \mathbf{v}}_{0}A_0B_0C_0
 -{\rm \mathbf{v}}_{1}A_0B_0C_1+2{\rm \mathbf{v}}_{1}A_0B_1C_0+{\rm \mathbf{v}}_{2}A_0B_1C_1
 -{\rm \mathbf{v}}_{2}A_1B_0C_0+{\rm \mathbf{v}}_{0}A_1B_0C_1+{\rm \mathbf{v}}_{0}A_1B_1C_0-
 {\rm \mathbf{v}}_{1}A_1B_1C_1}{3}\nonumber\\
  &=&\mathcal{I}_{3,3}(3)/.\{a_1\rightarrow {\rm Mod}[a_1+2,3],b_1\rightarrow {\rm Mod}[b_1+2,3],c_1\rightarrow {\rm Mod}[c_1+1,3],A_0\leftrightarrow A_1,C_0\leftrightarrow C_1\},\nonumber\\
  \mathcal{I}_{3,3}^{0,1}(14)&=&\frac{{\rm \mathbf{v}}_{1}A_0B_0C_0
 +2{\rm \mathbf{v}}_{0}A_0B_0C_1-{\rm \mathbf{v}}_{0}A_0B_1C_0+{\rm \mathbf{v}}_{2}A_0B_1C_1
 -{\rm \mathbf{v}}_{0}A_1B_0C_0+{\rm \mathbf{v}}_{2}A_1B_0C_1+{\rm \mathbf{v}}_{2}A_1B_1C_0-
 {\rm \mathbf{v}}_{1}A_1B_1C_1}{3}\nonumber\\
  &=&\mathcal{I}_{3,3}(3)/.\{a_1\rightarrow {\rm Mod}[a_1+2,3],b_1\rightarrow {\rm Mod}[b_1+2,3],c_0\rightarrow {\rm Mod}[c_0+2,3],A_0\leftrightarrow A_1,B_0\leftrightarrow B_1\},\nonumber\\
  \mathcal{I}_{3,3}^{0,1}(15)&=&\frac{{\rm \mathbf{v}}_{1}A_0B_0C_0
 -{\rm \mathbf{v}}_{2}A_0B_0C_1-{\rm \mathbf{v}}_{0}A_0B_1C_0+{\rm \mathbf{v}}_{1}A_0B_1C_1
 +2{\rm \mathbf{v}}_{1}A_1B_0C_0+{\rm \mathbf{v}}_{2}A_1B_0C_1+{\rm \mathbf{v}}_{0}A_1B_1C_0-
 {\rm \mathbf{v}}_{1}A_1B_1C_1}{3}\nonumber\\
  &=&\mathcal{I}_{3,3}(3)/.\{a_1\rightarrow {\rm Mod}[a_1+1,3],b_0\rightarrow {\rm Mod}[b_0+1,3],c_1\rightarrow {\rm Mod}[c_1+1,3],B_0\leftrightarrow B_1,C_0\leftrightarrow C_1\},\nonumber\\
 \mathcal{I}_{3,3}^{0,1}(16)&=&\frac{{\rm \mathbf{v}}_{1}A_0B_0C_0
 -{\rm \mathbf{v}}_{2}A_0B_0C_1+2{\rm \mathbf{v}}_{1}A_0B_1C_0+{\rm \mathbf{v}}_{2}A_0B_1C_1
 -{\rm \mathbf{v}}_{0}A_1B_0C_0+{\rm \mathbf{v}}_{1}A_1B_0C_1+{\rm \mathbf{v}}_{0}A_1B_1C_0-
 {\rm \mathbf{v}}_{1}A_1B_1C_1}{3}\nonumber\\
  &=&\mathcal{I}_{3,3}(3)/.\{a_1\rightarrow {\rm Mod}[a_1+2,3],b_0\rightarrow {\rm Mod}[b_0+2,3],c_0\rightarrow {\rm Mod}[c_0+2,3],A_0\leftrightarrow A_1,C_0\leftrightarrow C_1\}.
 \end{eqnarray}

The fist $(4,2,3)$-scenario Bell inequality in Ref.\cite{n23} is
\begin{eqnarray}\label{423-A1}
\mathcal{I}_{4}^{(1)}&=&-P(A_0+B_0+C_0+D_0=2)-P(A_0+B_0+C_0+D_1=1)-P(A_0+B_0+C_1+D_1=1)\nonumber\\
& &-P(A_0+B_1+C_0+D_0=2)-P(A_0+B_1+C_1+D_0=2)-P(A_0+B_1+C_1+D_1=1)\nonumber\\
& &+P(A_1+B_0+C_0+D_0=0)+P(A_1+B_0+C_0+D_1=1)+P(A_1+B_0+C_1+D_1=0)\nonumber\\
& &+P(A_1+B_1+C_0+D_0=1)+P(A_1+B_1+C_1+D_0=2)+P(A_1+B_1+C_1+D_1=2)\overset{{\rm{LHV}}}{\leq}2,
\end{eqnarray}
for which the corresponding multi-component Bell function is
\begin{eqnarray}
\mathcal{I}_{4,3}(17)&=&\frac{-{\rm \mathbf{v}}_{1}}{3}A_0B_0C_0D_0+\frac{-{\rm \mathbf{v}}_{2}}{3}A_0B_0C_0D_1
+\frac{-{\rm \mathbf{v}}_{2}}{3}A_0B_0C_1D_1\nonumber\\
& &+\frac{-{\rm \mathbf{v}}_{1}}{3}A_0B_1C_0D_0+\frac{-{\rm \mathbf{v}}_{1}}{3}A_0B_1C_1D_0
+\frac{-{\rm \mathbf{v}}_{2}}{3}A_0B_1C_1D_1+\frac{{\rm \mathbf{v}}_{0}}{3}A_1B_0C_0D_0+
\frac{{\rm \mathbf{v}}_{2}}{3}A_1B_0C_0D_1+
\frac{{\rm \mathbf{v}}_{0}}{3}A_1B_0C_1D_1\nonumber\\
& &+\frac{{\rm \mathbf{v}}_{2}}{3}A_1B_1C_0D_0+\frac{{\rm \mathbf{v}}_{1}}{3}A_1B_1C_1D_0+
\frac{{\rm \mathbf{v}}_{1}}{3}A_1B_1C_1D_1\nonumber\\
&=&\mathcal{I}_{4,3}(3)/.\{a_1\rightarrow {\rm Mod}[a_1+2,3],b_1\rightarrow {\rm Mod}[b_1+2,3],c_1\rightarrow {\rm Mod}[c_1+1,3]\}.
\end{eqnarray}
When $c_1=c_2={\rm \mathbf{v}}_{0}$ and $c_1={\rm \mathbf{v}}_{0},c_2={\rm \mathbf{v}}_{1}$, it reduces to $(\mathcal{I}_{3,3}(17))^{0,0}$ and $(\mathcal{I}_{3,3}(17))^{0,1}$, respectively. Considering that
\begin{eqnarray}
(\mathcal{I}_{3,3}(17))^{0,0}&=&\frac{{\rm \mathbf{v}}_{0}}{3}A_0B_0C_0
+\frac{-{\rm \mathbf{v}}_{2}}{3}A_0B_0C_1+\frac{-{\rm \mathbf{v}}_{1}}{3}A_0B_1C_0+\frac{{\rm \mathbf{v}}_{0}}{3}A_0B_1C_1\nonumber\\
& &
+\frac{-{\rm \mathbf{v}}_{1}}{3}A_1B_0C_0+
\frac{{\rm \mathbf{v}}_{0}}{3}A_1B_0C_1+\frac{{\rm \mathbf{v}}_{2}}{3}A_1B_1C_0+\frac{2{\rm \mathbf{v}}_{1}}{3}A_1B_1C_1\nonumber\\
  &=&\mathcal{I}_{3,3}(3)/.\{ABCD\rightarrow CBAD,b_1\rightarrow {\rm Mod}[b_1+2,3]\},\nonumber\end{eqnarray}
\begin{eqnarray}
(\mathcal{I}_{3,3}(17))^{0,1}&=&\frac{{\rm \mathbf{v}}_{2}}{3}A_0B_0C_0
+\frac{-{\rm \mathbf{v}}_{0}}{3}A_0B_0C_1+\frac{-{\rm \mathbf{v}}_{1}}{3}A_0B_1C_0+\frac{{\rm \mathbf{v}}_{2}}{3}A_0B_1C_1\nonumber\\
& &
+\frac{2{\rm \mathbf{v}}_{0}}{3}A_1B_0C_0+
\frac{{\rm \mathbf{v}}_{1}}{3}A_1B_0C_1+\frac{{\rm \mathbf{v}}_{2}}{3}A_1B_1C_0+\frac{-{\rm \mathbf{v}}_{0}}{3}A_1B_1C_1\nonumber\\
  &=&\mathcal{I}_{3,3}(3)/.\{a_1\rightarrow {\rm Mod}[a_1+2,3],b_1\rightarrow {\rm Mod}[b_1+2,3],c_0\rightarrow {\rm Mod}[c_0+2,3],B_0\leftrightarrow B_1,C_0\leftrightarrow C_1\}
\end{eqnarray}
we have that both of them are equivalent to $\mathcal{I}_{3,3}$.

  The second $(4,2,3)$-scenario Bell inequality with probability form in Ref.\cite{n23} is
\begin{eqnarray}\label{423-A2}
&&\mathcal{I}_{4}^{(2)}=P(A_0+B_0+C_0+D_0=0)-P(A_0+B_0+C_0+D_0=1)+P(A_0+B_0+C_0+D_1=0)\nonumber\\
&&-P(A_0+B_0+C_0+D_1=2)-P(A_0+B_0+C_1+D_0=1)+P(A_0+B_0+C_1+D_0=2)\nonumber\\
&&+P(A_0+B_0+C_1+D_1=0)-P(A_0+B_0+C_1+D_1=1)+P(A_0+B_1+C_0+D_0=0)\nonumber\\
&&-P(A_0+B_1+C_0+D_0=2)+P(A_0+B_1+C_0+D_1=1)-P(A_0+B_1+C_0+D_1=2)\nonumber\\
&&+P(A_0+B_1+C_1+D_0=0)-P(A_0+B_1+C_1+D_0=1)+P(A_0+B_1+C_1+D_1=0)\nonumber\\
&&-P(A_0+B_1+C_1+D_1=2)+P(A_1+B_0+C_0+D_0=0)-P(A_1+B_0+C_0+D_0=2)\nonumber\\
&&+P(A_1+B_0+C_0+D_1=1)-P(A_1+B_0+C_0+D_1=2)+P(A_1+B_0+C_1+D_0=1)\nonumber\\
&&-4P(A_1+B_0+C_1+D_0=2)-4P(A_1+B_0+C_1+D_1=1)+P(A_1+B_0+C_1+D_1=2)\nonumber\\
&&+P(A_1+B_1+C_0+D_0=1)-P(A_1+B_1+C_0+D_0=2)-P(A_1+B_1+C_0+D_1=0)\nonumber\\
&&+P(A_1+B_1+C_0+D_1=1)-4P(A_1+B_1+C_1+D_0=1)+P(A_1+B_1+C_1+D_0=2)\nonumber\\
&&-4P(A_1+B_1+C_1+D_1=0)+P(A_1+B_1+C_1+D_1=2)\overset{{\rm{LHV}}}{\leq} 2,
\end{eqnarray}
and its corresponding multi-component Bell function is
\begin{eqnarray}
\mathcal{I}_{4,3}(18)
&=&\frac{-{\rm \mathbf{v}}_{1}-2{\rm \mathbf{v}}_{2}}{9}A_0B_0C_0D_0+
\frac{-2{\rm \mathbf{v}}_{1}-{\rm \mathbf{v}}_{2}}{9}A_0B_0C_0D_1+
\frac{{\rm \mathbf{v}}_{1}-{\rm \mathbf{v}}_{2}}{9}A_0B_0C_1D_0+
\frac{-{\rm \mathbf{v}}_{1}-2{\rm \mathbf{v}}_{2}}{9}A_0B_0C_1D_1\nonumber\\
& &+\frac{-2{\rm \mathbf{v}}_{1}-{\rm \mathbf{v}}_{2}}{9}A_0B_1C_0D_0+
\frac{-{\rm \mathbf{v}}_{1}+{\rm \mathbf{v}}_{2}}{9}A_0B_1C_0D_1+
\frac{-{\rm \mathbf{v}}_{1}-2{\rm \mathbf{v}}_{2}}{9}A_0B_1C_1D_0+
\frac{-2{\rm \mathbf{v}}_{1}-{\rm \mathbf{v}}_{2}}{9}A_0B_1C_1D_1\nonumber\\
& &+\frac{-2{\rm \mathbf{v}}_{1}-{\rm \mathbf{v}}_{2}}{9}A_1B_0C_0D_0+
\frac{-{\rm \mathbf{v}}_{1}+{\rm \mathbf{v}}_{2}}{9}A_1B_0C_0D_1+
\frac{-4{\rm \mathbf{v}}_{1}+{\rm \mathbf{v}}_{2}}{9}A_1B_0C_1D_0+
\frac{{\rm \mathbf{v}}_{1}-4{\rm \mathbf{v}}_{2}}{9}A_1B_0C_1D_1\nonumber\\
& &+\frac{-{\rm \mathbf{v}}_{1}+{\rm \mathbf{v}}_{2}}{9}A_1B_1C_0D_0+
\frac{{\rm \mathbf{v}}_{1}+2{\rm \mathbf{v}}_{2}}{9}A_1B_1C_0D_1+
\frac{{\rm \mathbf{v}}_{1}-4{\rm \mathbf{v}}_{2}}{9}A_1B_1C_1D_0+
\frac{5{\rm \mathbf{v}}_{1}+4{\rm \mathbf{v}}_{2}}{9}A_1B_1C_1D_1\nonumber\\
&=&\mathcal{I}_{4,3}(2)/.\{ABCD\rightarrow CBAD,b_1\rightarrow {\rm Mod}[b_1+2,3]\}.
\end{eqnarray}
When $c_1=c_2={\rm \mathbf{v}}_{0}$ and $c_1={\rm \mathbf{v}}_{0},c_2={\rm \mathbf{v}}_{1}$, it reduces to $(\mathcal{I}_{3,3}(18))^{0,0}$ and $(\mathcal{I}_{2,3}(18))^{0,1}$, respectively. Since
\begin{eqnarray}
(\mathcal{I}_{3,3}(18))^{0,0}&=&\frac{{\rm \mathbf{v}}_{0}}{3}A_0B_0C_0+
\frac{-{\rm \mathbf{v}}_{2}}{3}A_0B_0C_1+\frac{-{\rm \mathbf{v}}_{1}}{3}A_0B_1C_0+
\frac{{\rm \mathbf{v}}_{0}}{3}A_0B_1C_1\nonumber\\
& &+\frac{-{\rm \mathbf{v}}_{1}}{3}A_1B_0C_0+
\frac{{\rm \mathbf{v}}_{0}}{3}A_1B_0C_1+\frac{{\rm \mathbf{v}}_{2}}{3}A_1B_1C_0+
\frac{2{\rm \mathbf{v}}_{1}}{3}A_1B_1C_1=(\mathcal{I}_{3,3}(17))^{0,0},\nonumber\\
(\mathcal{I}_{3,3}(18))^{0,1}
&=&\frac{-{\rm \mathbf{v}}_{2}}{3}A_0B_0C_0+
\frac{{\rm \mathbf{v}}_{1}}{3}A_0B_0C_1+\frac{{\rm \mathbf{v}}_{0}}{3}A_0B_1C_0+
\frac{-{\rm \mathbf{v}}_{2}}{3}A_0B_1C_1\nonumber\\
& &+\frac{{\rm \mathbf{v}}_{0}}{3}A_1B_0C_0+
\frac{2{\rm \mathbf{v}}_{2}}{3}A_1B_0C_1+\frac{-{\rm \mathbf{v}}_{1}}{3}A_1B_1C_0+
\frac{{\rm \mathbf{v}}_{0}}{3}A_1B_1C_1\nonumber\\
  &=&\mathcal{I}_{3,3}(3)/.\{a_1\rightarrow {\rm Mod}[a_1+2,3],c_1\rightarrow {\rm Mod}[c_1+1,3],B_0\leftrightarrow B_1\},
\end{eqnarray}
we obtain that both of them are equivalent to $\mathcal{I}_{3,3}$.

In the case $n=5$, the corresponding Bell inequalities of $\mathcal{I}_{5,3}(1)$ and $\mathcal{I}_{5,3}(2)$ are
\begin{eqnarray}
& &\mathcal{B}_{5,3}(1)\nonumber\\
&=&\frac{1}{6}[-4P(A_0+B_0+C_0+D_0+E_0=2)+P(A_0+B_0+C_0+D_0+E_1=1)-2P(A_0+B_0+C_0+D_1+E_0=1)\nonumber\\
& &+P(A_0+B_0+C_0+D_1+E_1=2)-2P(A_0+B_0+C_0+D_1+E_1=1)-2P(A_0+B_0+C_1+D_0+E_0=1)\nonumber\\
& &-2P(A_0+B_0+C_1+D_0+E_1=1)+P(A_0+B_0+C_1+D_0+E_1=2)-P(A_0+B_0+C_1+D_1+E_0=0)\nonumber\\
& &-2P(A_0+B_0+C_1+D_1+E_1=2)-2P(A_0+B_1+C_0+D_0+E_0=1)-P(A_0+B_1+C_0+D_0+E_1=0)\nonumber\\
& &+2P(A_0+B_1+C_0+D_1+E_0=0)+P(A_0+B_1+C_0+D_1+E_1=2)+2P(A_0+B_1+C_1+D_0+E_0=0)\nonumber\\
& &+2P(A_0+B_1+C_1+D_0+E_0=0)+P(A_0+B_1+C_1+D_0+E_1=2)-2P(A_0+B_1+C_1+D_1+E_0=2)\nonumber\\
& &-P(A_0+B_1+C_1+D_1+E_1=1)+P(A_1+B_0+C_0+D_0+E_0=2)-P(A_1+B_0+C_0+D_0+E_1=1)\nonumber\\
& &-P(A_1+B_0+C_0+D_1+E_0=1)+P(A_1+B_0+C_0+D_1+E_1=0)-P(A_1+B_0+C_1+D_0+E_0=1)\nonumber\\
& &+P(A_1+B_0+C_1+D_0+E_1=0)+P(A_1+B_0+C_1+D_1+E_0=0)-P(A_1+B_0+C_1+D_1+E_1=2)\nonumber\\
& &-P(A_1+B_1+C_0+D_0+E_0=1)-2P(A_1+B_1+C_0+D_0+E_1=0)-2P(A_1+B_1+C_0+D_1+E_0=0)\nonumber\\
& &-P(A_1+B_1+C_0+D_1+E_1=2)-3P(A_1+B_1+C_0+D_1+E_1=1)-2P(A_1+B_1+C_1+D_0+E_0=0)\nonumber\\
& &+2P(A_1+B_1+C_1+D_0+E_1=2)+3P(A_1+B_1+C_1+D_0+E_1=1)-4P(A_1+B_1+C_1+D_1+E_0=2)\nonumber\\
& &+P(A_1+B_1+C_1+D_1+E_1=1)]+\frac{7}{6}\overset{{\rm LHV}}{\leq}1,
\end{eqnarray}
and
\begin{eqnarray}
& &\mathcal{B}_{5,3}(2)\nonumber\\
&=&\frac{1}{6}[-P(A_0+B_0+C_0+D_0+E_0=2)-2P(A_0+B_0+C_0+D_0+E_1=1)+P(A_0+B_0+C_0+D_1+E_0=1)\nonumber\\
& &+2P(A_0+B_0+C_0+D_1+E_1=0)-4P(A_0+B_0+C_1+D_0+E_0=2)-P(A_0+B_0+C_1+D_0+E_0=1)\nonumber\\
& &+2P(A_0+B_0+C_1+D_0+E_1=2)+2P(A_0+B_0+C_1+D_1+E_0=2)-3P(A_0+B_0+C_1+D_1+E_1=2)\nonumber\\
& &+P(A_0+B_0+C_1+D_1+E_1=1)-2P(A_0+B_1+C_0+D_0+E_0=2)-P(A_0+B_1+C_0+D_0+E_1=0)\nonumber\\
& &+2P(A_0+B_1+C_0+D_1+E_0=0)+P(A_0+B_1+C_0+D_1+E_1=2)-P(A_0+B_1+C_1+D_0+E_0=2)\nonumber\\
& &+3P(A_0+B_1+C_1+D_0+E_0=1)-3P(A_0+B_1+C_1+D_0+E_1=2)-2P(A_0+B_1+C_1+D_0+E_1=1)\nonumber\\
& &-2P(A_0+B_1+C_1+D_1+E_0=1)+P(A_0+B_1+C_1+D_1+E_1=2)-2P(A_0+B_1+C_1+D_1+E_1=1)\nonumber\\
& &+P(A_1+B_0+C_0+D_0+E_0=2)-P(A_1+B_0+C_0+D_0+E_1=1)-P(A_1+B_0+C_0+D_1+E_0=1)\nonumber\\
& &+P(A_1+B_0+C_0+D_1+E_1=0)+P(A_1+B_0+C_1+D_0+E_0=2)-2P(A_1+B_0+C_1+D_0+E_0=1)\nonumber\\
& &-2P(A_1+B_0+C_1+D_0+E_1=2)+P(A_1+B_0+C_1+D_1+E_0=2)+3P(A_1+B_0+C_1+D_1+E_0=1)\nonumber\\
& &-3P(A_1+B_0+C_1+D_1+E_1=2)-4(A_1+B_0+C_1+D_1+E_1=1)\nonumber\\
& &-P(A_1+B_1+C_0+D_0+E_0=1)-2P(A_1+B_1+C_0+D_0+E_1=0)+P(A_1+B_1+C_0+D_1+E_0=0)\nonumber\\
& &+2P(A_1+B_1+C_0+D_1+E_1=2)+4P(A_1+B_1+C_1+D_0+E_0=2)+3P(A_1+B_1+C_1+D_0+E_0=1)\nonumber\\
& &+2P(A_1+B_1+C_1+D_0+E_1=1)+2P(A_1+B_1+C_1+D_1+E_0=1)-4P(A_1+B_1+C_1+D_1+E_1=2)\nonumber\\
& &-P(A_1+B_1+C_1+D_1+E_1=1)]+\frac{1}{2}\overset{{\rm LHV}}{\leq}1,
\end{eqnarray}
respectively.

In Example 3, we focus on the case $d=5$.  Employ the root method in \cite{root method} and the generic form of full-correlated multi-component Bell function, we give a new Bell function $\mathcal{I}_{2,5}(2)$, whose corresponding Bell inequality  is as robust as  the CGLMP inequality, but is not equivalent to it.   The corresponding Bell inequality of $\mathcal{I}_{2,5}(2)$ is
\begin{eqnarray}
\mathcal{B}_{2,5}(2)&=&\frac{1}{4}[-P(A_0+B_0=0)+P(A_0+B_0=1)-2P(A_0+B_0=2)+2P(A_0+B_0=4)\nonumber\\
& &+2P(A_0+B_1=0)-2P(A_0+B_1=2)+P(A_0+B_1=3)-P(A_0+B_1=4)\nonumber\\
& &+2P(A_1+B_0=0)-2P(A_1+B_0=2)+P(A_1+B_0=3)-P(A_1+B_0=4)\nonumber\\
& &+P(A_1+B_1=0)-2P(A_1+B_1=1)+2P(A_1+B_1=3)-P(A_1+B_1=4)]\overset{{\rm{LHV}}}{\leq}1.\ \ \
\end{eqnarray}

In the case $n=3,d=5$.

(i) If $\mathcal{I}_{2,5}^{0,0}=\mathcal{I}_{2,5}(2)$, and $\mathcal{I}_{2,5}^{0,1}$ goes through the $250$ ones equivalent to $\mathcal{I}_{2,5}$, when $\mathcal{I}_{2,5}^{0,1}=\mathcal{I}_{2,5}^{0,1}(1)$,
\begin{eqnarray}
\mathcal{I}_{2,5}^{0,1}(1)&=&\frac{4{\rm \mathbf{v}}_{1}+3{\rm \mathbf{v}}_{2}+2{\rm \mathbf{v}}_{3}+{\rm \mathbf{v}}_{4}}{5}A_0B_0+
\frac{{\rm \mathbf{v}}_{1}-3{\rm \mathbf{v}}_{2}-2{\rm \mathbf{v}}_{3}-{\rm \mathbf{v}}_{4}}{5}(A_0B_1+A_1B_0)
+\frac{-{\rm \mathbf{v}}_{1}+3{\rm \mathbf{v}}_{2}+2{\rm \mathbf{v}}_{3}+{\rm \mathbf{v}}_{4}}{5}A_1B_1\nonumber\\
&=&\mathcal{I}_{2,5}/.\{a_0\rightarrow {\rm Mod}[a_0+1,5],a_1\rightarrow {\rm Mod}[a_1+2,5],B_0\leftrightarrow B_1\},
\end{eqnarray}
we obtain one most robust and party symmetric $\mathcal{I}_{3,5}(1)$,
\begin{eqnarray}
\mathcal{I}_{3,5}(1)&=&\frac{3{\rm \mathbf{v}}_{1}+2{\rm \mathbf{v}}_{2}+2{\rm \mathbf{v}}_{3}+3{\rm \mathbf{v}}_{4}}{5}A_0B_0C_0+
\frac{-{\rm \mathbf{v}}_{2}-3{\rm \mathbf{v}}_{3}-{\rm \mathbf{v}}_{4}}{5}(A_0B_0C_1+A_0B_1C_0+A_1B_0C_0)\nonumber\\
& &
+\frac{-3{\rm \mathbf{v}}_{1}-{\rm \mathbf{v}}_{3}-{\rm \mathbf{v}}_{4}}{5}(A_0B_1C_1+A_1B_0C_1+A_1B_1C_0)+
\frac{{\rm \mathbf{v}}_{1}+{\rm \mathbf{v}}_{2}-2{\rm \mathbf{v}}_{4}}{5}A_1B_1C_1,
\end{eqnarray}
whose coincidence Bell inequality  is
\begin{eqnarray}
\mathcal{B}_{3,5}&=&\frac{1}{4}[-2P(A_0+B_0+C_0=0)+P(A_0+B_0+C_0=1)+P(A_0+B_0+C_0=4)+P(A_0+B_0+C_1=0)\nonumber\\
& &+P(A_0+B_1+C_0=0)+P(A_1+B_0+C_0=0)-2(P(A_0+B_0+C_1=2)+P(A_0+B_1+C_0=2)\nonumber\\
& &+P(A_1+B_0+C_0=2))+P(A_0+B_0+C_1=4)+P(A_0+B_1+C_0=4)+P(A_1+B_0+C_0=4)\nonumber\\
& &+P(A_0+B_1+C_1=0)+P(A_1+B_0+C_1=0)+P(A_1+B_1+C_0=0)+P(A_0+B_1+C_1=3)\nonumber\\
& &+P(A_1+B_0+C_1=3)+P(A_1+B_1+C_0=3)-2(P(A_0+B_1+C_1=4)+P(A_1+B_0+C_1=4)\nonumber\\
& &+P(A_1+B_1+C_0=4))-2P(A_1+B_1+C_1=1)+P(A_1+B_1+C_1=3)+P(A_1+B_1+C_1=4)]\overset{{\rm LHV}}{\leq}1.\ \ \ \ \ \ \ \
\end{eqnarray}
In fact, it is just the Bell inequality in Ref.\cite{Chen324} with $v_c=0.595047$.

(ii) If $\mathcal{I}_{2,5}^{0,0}=\mathcal{I}_{2,5}(2)$, and $\mathcal{I}_{2,5}^{0,1}$ goes through the $250$ ones equivalent to $\mathcal{I}_{2,5}(2)$, then in the case that $\mathcal{I}_{2,5}^{0,1}=\mathcal{I}_{2,5}^{0,1}(2)$,
\begin{eqnarray}
\mathcal{I}_{2,5}^{0,1}(2)&=&\frac{-3{\rm \mathbf{v}}_{1}-{\rm \mathbf{v}}_{2}-4{\rm \mathbf{v}}_{3}-2{\rm \mathbf{v}}_{4}}{5}A_0B_0+
\frac{-2{\rm \mathbf{v}}_{1}+{\rm \mathbf{v}}_{2}-{\rm \mathbf{v}}_{3}-3{\rm \mathbf{v}}_{4}}{5}(A_0B_1+A_1B_0)
+\frac{2{\rm \mathbf{v}}_{1}-{\rm \mathbf{v}}_{2}+{\rm \mathbf{v}}_{3}-2{\rm \mathbf{v}}_{4}}{5}A_1B_1\nonumber\\
&=&\mathcal{I}_{2,5}(2)/.\{b_0\rightarrow {\rm Mod}[b_0+1,5],B_0\leftrightarrow B_1\},
\end{eqnarray}
 we obtain one most robust $\mathcal{I}_{3,5}(2)$,
\begin{eqnarray}
\mathcal{I}_{3,5}(2)&=&\frac{-{\rm \mathbf{v}}_{2}-3{\rm \mathbf{v}}_{3}-{\rm \mathbf{v}}_{4}}{5}A_0B_0C_0+
\frac{3{\rm \mathbf{v}}_{1}+2{\rm \mathbf{v}}_{2}+2{\rm \mathbf{v}}_{3}+3{\rm \mathbf{v}}_{4}}{5}A_0B_0C_1+
\frac{-3{\rm \mathbf{v}}_{1}-{\rm \mathbf{v}}_{3}-{\rm \mathbf{v}}_{4}}{5}A_0B_1C_0\nonumber\\
& &+
\frac{-{\rm \mathbf{v}}_{2}-3{\rm \mathbf{v}}_{3}-{\rm \mathbf{v}}_{4}}{5}A_0B_1C_1
+\frac{-3{\rm \mathbf{v}}_{1}-{\rm \mathbf{v}}_{3}-{\rm \mathbf{v}}_{4}}{5}A_1B_0C_0+
\frac{-{\rm \mathbf{v}}_{2}-3{\rm \mathbf{v}}_{3}-{\rm \mathbf{v}}_{4}}{5}A_1B_0C_1\nonumber\\
& &+
\frac{{\rm \mathbf{v}}_{1}+{\rm \mathbf{v}}_{2}-2{\rm \mathbf{v}}_{4}}{5}A_1B_1C_0
+
\frac{-3{\rm \mathbf{v}}_{1}-{\rm \mathbf{v}}_{3}-{\rm \mathbf{v}}_{4}}{5}A_1B_1C_1=\mathcal{I}_{3,5}(1)/.\{C_0\leftrightarrow C_1\},
\end{eqnarray}
whose coincidence Bell inequality also has $v_c=0.595047$.

(iii) If $\mathcal{I}_{2,5}^{0,0}=\mathcal{I}_{2,5}$, and $\mathcal{I}_{2,5}^{0,1}$ goes through the $250$ ones equivalent to $\mathcal{I}_{2,5}(2)$,  then for $\mathcal{I}_{2,5}^{0,1}=\mathcal{I}_{2,5}^{0,1}(3)$,
\begin{eqnarray}
\mathcal{I}_{2,5}^{0,1}(3)&=&\frac{-3{\rm \mathbf{v}}_{1}-{\rm \mathbf{v}}_{2}-4{\rm \mathbf{v}}_{3}-2{\rm \mathbf{v}}_{4}}{5}A_0B_0+
\frac{3{\rm \mathbf{v}}_{1}+{\rm \mathbf{v}}_{2}-{\rm \mathbf{v}}_{3}+2{\rm \mathbf{v}}_{4}}{5}(A_0B_1+A_1B_0)
+\frac{2{\rm \mathbf{v}}_{1}-{\rm \mathbf{v}}_{2}+{\rm \mathbf{v}}_{3}+3{\rm \mathbf{v}}_{4}}{5}A_1B_1\nonumber\\
&=&\mathcal{I}_{2,5}(2)/.\{a_1\rightarrow {\rm Mod}[a_1+4,5],B_0\leftrightarrow B_1\},
\end{eqnarray}
we obtain one most robust $\mathcal{I}_{3,5}(3)$,
\begin{eqnarray}
\mathcal{I}_{3,5}(3)&=&\frac{-3{\rm \mathbf{v}}_{1}-{\rm \mathbf{v}}_{3}-{\rm \mathbf{v}}_{4}}{5}A_0B_0C_0+
\frac{-{\rm \mathbf{v}}_{1}-3{\rm \mathbf{v}}_{2}-{\rm \mathbf{v}}_{3}}{5}A_0B_0C_1+
\frac{-{\rm \mathbf{v}}_{2}-3{\rm \mathbf{v}}_{3}-{\rm \mathbf{v}}_{4}}{5}A_0B_1C_0+
\frac{-{\rm \mathbf{v}}_{1}-{\rm \mathbf{v}}_{2}-3{\rm \mathbf{v}}_{4}}{5}A_0B_1C_1\nonumber\\
& &
+\frac{-{\rm \mathbf{v}}_{2}-3{\rm \mathbf{v}}_{3}-{\rm \mathbf{v}}_{4}}{5}A_1B_0C_0+
\frac{-{\rm \mathbf{v}}_{1}-{\rm \mathbf{v}}_{2}-3{\rm \mathbf{v}}_{4}}{5}A_1B_0C_1+
\frac{3{\rm \mathbf{v}}_{1}+2{\rm \mathbf{v}}_{2}+2{\rm \mathbf{v}}_{3}+3{\rm \mathbf{v}}_{4}}{5}A_1B_1C_0
\nonumber\\
& &+
\frac{-2{\rm \mathbf{v}}_{1}+{\rm \mathbf{v}}_{3}+{\rm \mathbf{v}}_{4}}{5}A_1B_1C_1=\mathcal{I}_{3,5}(1)/.\{c_1\rightarrow {\rm Mod}[c_1+1,5],C_0\leftrightarrow C_1\}.
\end{eqnarray}

In the case that $n=2,d=7$, we find three full-correlated multi-component Bell functions with the popular properties: (i) their corresponding Bell inequalities have the save possible values in LHV theory as the CGLMP one for case $d=7$; (ii) they have the same $v_c$ as the CGLMP one; (iii) they are not equivalent to each other, and are not equivalent to the CGLMP one. Now, we list the corresponding Bell inequalities $\mathcal{I}_{2,7}(2),\mathcal{I}_{2,7}(3),$ and $\mathcal{I}_{2,7}(4)$,
\begin{eqnarray}
\mathcal{B}_{2,7}(2)
&=&\frac{1}{6}[P(A_0+B_0=0)-P(A_0+B_0=1)-3P(A_0+B_0=2)+2P(A_0+B_0=3)\nonumber\\
& &-2P(A_0+B_0=5)+3P(A_0+B_0=6)+3P(A_0+B_1=0)-3P(A_0+B_1=1)\nonumber\\
& &-P(A_0+B_1=2)+P(A_0+B_1=3)+3P(A_0+B_1=4)-2P(A_0+B_1=5)\nonumber\\
& &+3P(A_1+B_0=0)-3P(A_1+B_0=1)-P(A_1+B_0=2)+P(A_1+B_0=3)\nonumber\\
& &+3P(A_1+B_0=4)-2P(A_1+B_0=5)+P(A_1+B_1=0)+P(A_1+B_1=1)\nonumber\\
& &-3P(A_1+B_1=2)+2P(A_1+B_1=3)-2P(A_1+B_1=5)+3P(A_1+B_1=6)]\overset{{\rm{LHV}}}{\leq}1,\ \ \ \ \nonumber\\
\mathcal{B}_{2,7}(3)
&=&\frac{1}{6}[3P(A_0+B_0=0)-3P(A_0+B_0=1)-P(A_0+B_0=2)+P(A_0+B_0=3)\nonumber\\
& &+3P(A_0+B_0=4)-2P(A_0+B_0=5)+2P(A_0+B_1=0)-2P(A_0+B_1=2)\nonumber\\
& &+3P(A_0+B_1=3)+P(A_0+B_1=4)-P(A_0+B_1=5)-3P(A_0+B_1=6)\nonumber\\
& &+2P(A_1+B_0=0)-2P(A_1+B_0=2)+3P(A_1+B_0=3)+P(A_1+B_0=4)\nonumber\\
& &-P(A_1+B_0=5)-3P(A_1+B_0=6)+2P(A_1+B_1=1)-3P(A_1+B_1=2)\nonumber\\
& &-P(A_1+B_1=3)+P(A_1+B_1=4)+3P(A_1+B_1=5)-2P(A_1+B_1=6)]\overset{{\rm{LHV}}}{\leq}1,\ \ \ \ \nonumber\\
\mathcal{B}_{2,7}(4)
&=&\frac{1}{6}[P(A_0+B_0=0)-2P(A_0+B_0=1)+2P(A_0+B_0=2)-P(A_0+B_0=3)\nonumber\\
& &+3P(A_0+B_0=4)-3P(A_0+B_0=6)-3P(A_0+B_1=0)+3P(A_0+B_1=2)\nonumber\\
& &-P(A_0+B_1=3)+2P(A_0+B_1=4)-2P(A_0+B_1=5)+P(A_0+B_1=6)\nonumber\\
& &-3P(A_1+B_0=0)+3P(A_1+B_0=2)-P(A_1+B_0=3)+2P(A_1+B_0=4)\nonumber\\
& &-2P(A_1+B_0=5)+P(A_1+B_0=6)-3P(A_1+B_1=0)+P(A_1+B_1=1)\nonumber\\
& &-2P(A_1+B_1=2)+P(A_1+B_1=3)-P(A_1+B_1=4)+3P(A_1+B_1=5)]\overset{{\rm{LHV}}}{\leq}1.\ \ \ \
\end{eqnarray}


\begin{thebibliography}{10}

\bibitem{RMP} N. Brunner, D. Cavalcanti, S. Pironio, V. Scarani, and S.
Wehner,  \href{https://doi.org/10.1103/RevModPhys.86.419}{Rev. Mod. Phys. \textbf{86}, 419 (2014)}.

\bibitem{Nonlocality} V. Scarani, (Oxford University Press,
New York, 2019).

\bibitem{Bell2} J. S. Bell,  \href{https://doi.org/10.1103/PhysicsPhysiqueFizika.1.195}{Phys. Phys. Fiz. \textbf{1}, 195 (1964)}.

\bibitem{CHSH} J. F. Clauser, M. A. Horne, A. Shimony, and R. A. Holt,
\href{https://doi.org/10.1103/PhysRevLett.23.880}{Phys. Rev. Lett. \textbf{23}, 880 (1969)}.

\bibitem{M} N. D. Mermin, \href{https://doi.org/10.1103/PhysRevLett.65.1838}{Phys. Rev. Lett. \textbf{65}, 1838 (1990)}.

\bibitem{A} M. Ardehali,
         \href{https://link.aps.org/doi/10.1103/PhysRevA.46.5375}{{Phys. Rev. A} \textbf{46}, 5375 (1992)}.

\bibitem{BK} A.V. Belinski{\u{\i}} and D. N. Klyshko, \href{https://doi.org/10.1070/PU1993v036n08ABEH002299}{Phys. Usp. \textbf{36}, 653
(1993)}.



\bibitem{Bell's theorem} M. \.{Z}ukowski, and ${\rm\check{C}}$. Brukner,  \href{https://journals.aps.org/prl/pdf/10.1103/PhysRevLett.88.210401}{Phys. Rev. Lett. \textbf{88} 210401 (2002)}.

\bibitem{CGLMP} D. Collins, N. Gisin, N. Linden, S. Massar, and S. Popescu, \href{https://doi.org/10.1103/PhysRevLett.88.040404}{Phys. Rev. Lett. \textbf{88}, 040404 (2002)}.


\bibitem{qutrits1} J. L. Chen, D. Kaszlikowski, L. C. Kwek , and C. H. Oh,  \href{https://www.worldscientific.com/doi/abs/10.1142/S0217732302008885}{Mod. Phys. Lett. A \textbf{17} 2231 (2002)}.


\bibitem{Coincidence}
A. Ac\'{\i}n, J. L. Chen, N. Gisin, D. Kaszlikowski, L. C. Kwek, C. H. Oh, and M. ${\rm\dot{Z}}$ukowski,
\href{https://doi.org/10.1103/PhysRevLett.92.250404}{Phys. Rev. Lett. \textbf{92}, 250404 (2004)}.


\bibitem{Multicomponent} J. L. Chen, Chunfeng Wu,  L. C. Kwek,  D. Kaszlikowski,  M. ${\rm{\dot{Z}}}$ukowski,  and C. H. Oh,  \href{https://doi.org/10.1103/PhysRevA.71.032107}{Phys. Rev. A \textbf{71}, 032107 (2005)}.


\bibitem{Chen324} J. L. Chen, Chunfeng Wu, L. C. Kwek, and C. H. Oh,
\href{https://doi.org/10.1103/PhysRevA.78.032107}{Phys. Rev. A \textbf{78}, 032107 (2008)}.


\bibitem{Mermin3} J. Lawrence, \href{https://doi.org/10.1103/PhysRevA.95.042123}{Phys. Rev. A, \textbf{95}, 042123 (2017)}.

 \bibitem{Platonic} K. F. P${\rm\acute{a}}$l, and T. V${\rm\acute{e}}$rtesi,
 \href{https://doi.org/10.22331/q-2022-07-07-756}{Quantum, \textbf{6}, 756 (2022)}.



\bibitem{Questions}    N. Gisin,
\href{https://doi.org/10.48550/arXiv.quant-ph/0702021}{arXiv:quant-ph/0702021 (2007)}.

\bibitem{structure} G. Schachner,  \href{https://browse.arxiv.org/pdf/quant-ph/0312117.pdf}{arXiv:quant-ph/0312117 (2003)}.

\bibitem{L. B. Fu} L. B. Fu,  \href{https://doi.org/10.1103/PhysRevLett.92.130404}{Phys. Rev. Lett. \textbf{92}, 130404 (2004)}.

\bibitem{SLK} W. Son, Jinhyoung Lee, and M.S. Kim, \href{https://doi.org/10.1103/PhysRevLett.96.060406}{Phys. Rev. Lett. \textbf{96}, 060406 (2006)}.


\bibitem{SCL}    S. W. Lee,  Y. W. Cheong, and Jinhyoung Lee,  \href{https://doi.org/10.1103/PhysRevA.76.032108}{Phys. Rev. A \textbf{76}, 032108 (2007)}.

    \bibitem{Avenues} M. Karczewski, G. Scala, A. Mandarino, A. B. Sainz, M. Zukowski,  \href{https://doi.org/10.1088/1751-8121}{J. Phys. A-Math. Theor.  \textbf{55}, 384011 (2022)}.

\bibitem{XYFan} X. Y. Fan, Z. P. Xu, J. L. Miao, H. Y. Liu, Y. J. Liu, W. M. Shang, J. Zhou, H. X. Meng, O. G${\rm {\ddot{u}}}$hne, J. L. Chen,
\href{https://doi.org/10.48550/arXiv.2109.05521}{quant-ph/2109.05521 (2023)}.


 \bibitem{Tightn22} R. F. Werner, and M. M. Wolf,  \href{https://doi.org/10.1103/PhysRevA.64.032112}{Phys. Rev. A \textbf{64}, 032112 (2001)}.

 \bibitem{Lluis Masanes} L. Masanes, \href{https://dblp.org/rec/journals/qic/Masanes03.html}{Quantum Inf. Comput. \textbf{3}, 345 (2003)}.

        \bibitem{Tightn2d} J. L. Chen, and D. L. Deng,
\href{https://doi.org/10.1103/PhysRevA.79.012111}{Phys. Rev. A  \textbf{79}, 012111 (2009)}


\bibitem{completed} F. Arnault,
\href{https://doi.org/10.1088/1751-8113/45/25/255304}{J. Phys. A-Math. Theor. \textbf{45}, 255304 (2012)}.


 \bibitem{relevant22} D. Collins, and N. Gisin,
  \href{https://doi.org/10.1088/0305-4470/37/5/021}{J. Phys. A-Math. Theor. \textbf{37}, 1775 (2004)}.

 \bibitem{relevantk23} D. L. Deng, Z. S. Zhou, and J. L. Chen, \href{https://www.sciencedirect.com/science/article/pii/S0003491609000980}{Anna. Phys.  \textbf{324}, 1996 (2009)}.

     \bibitem{3n2} T. V${\rm{\acute{e}}}$rtesi and K. F. P${\rm{\acute{a}}}$l,  \href{https://doi.org/10.1103/PhysRevA.84.042122}{Phys. Rev. A \textbf{84}, 042122 (2011)}.

\bibitem{n23} H. X. Meng, Z. Y. Li, X. Y. Fan, J. L. Miao, H. Y. Liu, Y. J. Liu, W. M. Shang, J. Zhou, and J. L. Chen,  \href{https://doi.org/10.1103/PhysRevA.105.062215}{Phys. Rev. A \textbf{105}, 062215 (2022)}.


 \bibitem{Quantum cryptography}   D. Kaszlikowski, D. K. L. Oi, M. Christandl, K. Chang, A. Ekert, L. C. Kwek, and C. H. Oh,  \href{https://doi.org/10.1103/PhysRevA.67.012310}{Phys. Rev. A \textbf{67}, 012310 (2003)}.

        \bibitem{QKD}  A. Ac\'{\i}n, N. Gisin, and  L. Masanes,  \href{https://doi.org/10.1103/PhysRevLett.97.120405}{Phys. Rev. Lett. \textbf{97}, 120405 (2006)}.

   \bibitem{3bit2}H. Chau Nguyen, and O. G\"{u}hne,  \href{https://doi.org/10.1103/PhysRevLett.125.230402}{Phys. Rev. Lett. \textbf{125}, 230402 (2020)}.


\bibitem{Strong nonlocality} F. Shi, Z. Ye, L. Chen, and X. D. Zhang,    \href{https://doi.org/10.1103/PhysRevA.105.022209}{Phys. Rev. A \textbf{105}, 022209 (2022)}.


\bibitem{Optimal measurements}S. Schwarz, S. Wolf, and A. Montina, \href{https://doi.org/10.1103/PhysRevA.94.022322}{Phys. Rev. A \textbf{94}, 022322 (2016)}.

\bibitem{splitter1} M. ${\rm\dot{Z}}$ukowski, A. Zeilinger, and M. A. Horne,
 \href{https://doi.org/10.1103/PhysRevA.55.2564}{Phys. Rev. A \textbf{55}, 2564 (1997)}.

\bibitem{splitter2} R. T. Thew, A. Ac\'{\i}n, H. Zbinden, and N. Gisin, \href{https://arxiv.org/abs/quant-ph/0307122}{quant-ph/0307122 (2003)}.


\bibitem{root method}  J. L. Chen, and D. L. Deng, \href{https://doi.org/10.1103/PhysRevA.79.012115}{Phys. Rev. A \textbf{79}, 012115 (2009)}.


\bibitem{Hardy1}D. Boschi, S. Branca, F. De Martini, and L. Hardy, \href{https://journals.aps.org/prl/abstract/10.1103/PhysRevLett.79.2755}{Phys. Rev. Lett. \textbf{79}, 2755 (1997)}.


\bibitem{Hardy3} J. L. Chen, A. Cabello, Z. P. Xu, H. Y. Su, C. Wu, and L. C.
Kwek, \href{https://doi.org/10.1103/PhysRevA.88.062116}{Phys.
Rev. A \textbf{88}, 062116 (2013)}.

\bibitem{Hardy4} S. H. Jiang, Z. P. Xu, H. Y. Su, A. K. Pati, and J. L. Chen,
 \href{https://doi.org/10.1103/PhysRevA.120.050403}{Phys. Rev. Lett. \textbf{120}, 050403 (2018)}.

\bibitem{Hardy's paradox1} H. X. Meng, J. Zhou, Z. P. Xu, H. Y. Su, T. Gao,  F. L. Yan,  and J. L. Chen, \href{https://doi.org/10.1103/PhysRevA.98.062103}{Phys. Rev. A \textbf{98}, 062103 (2018)}.

\bibitem{Stronger Hardy-type paradox} M. Yang, H. X. Meng, J. Zhou, Z. P. Xu,  Y. Xiao, K. Sun, J. L. Chen,
J. S. Xu, C. F. Li, and G. C. Guo, \href{https://doi.org/10.1103/PhysRevA.99.032103}{Phys. Rev. A \textbf{99}, 032103 (2019)}.

\bibitem{Hardy2} S. Abramsky, and L. Hardy, \href{https://doi.org/10.1103/PhysRevA.85.062114}{Phys. Rev. A \textbf{85}, 062114 (2012)}.



\bibitem{Quantum Communication} D. Cozzolino, B. D. Lio, D. Bacco, L. K. Oxenlwe,  \href{https://onlinelibrary.wiley.com/doi/10.1002/qute.201900038}{Adv. Quantum Technol. \textbf{2}, 1900038 (2019)
}.


\bibitem{quantum computing} J. Mackeprang, D. Bhatti, M. J. Hoban, S. Barz,  \href{https://doi.org/10.1088/1367-2630/acdf77}{New J. Phys. \textbf{25}, 073007 (2023)}.


\bibitem{machine learning} Y. C.  Ma, and M. H. Yung,  \href{https://doi.org/10.1038/s41534-018-0081-3}{npj Quantum Information \textbf{4}:34 (2018)}.

\bibitem{Quantum Networks1}P. Contreras-Tejada, C. Palazuelos, and J. I. de Vicente,     \href{https://doi.org/10.1103/PhysRevLett.126.040501}{Phys. Rev. Lett. \textbf{126}, 040501 (2021)}.

    \bibitem{Quantum Networks2} L. H. Yang, X. F. Qi, and J. H. Hou, \href{https://doi.org/10.1103/PhysRevA.104.042405}{Phys. Rev. A \textbf{104}, 042405 (2021)}.

\bibitem{Quantum Networks3} A. G. Lamas, and E. Chitambar, \href{https://doi.org/10.1103/PhysRevLett.130.240802}{Phys. Rev. Lett. \textbf{130}, 240802 (2023)}.

\bibitem{Quantum Networks4}Q. Zhou, X. Y. Xu, S. M. Hu, S. Zhao, S. X. Yu, L. Li,
N. L. Liu,  and K. Chen,    \href{https://doi.org/10.1103/PhysRevA.107.052416}{Phys. Rev. A \textbf{107}, 052416 (2023)}.

	
\end{thebibliography}
\end{document}